\documentclass[twocolappendix, numberedappendix, iop]{emulateapj}
\usepackage{amsmath,amssymb}
\usepackage{xspace}
\usepackage{graphicx}
\usepackage{graphicx}
\usepackage{epsfig}
\usepackage{hyperref} %hyperlinks
\usepackage{xcolor}
\usepackage{array,longtable}
\usepackage{threeparttable}
\usepackage{gensymb}

\slugcomment{}
\shorttitle{X-ray Variability of Typical Distant AGNs}
\shortauthors{Yang et~al.}

\def\gsim{\mathrel{\rlap{\lower4pt\hbox{\hskip1pt$\sim$}}
    \raise1pt\hbox{$>$}}}
\def\lsim{\mathrel{\rlap{\lower4pt\hbox{\hskip1pt$\sim$}}
    \raise1pt\hbox{$<$}}}
\def\nh{$N_{\rm{H}}$}
\def\lx{$L_{\rm{X}}$}
\def\sigmaexc{${\sigma^2_{\rm exc}}$}

\def\sigmaexcL{${\sigma^2_{\mathrm{exc}, L}}$}
\def\sigmaexcLr{${\sigma^2_{\mathrm{exc}, L, \mathrm{rest}}}$}
\def\sigmaexcN{${\sigma^2_{\mathrm{exc}, N}}$}

\def\trest{\Delta t_{\rm rest}}
\def\nubf{\nu_{\rm bf}}
\def\psdamp{\mathrm{PSD_{amp}}}

\def\chandra{{\it Chandra\/}}  
  
\def\xmm{\hbox{\it XMM-Newton\/}}
\def\nustar{{\it NuSTAR\/}}
\def\athena{{\it Athena\/}}

\def\xray{\hbox{X-ray}}  %box words
\def\cdfs{\hbox{CDF-S}}

\def\zs{\textit{z$_{\rm spec}$}}
\def\zp{\textit{z$_{\rm phot}$}}

\def\xidbal{\hbox{J033209.4--274806}}
\def\xidone{\hbox{J033226.5--274035}}
\def\xidtwo{\hbox{J033259.7--274626}}
\def\xidthr{\hbox{J033229.9--274530}}
\def\xidctk{\hbox{J033218.3--275055}}

\def\Xidone{J033226.5--274035}
\def\Xidtwo{J033259.7--274626}
\def\Xidthr{J033229.9--274530}
\def\Xidctk{J033218.3--275055}

\begin{document}

\title{Long-Term X-RAY Variability of Typical Active Galactic Nuclei in the Distant Universe}
\author{G. Yang\altaffilmark{1,2},
W.~N. Brandt\altaffilmark{1,2,3},
B. Luo\altaffilmark{4,1},
Y.~Q. Xue\altaffilmark{5},
F.~E. Bauer\altaffilmark{6, 7, 8, 9},
M.~Y. Sun\altaffilmark{10,1,5},
S. Kim\altaffilmark{6},
S. Schulze\altaffilmark{7,6},
X.~C. Zheng\altaffilmark{5}, 
M. Paolillo\altaffilmark{11,12},
O. Shemmer\altaffilmark{13}, 
T. Liu\altaffilmark{5},
D.~P. Schneider\altaffilmark{1,2},
C. Vignali\altaffilmark{14},
F. Vito\altaffilmark{1,2},
J.-X. Wang\altaffilmark{5}
}

\altaffiltext{1}{Department of Astronomy and Astrophysics, 525 Davey Lab, The Pennsylvania State University, University Park, PA 16802, USA; gxy909@psu.edu}
\altaffiltext{2}{Institute for Gravitation and the Cosmos, The Pennsylvania State University, University Park, PA 16802, USA}
\altaffiltext{3}{Department of Physics, 104 Davey Laboratory, The Pennsylvania State University, University Park, PA 16802, USA}
\altaffiltext{4}{School of Astronomy \& Space Science, Nanjing University, Nanjing 210093, China}
\altaffiltext{5}{CAS Key Laboratory for Researches in Galaxies and Cosmology, Center for Astrophysics, Department of Astronomy, University of Science and Technology of China, Chinese Academy of Sciences, Hefei, Anhui 230026, China}
\altaffiltext{6}{Instituto de Astrof\'{\i}sica, Facultad de F\'{i}sica, Pontificia Universidad Cat\'{o}lica de Chile, Casilla 306, Santiago 22, Chile}
\altaffiltext{7}{Millennium Institute of Astrophysics, Nuncio Monse\~{n}or S\'{o}tero Sanz 100, Providencia, Santiago, Chile}
\altaffiltext{8}{Space Science Institute, 4750 Walnut Street, Suite 205, Boulder, Colorado 80301}
\altaffiltext{9}{EMBIGGEN Anillo, Concepci\'{o}n, Chile}
\altaffiltext{10}{Department of Astronomy and Institute of Theoretical Physics and Astrophysics, Xiamen University, Xiamen, Fujian 361005, China}
\altaffiltext{11}{Dipartimento di Fisica, Universitá Federico II, Napoli 80126, Italy}
\altaffiltext{12}{ASI Science Data Center, via del Politecnico snc, Roma 80126, Italy}
\altaffiltext{13}{Department of Physics, University of North Texas, Denton, TX 76203, USA}
\altaffiltext{14}{Universita di Bologn\'{a}, Via Ranzani 1, Bologna, Italy}

%======================Abstract=================================
\begin{abstract}
We perform long-term (\hbox{$\approx 15$~yr}, observed-frame) \xray\ 
variability analyses of the 68 brightest \hbox{radio-quiet} active 
galactic nuclei (AGNs) in the 6 Ms \chandra\ Deep Field-South 
(\cdfs) survey; 
the majority are in the redshift range of \hbox{0.6--3.1},
providing access to penetrating rest-frame \xray s up to 
\hbox{$\approx10-30$~keV}. 
Twenty-four of the 68 sources are optical spectral type~I 
AGNs, and the rest (44) are type II AGNs.
The time scales probed in this work are among the longest 
for \xray\ variability studies of distant AGNs. 
Photometric analyses reveal widespread \hbox{photon-flux} variability: 
$90\%$\ of AGNs are variable above a 95\% confidence level,
including many \xray\ obscured AGNs and several optically classified 
type II quasars.
We characterize the intrinsic \xray\ luminosity (\lx) 
and absorption (\nh) variability via spectral fitting.
Most (74\%) sources show \lx\ variability; 
the variability amplitudes are generally 
smaller for quasars. A Compton-thick candidate AGN shows variability 
of its high-energy \xray\ flux, indicating the size of reflecting 
material to be $\lsim 0.3$~pc.
\lx\ variability is also detected in a 
broad absorption line (BAL) quasar.
The \nh\ variability amplitude for our sample appears to rise 
as time separation increases.
About 16\% of sources show \nh\ variability. 
One source transitions from an \xray\ 
unobscured to obscured state while its optical classification 
remains type I; this behavior indicates the \xray\ eclipsing 
material is not large enough to obscure the whole 
broad-line region.
\end{abstract}

\keywords{galaxies: active -- galaxies: nuclei -- X-rays: galaxies --
          quasars: general -- X-rays: general -- methods: data analysis }

%======================Main Text=================================
\section{Introduction}\label{sec:intro}
Variability studies are valuable in probing the physical 
properties of active galactic nuclei (AGNs) 
\citep[e.g.,][]{ulrich97, peterson01}. 
Simple light-travel time arguments enable a 
first-order estimation of the sizes of the 
radiation-emitting regions. 
More detailed reverberation-based studies provide 
size estimates of different components; e.g.,
the correlations among multi-color light curves
give temperature profiles of accretion disks 
\citep[e.g., ][]{fausnaugh15}; time lags between the
continuum and emission lines indicate the radius of 
broad-line region (BLR) \citep[e.g.,][]{peterson14};
and delays of the near-infrared (NIR) dust emission 
compared to the optical disk emission constrain 
the inner size of the dusty torus \citep[e.g.,][]{koshida14}.
Changes in absorption provide insights into the absorbing matter: 
e.g., variability of broad absorption line (BAL) troughs 
reveals wind properties \citep[e.g.,][]{filiz13}, and
variations of optical reddening and \xray\
absorption indicate a clumpy nature of 
the ambient gas \citep[e.g.,][]{goodrich95, netzer15}.
In AGN jet studies, the rapid variability of blazars'
\hbox{$\gamma$-ray} emission often indicates small emitting regions
and the relativistically beamed nature of the 
radiation \citep[e.g.,][]{begelman08, abdo10}. 

X-ray variability is of great importance among
AGN variability studies. In AGN spectral energy 
distributions (SEDs), luminous \xray\ emission is almost 
universal and often a significant contributor
to the total source power \citep[e.g.,][]{gibson08}. \xray\ 
variability is generally of larger amplitude 
and more rapid compared 
to that at longer wavelengths \citep[e.g.,][]{ulrich97, peterson01},
indicating that the high-energy radiation is probing 
the immediate vicinity of the 
supermassive black hole (SMBH). For 
the majority population of obscured AGNs,
the penetrating nature of \xray s often allows 
variability studies of emission as well as 
absorption \citep[e.g.,][]{puccetti14, hernandez15}. 

Intensive \xray\ variability analyses have been performed for the 
radio-quiet AGNs that are the majority population. 
Studies of broadband \xray\ continuum variability have
found that local \hbox{Seyfert 1s} (i.e., unabsorbed 
AGNs) are highly \xray\ variable, and that the variation amplitude
generally decreases as luminosity increases 
\citep[e.g.,][]{nandra97, ponti12}.
This amplitude-luminosity relation might be a byproduct of 
a primary amplitude-SMBH mass relation 
\citep[e.g.,][and references therein]{ponti12}.
\hbox{Seyfert 2s}, especially the Compton-thick 
(obscuration column density \nh\ 
$\gsim 1.5\times10^{24}$~cm$^{-2}$) ones, tend to be 
less \xray\ variable than \hbox{Seyfert 1s}\ at least
on time scales of hours or less
\citep[e.g., ][]{turner97, awaki06, hernandez15}.
Studies of distant AGNs, such as optically selected 
quasars, have also reported prevalent 
\xray\ variability (e.g., \citealt{gibson12}, G12 hereafter; 
\citealt{shemmer14}) and a similar amplitude-luminosity 
relation as seen in local Seyfert galaxies 
\citep[e.g.,][]{almaini00, paolillo04, papadakis08, lanzuisi14}. 
Compared to the \xray\ continuum, 
the narrow Fe K emission-line components are 
less variable \citep[e.g.,][]{markowitz03}, consistent 
with the picture that they originate from the outer disk 
or torus. Recent works 
\citep[e.g.,][]{zoghbi12, kara13a}
reveal short-timescale (minutes) lags  
of the broad Fe K emission-line components 
relative to the continuum,
suggesting they are emitted from the inner disk illuminated by 
the continuum \xray s. 
\xray\ photoelectric absorption 
variability has also been found in many \hbox{Seyfert 2s}
\citep[e.g.,][]{risaliti02, risaliti07}.
Modeling of the observed absorption variations 
suggests this absorption originates in 
gas clouds in the BLR or torus orbiting the central source
\citep[e.g.,][]{maiolino10, markowitz14, 
torricelli14, netzer15}. \xray\ absorption variability
has also been widely investigated in warm absorbers, 
BAL quasar winds,
and ultra fast outflows (UFOs), usefully characterizing wind
properties \citep[e.g., ][]{chartas09, matt11, saez12, king15, 
scott15}. 

The \xray\ variability of AGNs generally shows a \hbox{red-noise} 
power spectral density \citep[PSD, e.g.,][]{uttley02}; i.e., 
AGNs vary by larger amplitudes 
on longer time scales. \xray\ variability analyses on longer 
time scales thus have a better chance of detecting variability 
in data of a given signal-to-noise ratio, thereby providing physical 
insights about the nature of the AGNs. Long-term variability 
studies are also helpful in evaluating the effects of AGN 
variability upon statistical inferences made about source 
populations in single-epoch X-ray surveys. Furthermore, long-term 
X-ray variability studies might capture novel AGN phenomena; 
e.g., ``changing-look'' events \citep[e.g.,][]{matt03, ricci16}, 
emission-state changes \citep[e.g.,][]{miniutti12}, and 
torus-eclipse events \citep[e.g.,][]{markowitz14}.
Variability analyses on time scales of years can 
probe the regime of mechanical instabilities of the
accretion disk (assuming a \hbox{$\sim10^8\ \rm M_{\sun}$}\ 
SMBH; see, e.g., \citealt{peterson01}).

\begin{figure}
\includegraphics[width=\linewidth]{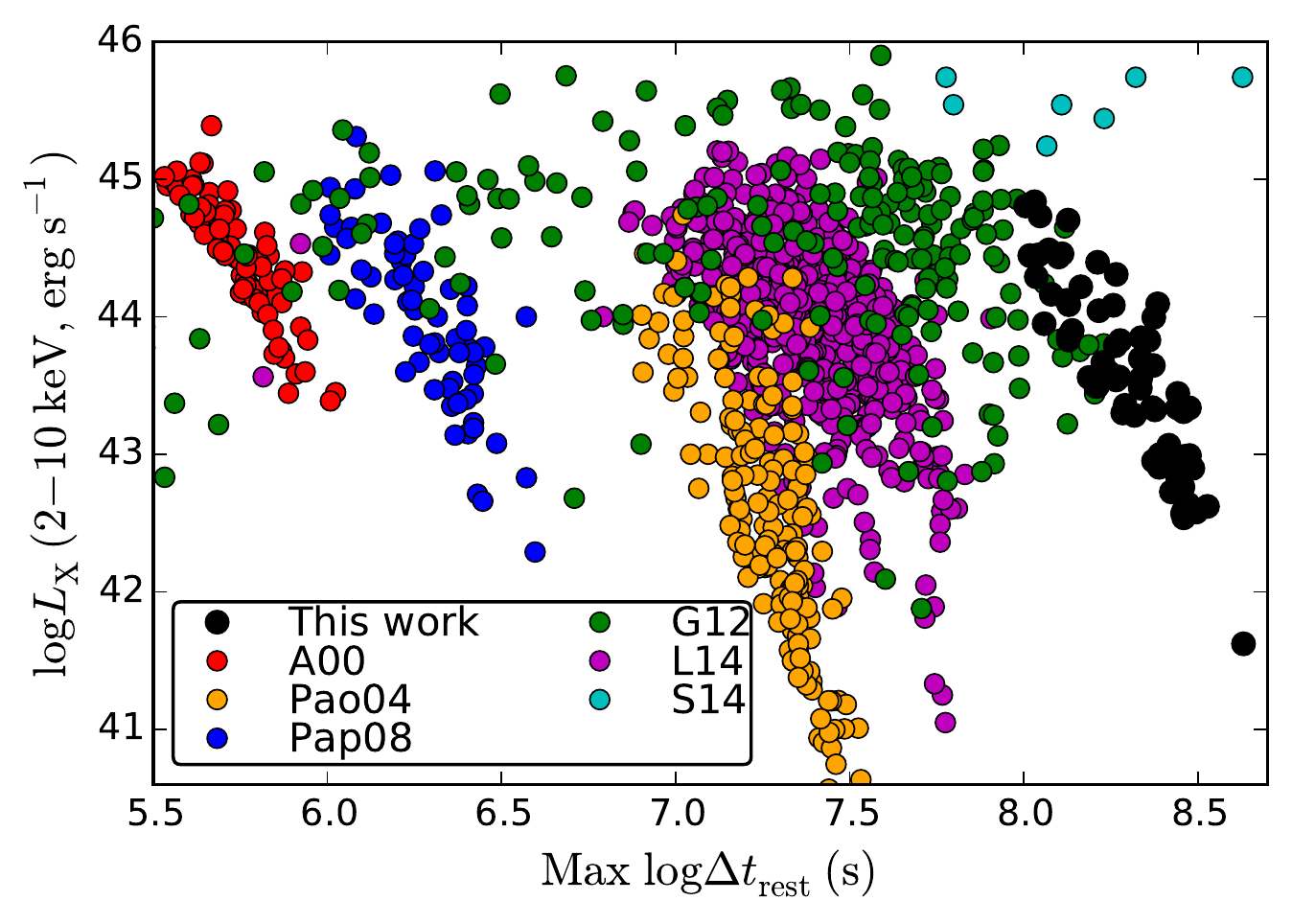}
\caption{
\xray\ luminosity as a function of maximum rest-frame time 
span. Different colors indicate samples from different 
studies (A00: \citealt{almaini00}; Pao04: \citealt{paolillo04}; 
Pap08: \citealt{papadakis08}; G12: \citealt{gibson12}, 
L14: \citealt{lanzuisi14}; S14: \citealt{shemmer14}). 
The \xray\ luminosities from the literature are derived assuming 
a power-law photon index of 1.8. 
The \xray\ luminosities of our sources are the mean values 
detailed in Section~\ref{sec:spec_fit_method}. 
The maximum rest-frame time spans for A00, Pao04, and Pap08 
are calculated assuming that each source is present in all 
observations; this assumption might overestimate time spans for 
some sources. The rest-frame time spans probed in 
this work are among the longest for \xray\ variability studies.
}
\label{fig:Lx_vs_det_t}
\end{figure}

Motivated by the importance of \xray\ variability studies, 
especially on long time scales, we here explore the 
\xray\ variability of the \xray\ brightest \hbox{radio-quiet} AGNs in 
the Chandra Deep \hbox{Field-South} (\cdfs; e.g., 
\citealt{xue11}; B.~Luo et al., in preparation, L16 hereafter).
We take advantage of the data products from the observations by 
\chandra\ of the \cdfs. The observations span \hbox{$\approx 15$~yr} and 
are well separated, enabling us to characterize long-term 
variability properties of distant AGNs. The rest-frame time
scales probed in this study are among the longest for \xray\ 
variability studies of distant AGNs (see Figure~\ref{fig:Lx_vs_det_t}),
and they are the longest for an \xray\ selected AGN sample.
Table~\ref{tab:var_study} summarizes the basic sample 
properties in this work and previous variability studies of 
distant AGNs. 
Furthermore, the long exposure times, low source-cell backgrounds, 
and state-of-the-art source-extraction 
techniques (\citealt{xue11, xue16}; L16) 
yield high-quality data products, 
allowing us to perform, in addition to photometric  
analyses, reliable basic spectral analyses. 
We assess variability of both 
intrinsic (absorption-corrected) \xray\ luminosity and 
absorption, and study their dependence on 
time scale and source properties. 

\begin{table}
\begin{center}
\caption{Summary of Variability Studies of Distant AGNs}
\label{tab:var_study}
\begin{tabular}{cccccccc}\hline\hline
Reference & Med.\ Max.\ & Med. & 
$z$\ & $\log$\lx\ & $N$ \\ 
  & $\log \Delta t_{\rm rest}$ & Counts & & & \\ \hline
(1) & (2) & (3) & (4) & (5) & (6) \\ \hline
This work &  8.3  &  1399 &  0.6--3.1  &  42.7--44.5  & 68 \\
A00       &  5.7  &  63   &  0.6--2.0  &  43.8--45.0  & 86 \\
Pao04     &  7.3  &  64   &  0.3--2.1  &  41.0--43.9  & 186 \\
Pap08     &  6.3  &  280  &  0.6--2.7  &  43.1--44.7  & 66 \\
G12 	  &  7.3  &  130  &  0.4--3.0  &  43.6--45.2  & 264 \\
L14 	  &  7.4  &  350  &  0.5--2.3  &  43.2--44.7  & 638 \\
S14 	  &  8.1  &  --   &  1.8--4.3  &  45.4--45.7  & 7 \\ \hline
\end{tabular}
\end{center}
{\sc Note.} --- \\
(1) A00: \cite{almaini00}; Pao04: \cite{paolillo04}; 
Pap08: \cite{papadakis08}; G12: \cite{gibson12}, 
L14: \cite{lanzuisi14}; S14: \cite{shemmer14}. 
For Pao04, we only include sources with redshift information.
(2) Sample median of maximum rest-frame time spans, where 
$\Delta t_{\rm rest}$\ is in units of seconds.
(3) Sample median of net counts. We do not calculate the median 
counts of S14 due to large fluctuations in counts among their 
seven sources. 
(4) Redshift range. We adopt the \hbox{10th--90th} percentile ranges.
(5) \xray\ luminosity range (\hbox{10th--90th} percentile).
(6) Number of sources.
\end{table}

The paper is structured as follows. We describe the observations, 
data reduction, and sample selection in Section~\ref{sec:data}.
In Section~\ref{sec:analysis}, we perform both photometric and 
spectral variability analyses and investigate variability
dependences on source properties and time scales. 
We discuss our results and draw conclusions in Section~
\ref{sec:conclusion}. 

Throughout this paper, we assume a cosmology with 
\hbox{$H_0=70$~km~s$^{-1}$~Mpc$^{-1}$}, \hbox{$\Omega_M=0.3$}, 
and \hbox{$\Omega_{\Lambda}=0.7$}. 
We adopt \hbox{\nh $=8.8\times10^{19}$~cm$^{-2}$} as the value of the column density
of Galactic absorption \citep{stark92}. Quoted 
uncertainties are at the 1$\sigma$ (68\%) confidence level, 
unless otherwise stated.
We adopt the standard naming convention for \chandra\ \cdfs\ 
sources, i.e., ``CXOCDFS \hbox{J033XXX.X--27XXXX}'';
for simplicity, we drop the ``CXOCDFS'' and quote as 
``\hbox{J033XXX.X--27XXXX}'' directly.

\section{Data and Sample}\label{sec:data}
\subsection{Observations and Data Reduction}
\label{sec:spec}
This work is based on the \chandra\ \cdfs\ data.
The observations were taken from October 1999 to January 
2015 with a total observation time of 5.7~Ms (referred as 
6~Ms, hereafter). In total, there are 84 observations
utilized with median exposure time \hbox{$\approx 60$~ks}. 
All of the 84 observations were performed using the Advanced 
CCD Imaging Spectrometer imaging array (ACIS-I; 
\citealt{garmire03}; see L16 for more observation details).
The data products were processed 
from level 1 files (L16) using CIAO v4.7 with CALDB v4.6.7. 
Spectra and photometry from each observation were extracted 
using ACIS Extract v4864 \citep[AE; ][]{broos10}. 
For each source, AE constructs observation-specific 
polygonal extraction apertures. The apertures
are chosen to maximize S/N, based on simulations of point 
spread functions (PSFs).

Since most \cdfs\ sources have low S/N in single observations, 
we bin data from neighboring observations as one 
``epoch'' to enhance the signal-to-noise ratio (S/N). 
We have binned the 
observations so that each epoch consists of observations 
totaling about \hbox{1--2}~Ms of exposure time, resulting in
a photometric/spectral set of four epochs for each source. 
Our bins were chosen to include as much data as possible in
the shortest possible span of time, in order to minimize 
variability effects within the bins.
The bin widths range from about several months to one year.
Due to cosmological time dilation, 
the rest-frame total time span and bin width are 
a factor of $1+z$\ shorter than the observed-frame values. 
Table~\ref{tab:obs} shows our observation-binning
approach. This binning process is carried out using
the MERGE\_OBSERVATIONS stage of AE. 
We do not include \xmm\ \cdfs\ data in our 
analyses, since cross-calibration between \chandra\ and \xmm\ 
spectra can be problematic \citep[e.g.,][]{iwasawa15}, and 
the \xmm\ data have substantially higher background. 
Also, most of the \xmm\ observations (\hbox{$\sim 85\%$} of the exposure time) 
were taken within about 1.5~years \citep{ranalli13}, 
and thus are not suitable for year-scale variability studies.

\begin{table*}
\begin{center}
\caption{Observation Binning}
\label{tab:obs}
\begin{tabular}{cccccc}\hline\hline
Epoch & Start Date & End Date & Bin Width (yr) & 
Exp.$^{\mathrm{a}}$ (Ms) & Obs.$^{\mathrm{b}}$ \\ \hline
1 & Oct.\ 1999 & Dec.\ 2000  & 1.19  & 0.94 & 11 \\  
2 & Sept.\ 2007 & Nov.\ 2007 & 0.12  & 0.97 & 12 \\ 
3 & Mar.\ 2010 & July\ 2010  & 0.34  & 1.98 & 31 \\
4 & June\ 2014 & Jan.\ 2015  & 0.57  & 1.85 & 30 \\\hline
\end{tabular}
\end{center}
{\sc Note.} --- \\
a.\ Total exposure time of observations in each bin.\\
b.\ Number of observations in each bin.
\end{table*}

\subsection{Sample Selection}\label{sec:sample}
To perform reliable analyses, we select our
sources based on the following criteria: \\
1.\ Classified as an AGN in the L16 catalog; \\
2.\ Not identified as a radio-loud AGN by \cite{bonzini13}. \\
3.\ More than 600 total net counts 
	(i.e., background-subtracted, \hbox{0.5--7}~keV) 
	in the full 6~Ms exposure; and \\
4.\ Off-axis angle $<8'$. \\
\\
The first criterion classifies AGNs based upon
the \xray\ luminosity, \xray\ spectral shape, 
\xray-to-optical flux ratio, \xray-to-radio 
luminosity ratio, and optical emission-line 
properties, and it is described in more detail in
Section 4.4 of \cite{xue11}.
The second criterion excludes the sources possibly 
affected by jet-linked emission in their \xray\ spectra, 
to avoid dealing with the significant additional complexity 
of such emission \citep[see, e.g.,][]{miller11}.
This criterion removes only 6 sources that 
satisfy the other three criteria.
The third constraint guarantees suitable photon
statistics for variability characterization and 
spectral fitting; given the observed
\xray\ variability of our sources, we do not have 
any problematic cases where, e.g., one bin contained
most of the counts and the others had very poor 
counting statistics (see Sections~\ref{sec:PFHR} and
\ref{sec:flux_15}). The fourth criterion 
discards sources with large off-axis angles, 
which have generally poorer S/N due to the
degraded PSF,
and it also guarantees  
each source is covered by almost all observations.

\begin{figure}
\includegraphics[width=\linewidth]{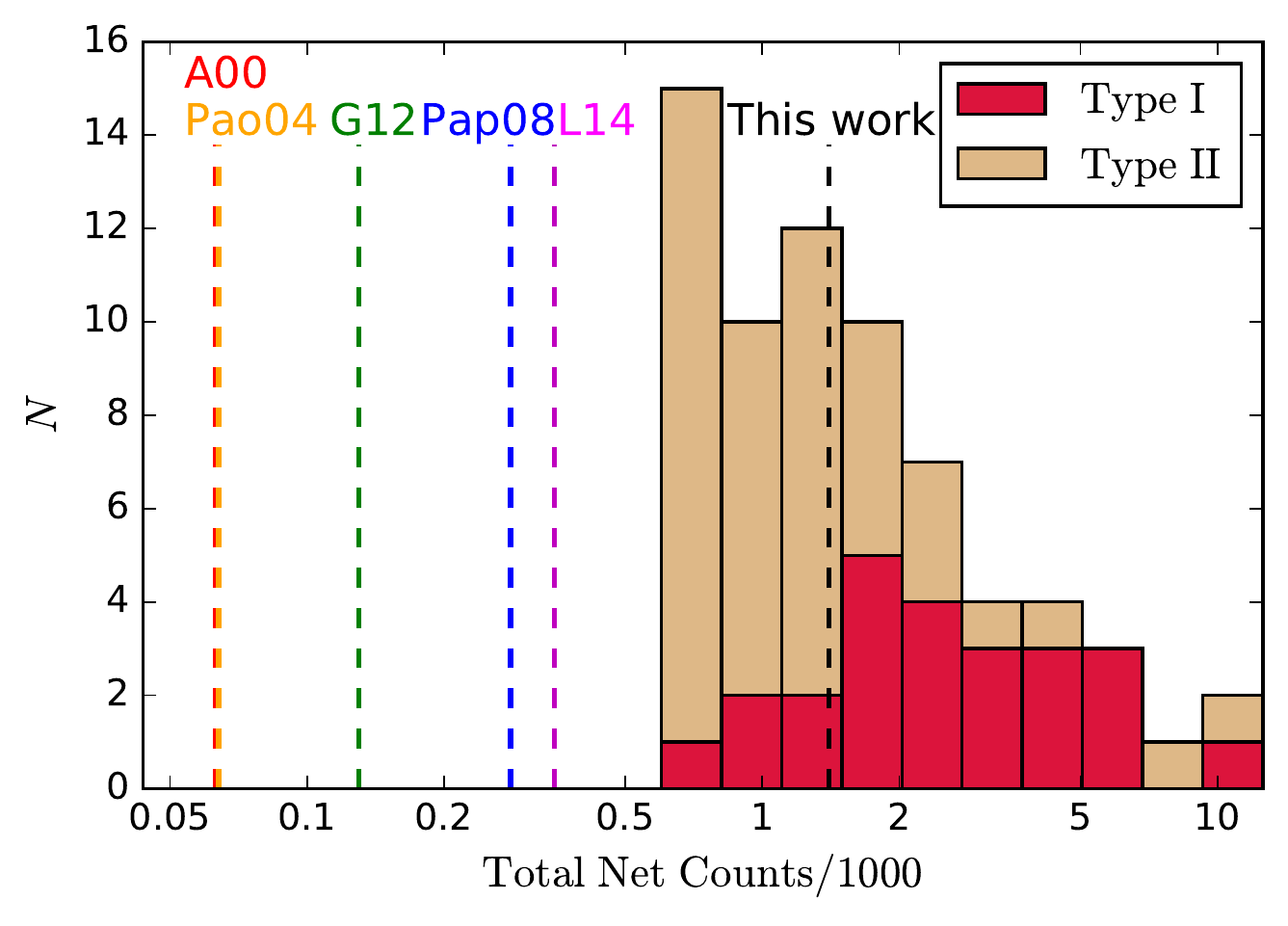}
\caption{
The histogram of total net counts of our 68
sources. The red and brown bars indicate type I
and type II AGNs. 
The vertical dashed lines indicate the median 
counts of this work and previous studies, respectively 
(A00: \citealt{almaini00}; Pao04: \citealt{paolillo04}; 
Pap08: \citealt{papadakis08}; G12: \citealt{gibson12}, 
L14: \citealt{lanzuisi14}).
Note that A00 and Pao04 have very similar median counts.
We do not show the median counts of \cite{shemmer14} 
due to large fluctuations in counts among their seven 
sources.
Our sample only includes bright 
sources with counts greater than 600.
Our type I AGNs generally have more available 
counts than their type II counterparts.
}
\label{fig:cnt_hist}
\end{figure}

Sixty-eight sources are selected with \hbox{649--11283}
counts; the median number of counts is 1399.
Figure~\ref{fig:cnt_hist} shows the distribution of 
our counts. Our sources generally have more counts
than those in previous studies, usually by substantial 
factors (Table~\ref{tab:var_study}), 
allowing improved source characterization.
The faintest selected sources have fluxes 
(observed-frame \hbox{0.5--7}~keV; 
see Section~\ref{sec:flux_15} for flux calculation) of 
\hbox{$\approx 1\times 10^{-15}$~$\mathrm{erg\ cm^{-2}\ s^{-1}}$},
similar to the \hbox{full-band} \hbox{source-detection} limit of 
the \chandra\ COSMOS-Legacy survey \citep[e.g.,][]{civano16}.
We are thus characterizing in this work the AGNs responsible for
producing much of cosmic accretion power; our measurements probe
\hbox{$\approx$\ \hbox{20--3}}~times below the knee luminosity, $L_{\rm X}^*$,
of the \xray\ luminosity function
at \hbox{$z=0.5-4$}\ \citep[e.g.,][]{ueda14, aird15}.

\begin{figure}
\includegraphics[width=\linewidth]{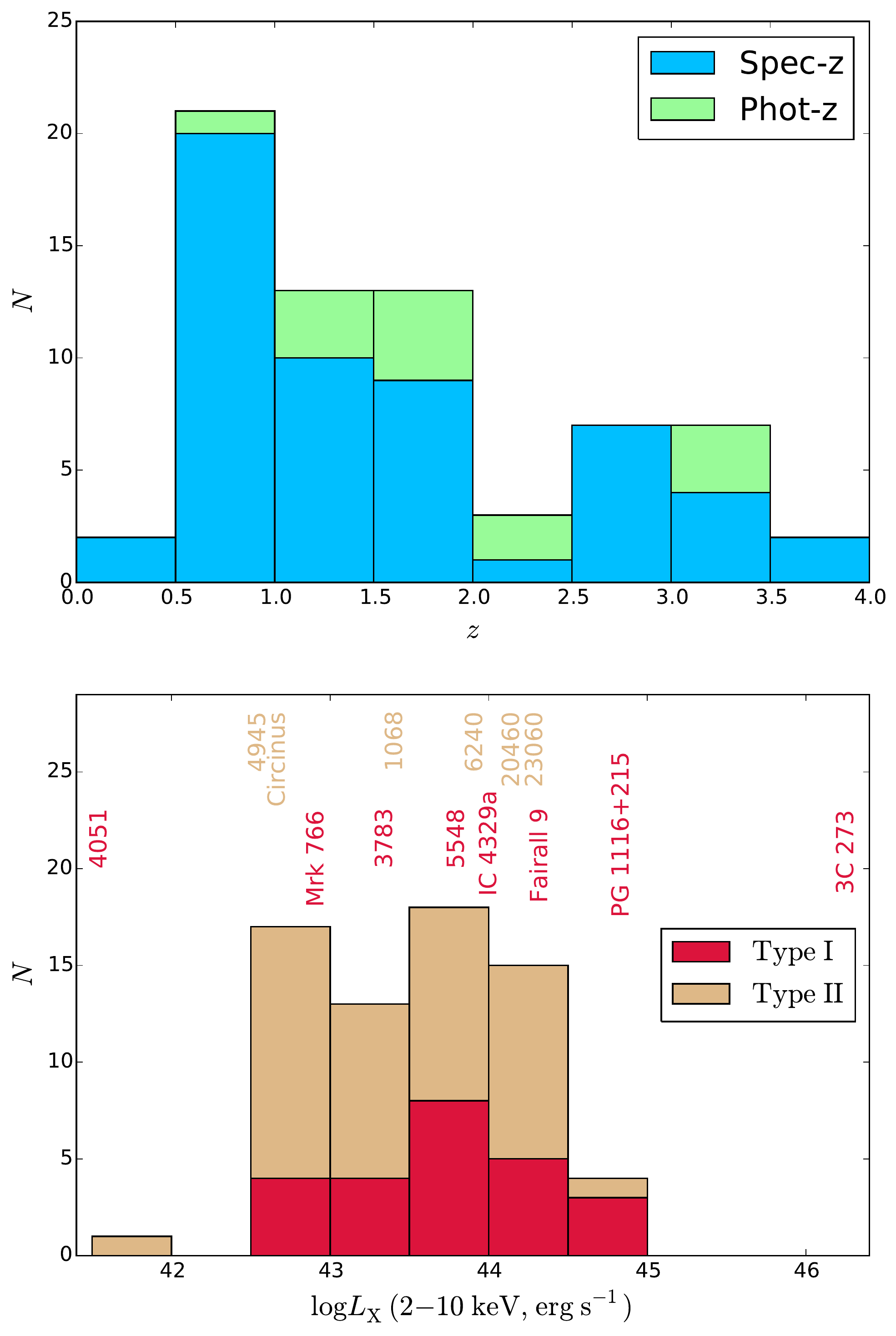}
\caption{
Upper panel: Redshift histogram for our 68 sources. 
The blue and green bars indicate sources with 
spectroscopic redshifts and photometric redshifts,
respectively.  
Lower panel: Absorption-corrected 
\xray\ luminosity (\hbox{2--10 keV}, rest-frame) histogram.
The red and brown bars indicate type I and type II
AGNs, respectively. 
The values are the epoch-mean \lx\ described in 
Section~\ref{sec:spec_fit_method}.
Along the top of the panel, 
we label the typical absorption-corrected
\lx\ values for some representative
type~I (red) and type~II (brown) AGNs in the local universe. 
The 4-digit and 5-digit numbers indicate NGC and IRAS 
sources, respectively. The \lx\ data were compiled 
from the literature \citep[e.g.,][]{brandt97, ogasaka97, 
reynolds97, george00, pounds03, arevalo14, 
puccetti14, puccetti16, bauer15}. 
Our sources cover wide ranges of both $z$\ and \lx.
In terms of \lx, our sources appear to be distant 
analogs of many local Seyfert galaxies and moderate-luminosity
quasars.
}
\label{fig:z_Lx_dis}
\end{figure}

The spectroscopic and photometric redshift data are compiled 
by L16. For each source, we adopt 
the spectroscopic redshift (\zs) either marked as ``secure'' 
by L16 or consistent with the photometric redshift (\zp) 
measurement (i.e., \zp\ differs from \zs\ by less than 10\%).
Otherwise, if \zs\ is not available or disagrees with
\zp, we adopt \zp\ because ``insecure'' \zs\ are less reliable
than \zp. Those insecure \zs\ are often based on spectra with few 
features, e.g., a tentative single absorption line, but \zp\ are 
derived from SED fitting of $\gsim 15$\ photometric bands
\citep[][]{luo10, hsu14}. 
In total, we have 55 \zs\ and 13 \zp. 
The \zp\ quality is generally high (fractional 
errors $\sim$\ a few percent) thanks to 
the wide multi-band photometric coverage used to derive \zp\ 
\citep[][]{luo10, hsu14}. The redshift and intrinsic \xray\ 
luminosity \lx\footnote{Absorption-corrected 
\xray\ luminosity in the rest-frame \hbox{2--10 keV} band; we use 
the epoch-mean value (see Section~\ref{sec:spec_fit_method}).} 
distributions of the sample are plotted in Figure~\ref{fig:z_Lx_dis},
with the \lx\ values for some representative local AGNs 
marked for comparison purposes.
The median \lx\ of our sources is 
$4\times 10^{43}\ \rm erg\ s^{-1}$, several times 
larger than that of the whole \cdfs\ AGN sample
(\hbox{$\approx 8\times 10^{42}\ \rm erg\ s^{-1}$}, L16);
the median redshifts of the two samples are similar 
(\hbox{$\approx$1.5}).
Many local well-known AGNs have \lx\ within our 
luminosity coverage, and thus our sources, at least in this sense, 
appear to be distant analogs of these local AGNs.

We classify a source as type I if any broad emission line is 
reported in the redshift literature (20 sources);
if only narrow emission lines/absorption lines are reported 
(35 sources) we assign a type II classification. 
We caution that there might be some intrinsic type I 
objects misclassified as type II due 
to, e.g., 
lack of spectral coverage of the hydrogen Balmer lines
\citep[e.g.,][]{khachikian74} or
the spectral S/N not being sufficient 
to identify broad lines.   
If the spectral classification of a source is not
available in the literature, it is classified based on 
fitting of the spectral energy distribution 
\citep[SED; data from][]{hsu14}:
if its rest-frame optical color is blue 
(i.e., rest-frame $u-g<0.8$)\footnote{This threshold 
generally separates our type I and type II AGNs classified
by spectral features; see, e.g., \cite{richards02} and \cite{barger03} 
for the effectiveness of optical color classifications.}
we classify it as type I, otherwise as type II.
This \hbox{color-classification} scheme results in 4 type 
I and 9 type II AGNs, in addition to the spectral classification. 
Therefore, our sample consists of 24 type I and 44 type II AGNs.
The optical classification is broadly consistent with 
\xray\ classification approaches (i.e., using the epoch-mean 
\hbox{\nh\ $=10^{22}\rm\ cm^{-2}$} as the threshold for type II; see 
Section~\ref{sec:spec_fit_method}) despite some exceptions 
(see Figure~\ref{fig:Lx_vs_z_and_Lx_vs_nH}). 
Type I sources generally 
have more counts than type II sources; their median 
counts are 2199 and 1098, respectively. 
Table~\ref{tab:src_info} lists the properties of individual 
sources in our sample.

BAL quasars often show heavy and 
complex \xray\ absorption despite their type~I nature 
\citep[e.g.,][]{gallagher02, gallagher06}.
It is of interest to investigate their \xray\ variability behavior 
\citep[e.g.,][]{gallaghper04, saez12}.
There are 8 type~I quasars in our sample, if we define 
quasars as AGNs with \hbox{\lx$ > 10^{44}\ \mathrm{erg\ s^{-1}}$}.\footnote{
Corresponding to \hbox{$L_{\nu}(2500\ \mathrm{\AA})\sim 10^{30}\ 
\mathrm{erg\ s^{-1} Hz^{-1}}$} calculated from the 
\hbox{$\alpha_{\mathrm OX}$-$L_{\nu}(2500\ \mathrm{\AA})$} relation presented
in \cite{just07}.} 
One of our objects has been reported as a BAL quasar 
(\hbox{J033209.4--274806}, e.g., \citealt{szokoly04}) at \hbox{$z=2.81$}; 
it is \xray\ luminous (\hbox{\lx $\approx 3\times 10^{44}$~erg~s$^{-1}$})
and highly obscured (\hbox{\nh $\approx 2\times 10^{23}$~cm$^{-1}$}) with
an intrinsic power-law photon index \hbox{$\Gamma=1.68$}\ 
(see Section~\ref{sec:spec_analyses}).
This small number of BAL quasars is expected considering only 
\hbox{$\approx$20\%} of type~I quasars are BAL quasars 
\citep[e.g.,][]{hewett03, gibson09}. 

Three of our sources, \hbox{J033247.8--274232} at \hbox{$z=0.98$}, 
\hbox{J033211.3--275213} at \hbox{$z=3.74$}, and
\hbox{J033212.9--275236} at \hbox{$z=2.56$}, 
are potentially detected in the \nustar\ hard band (\hbox{8--24}~keV; \citealt{mullaney15}). 
The matching between \chandra\ and \nustar\ sources was performed 
by \cite{mullaney15}. \hbox{J033247.8--274232}\ has an almost 
unambiguous \nustar\ counterpart, because it is the only bright \chandra\ source
within $23''$\ (\hbox{$\approx 3$}\ times the typical \nustar\ positional uncertainty) of 
the \nustar\ source position. \hbox{J033211.3--275213} 
and \hbox{J033212.9--275236} are matched to a single \nustar\
source, and are both \hbox{$\approx 15''$}\ from the \nustar\ counterpart. 
There are no other bright \chandra\ sources within $23''$\ of this 
\nustar\ source position. Therefore, one of these two high-redshift 
\chandra\ sources likely has a \nustar\ hard-band detection.  

\begin{table*}
\begin{center}
\caption{Source Properties}
\label{tab:src_info}
\begin{tabular}{ccccccccccccc}\hline\hline
Name (CXOCDFS) & RA  & DEC & $z$ & $\Delta z$ & Opt.\ type & Off-Axis      &
Counts                   & $P_{\rm PF}$ & $P_{\rm HR}$ & $E_{peg}$ & $goodness$ & $\Gamma$ \\ 
 (1) & (2) & (3) & (4) & (5) & (6) & (7) & (8) & (9) & (10) & (11) & (12) & (13) \\ \hline
J033158.1--274833 &52.99210 &$-$27.80943 &0.74 &n/a  &IIs &6.69 &1366.0 &0.00 &32.75 & 0.50 &10.2 &1.52    \\
J033158.2--275041 &52.99283 &$-$27.84490 &3.31 &0.09 &Ic  &7.05 &2684.8 &0.00 &86.68 & 0.96 &0.5  &1.72  \\
J033200.3--274319 &53.00150 &$-$27.72209 &1.04 &n/a  &Is  &7.96 &3598.1 &0.00 &60.04 & 0.50 &1.0  &1.78   \\
J033201.5--274327 &53.00663 &$-$27.72420 &2.73 &n/a  &Is  &7.67 &2772.0 &0.00 &15.30 & 0.66 &4.8  &1.72  \\
J033202.4--274600 &53.01026 &$-$27.76675 &1.62 &n/a  &Is  &6.18 &2012.5 &0.00 &61.65 & 0.50 &7.3  &1.65   \\ \hline
\end{tabular}
\end{center}
The full table contains 21 columns of information for 68 sources. 
(This table is available in its entirety in a machine-readable form in the online 
journal. A portion is shown here for guidance regarding its form and content.)\\
Column (1): Standard name of \chandra\ \cdfs\ sources. \\
Columns (2) and (3): J2000 coordinates. \\
Column (4): Adopted redshift. See L16 for redshift references. \\
Column (5): 1$\sigma$\ errors of photometric redshift. n/a indicates a spectroscopic 
			redshift has been used.\\
Column (6): Optical spectral type (see Section~\ref{sec:sample} for the classification scheme). 
			Is (IIs): type I (II) from spectral classification; 
			Ic (IIc): type I (II) from color classification. 
The spectral classifications are from 
\cite{szokoly04, mignoli05, ravikumar07, tozzi09, treister09, silverman10, mao12}. \\
Column (7): Off-axis angle in units of arcminutes. \\
Column (8): Full-band (\hbox{0.5--7 keV}) total available net counts (not aperture corrected).\\
Columns (9) and (10): $p$-value of photon-flux and hardness-ratio variability, respectively
					  (Section~\ref{sec:PFHR}). Both are in units of \%.\\
Column (11): The lower end of the normalization band used in $pegpwrlw$, 
		     in units of keV (Appendix~\ref{app:err}).  \\
Column (12): $goodness$\ of the spectral fitting 
  	     ($wabs \times zwabs \times pegpwrlw$\ model), in units 
	     of \% (Section~\ref{sec:spec_fit_method}).\\
Column (13): Power-law photon index. \\
Columns (14) and (15): Epoch-mean values of intrinsic \lx\ (rest-frame \hbox{2--10 keV, in units of erg~s$^{-1}$) and \nh\ (in units of cm$^{-2}$), respectively} (Section~\ref{sec:spec_fit_method}). \\
Columns (16) and (17); Metrics ($\Delta$AIC$_L$\ and $\Delta$AIC$_N$) 
			       of the significance of \lx\ and \nh\ variability, respectively 
			       (Section~\ref{sec:identify}).\\
Columns (18) and (19): Normalized excess variance of \lx\ variability (\sigmaexcL) and its uncertainty, respectively (Section~\ref{sec:var_scale}).  \\
Columns (20) and (21): Normalized excess variance of \nh\ variability (\sigmaexcN) and its uncertainty, respectively (Section~\ref{sec:var_scale}). We only calculated this quantity for the 35 sources with all four epochs having $\delta N_{\mathrm{H},i}/N_{\mathrm{H},i}<0.4$\ (see Appendix~\ref{app:err}); n/a indicates the other 33 sources. \\
\end{table*}

\section{Data Analyses}\label{sec:analysis}
\subsection{Photometric Variability}
\label{sec:PFHR}
\subsubsection{Method}\label{sec:method}
\begin{figure}
\includegraphics[width=\linewidth]{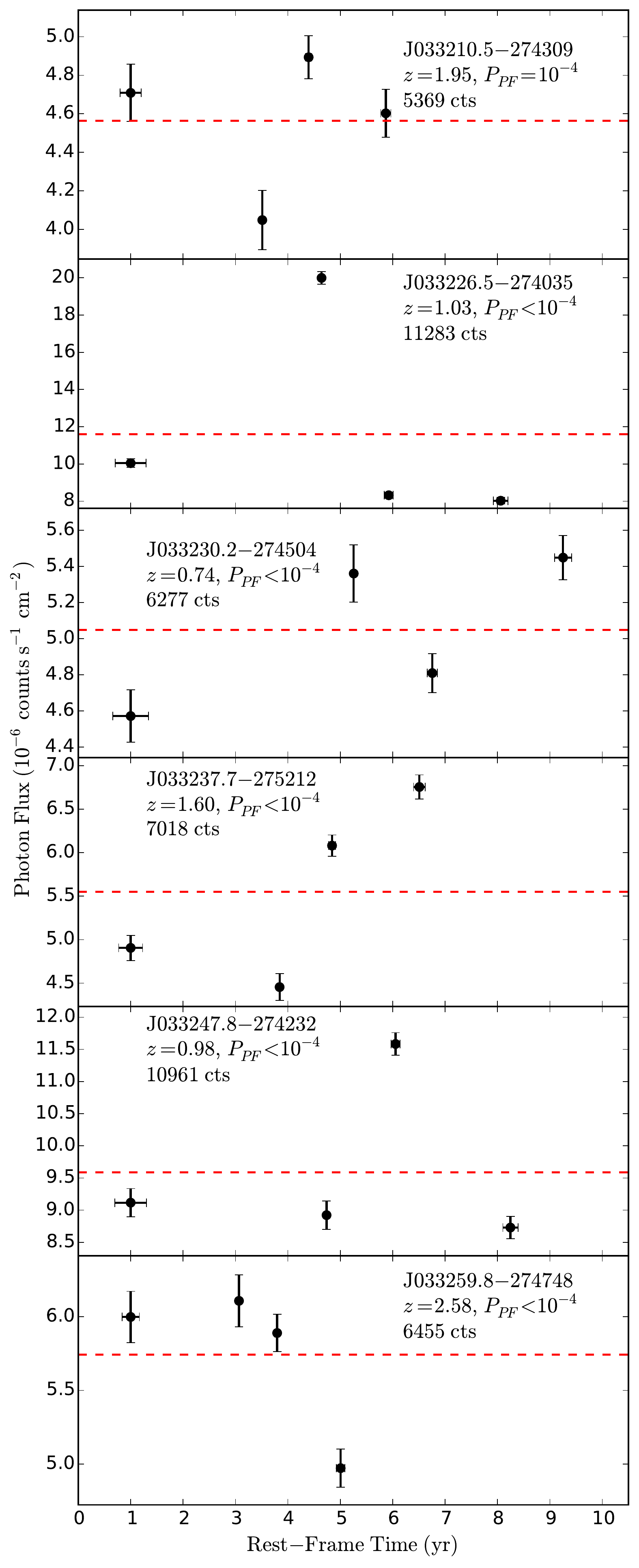}
\caption{
Light curves of the 6 sources with the most counts. 
The red dashed horizontal lines
indicate the unweighted mean of photon fluxes for each
source. The horizontal error bars indicate the bin
width of each epoch. The rest-frame time of epoch~1 
is set to 1~yr. Some source properties are labeled on
the corresponding panels. All of the 6 sources show 
photon-flux variability (i.e., $P_{PF}<5\%$).
}
\label{fig:light_curv}
\end{figure}

We use full-band (\hbox{0.5--7 keV}) photometry to analyze 
\hbox{photon-flux} ($PF$, i.e., count rate per unit area, in units of 
counts~s$^{-1}$~cm$^{-2}$) variability. For each epoch of a source, 
we calculate $PF$ as 
\begin{equation}
PF_{i} = \frac{ {net\_counts}_{i}}{ {effarea}_{i}
\times {exptime}_{i}},
\end{equation}
and its uncertainty
\begin{equation}
\delta PF_{i} = \frac{ {\delta net\_counts}_{i}}
{ {effarea}_{i}\times {exptime}_{i}},
\end{equation}
where the subscript $i$\ denotes the epoch; ${net\_counts}_{i}$ 
and ${\delta net\_counts}_{i}$\footnote{We use the average of 
upper and lower 1$\sigma$\ errors here, which are calculated
by AE using \cite{gehrels86}.}\ 
are background-subtracted 
counts and corresponding error, respectively; $effarea_i$\ and 
$exptime_i$\ are the effective area and exposure 
time, respectively. The effective area is approximated as the 
average ancillary response file (ARF) weighted by the 
average \cdfs\ AGN spectrum
\citep[i.e., a \hbox{$\Gamma=1.4$}\ power law with Galactic 
absorption; see, e.g.,][]{paolillo04}. This procedure accounts 
for the varying sensitivity of \chandra\ among different epochs 
due to, e.g., quantum-efficiency degradation and gaps between CCDs,
since these factors are considered by AE when calculating the ARF.
Figure~\ref{fig:light_curv} shows light curves of the 6 
sources with the most counts.

To identify variable sources, we calculate 
the statistic
\begin{equation}
X^2_{PF} = \sum_{i=1}^4 \frac
{ (PF_i - \langle PF \rangle)^2 }
{ (\delta PF_i)^2 },
\end{equation}
where $\langle PF \rangle$\ is the unweighted mean of $PF_i$\
\citep[e.g.,][]{turner99}.\footnote{
We have tested using the mean weighted by $1/(\delta PF_i)^2$,
and obtained similar results. This is also true for the 
hardness-ratio variability. } 
For sources with large numbers of counts, 
the probability distribution of $PF$\ approaches 
the normal distribution, and thus 
$X^2_{PF}$\ should follow the $\chi^2$\ distribution
with three degrees of freedom ($\chi^2_{\rm dof=3}$). 
Our sources have at least 600 total net
counts (at least
\hbox{$\sim 150$~counts} or \hbox{$\mathrm{S/N}\sim 12$} per bin). 
Therefore $\chi^2_{\rm dof=3}$\ 
should be a good approximation of $X^2_{PF}$. To verify this 
point and find an accurate $p$-value ($P_{PF}$) from $X^2_{PF}$, 
we adopt a Monte Carlo simulation strategy similar to that in 
\cite{paolillo04} and \cite{young12}.
The null hypothesis is that $PF$\ remains constant in the 
four epochs at $PF_{\mathrm{mean}}$.
To perform the simulations, we need to know
model counts in each epoch.
We convert $PF_{\mathrm{mean}}$\ to the  
net counts expected in each epoch as 
\begin{equation}
net\_counts_{i}^{\mathrm{model}}
= PF_{\mathrm{mean}} \times {effarea}_{i}\times 
exptime_{i}.
\end{equation} 
We obtain the model source counts 
(not background subtracted) and model background 
counts as
\begin{equation}\label{equ:src_cnt}
\begin{gathered}
{src\_counts}_{i}^{\mathrm{model}}
= {net\_counts}_{i}^{\mathrm{model}} + \frac{
{bkg\_counts}_{i}^{\mathrm{observed}}} {backscal_i}, \\
{bkg\_counts}_{i}^{\mathrm{model}} = 
{bkg\_counts}_{i}^{\mathrm{observed}},
\end{gathered}
\end{equation} 
where $backscal$\ is the factor produced by AE 
that scales background counts (usually obtained 
in a large aperture) to the source-extraction 
aperture. We use ${src\_counts}_{i}^{\mathrm{model}}$\
(${bkg\_counts}_{i}^{\mathrm{model}}$) as the mean in a
Poisson distribution to simulate ${src\_counts}_{i}$\
(${bkg\_counts}_{i}$), and extract photometry following  
the algorithm in AE.\footnote{See Section 5.10 of the AE manual
available at 
http://www2.astro.psu.edu/xray/docs/TARA/ae\_users\_guide.}
We obtain the $X^2_{PF}$\ for a simulated data set (10000 simulations) 
following the same procedures as described above. 
We perform \hbox{Kolmogorov-–Smirnov} (KS) tests 
between the simulated $X^2_{PF}$\ distributions and 
the $\chi^2_{\rm dof=3}$ distribution. 
The resulting KS statistics for our sources 
have median 0.015; this small
value indicates our simulated $X^2_{PF}$\ distributions are 
very similar to the $\chi^2_{\rm dof=3}$\ distribution. 

We define the hardness ratio ($HR$) for a source in an epoch as
\begin{equation}\label{equ:hr}
HR_{i} = \frac{ PF_{i, hard} - PF_{i, soft} }
		  { PF_{i, hard} + PF_{i, soft} },
\end{equation}
where $PF_{i, hard}$\ and $PF_{i, soft}$ are photon fluxes as
defined above but for the hard band (\hbox{2--7 keV}) and soft band 
(\hbox{0.5--2 keV}), respectively. We estimate the error of $HR$\ 
from error propagation as
\begin{equation}
\begin{gathered}
\delta HR_{i} = \frac{ 2PF_{i, hard} PF_{i, soft}  }
					 { (PF_{i, hard}+PF_{i, soft})^2 } 
			    \times \\
				\sqrt{ \left(\frac{\delta PF_{i, hard}}
						     {PF_{i, hard}       }\right)^2
					  +\left(\frac{\delta PF_{i, soft}}
							 {PF_{i, soft}       }\right)^2}.
\end{gathered}
\end{equation}
To identify $HR$-variable sources, 
we calculate 
\begin{equation}
X^2_{HR} = \sum_{i=1}^4 \frac
{ (HR_i - \langle HR \rangle )^2 }
{ (\delta HR_i)^2 },
\end{equation}
where $\langle HR \rangle$\ is the unweighted mean of $HR_i$.

We perform similar simulations as above to convert $X^2_{HR}$\
to a $p$-value ($P_{HR}$). 
The null-hypothesis is that $HR_i$\ is constant and equals 
$HR_{\mathrm{mean}}$\ over the four epochs.
We approximate the model full-band net counts as the 
sum of the observed hard-band and soft-band net counts,
\begin{flalign}
{net\_}&{counts}_{i, full}^{\mathrm{model}} = &&\\\nonumber
& {net\_counts}_{i, hard}^{\mathrm{observed}} +
  {net\_counts}_{i, soft}^{\mathrm{observed}}. &&
\end{flalign}
The model hard-band and soft-band net counts are calculated by
\begin{flalign}
{net\_}&{counts}_{i, hard}^{\mathrm{model}} = &&\\\nonumber
& {net\_counts}_{i, full}^{\mathrm{model}} \times &&\\\nonumber
& ( \frac{ {net\_counts}_{i, hard} }  
              { {net\_counts}_{i, hard} + {net\_counts}_{i, soft} }
  )^{\mathrm{model}} &&\\\nonumber
= & {net\_counts}_{i, full}^{\mathrm{model}} \times &&\\\nonumber
& \frac{1}{2} ( 1 + 
(\frac{ {net\_counts}_{i, hard} - {net\_counts}_{i, hard}}  
              { {net\_counts}_{i, hard} + {net\_counts}_{i, soft} })
			  ^{\mathrm{model}}) &&\\\nonumber
= & {net\_counts}_{i, full}^{\mathrm{model}} \times 
	\frac{1 + HR_{\mathrm{mean}}}{2} &&
\end{flalign}  
and
\begin{flalign}
{net\_}&{counts}_{i, soft}^{\mathrm{model}} = &&\\\nonumber
& {net\_counts}_{i, full}^{\mathrm{model}} 
-{net\_counts}_{i, hard}^{\mathrm{model}}, &&
\end{flalign}  
respectively.

Knowing ${net\_counts}_{i, hard(soft)}^{\mathrm{model}}$,
we obtain ${src\_counts}_{i, hard(soft)}^{\mathrm{model}}$\
and ${bkg\_counts}_{i, hard(soft)}^{\mathrm{model}}$\ using
Equation~\ref{equ:src_cnt}. 
Similar to the case of $X^2_{PF}$, the simulated $X^2_{HR}$\ 
distributions are close to the $\chi^2_{\rm dof=3}$ distribution.
The KS statistics derived from comparing simulated $X^2_{HR}$\
distributions and $\chi^2_{\rm dof=3}$\ have median 0.023.

\subsubsection{Results}\label{sec:photo_res}
\begin{figure}
\includegraphics[width=\linewidth]{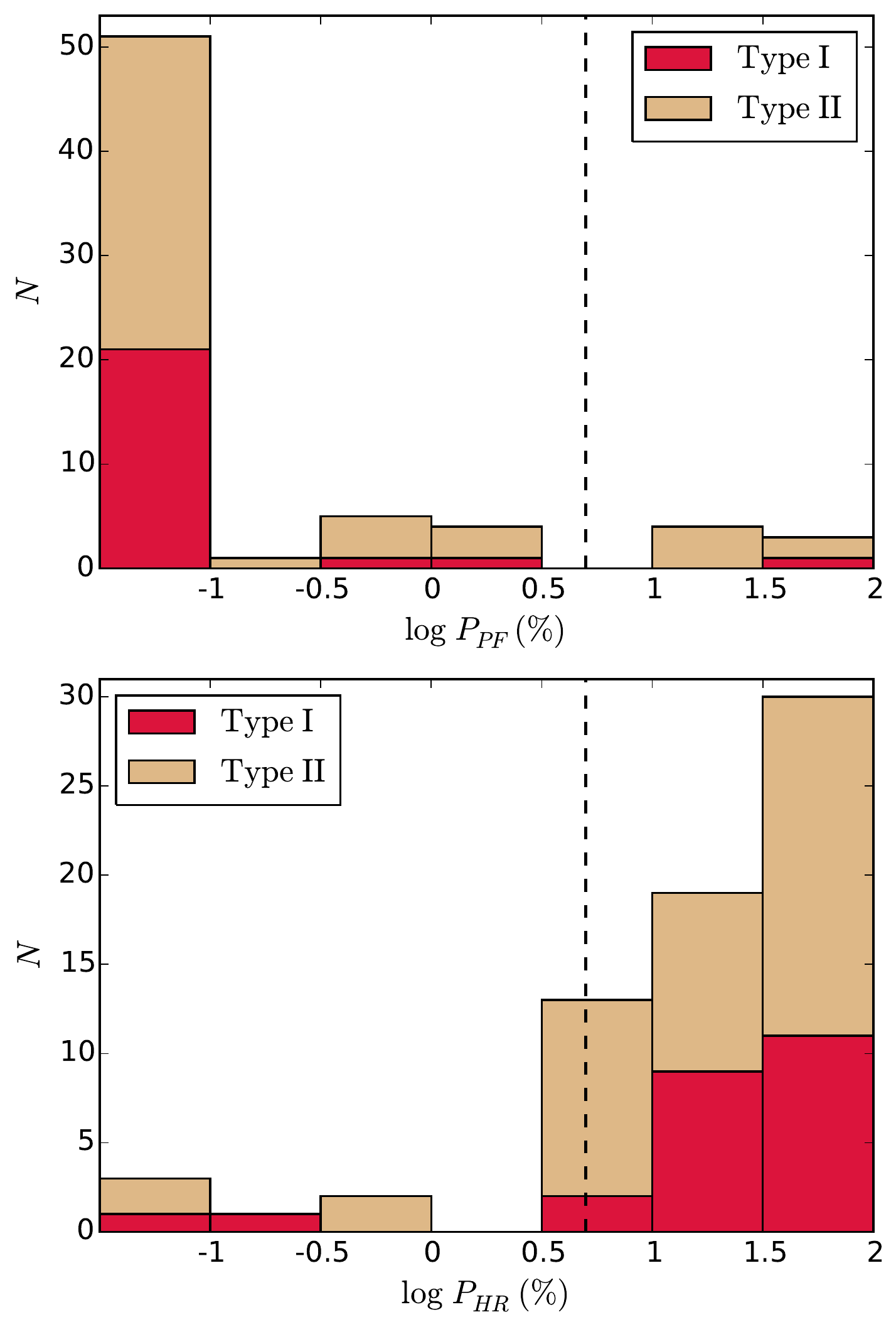}
\caption{
The histograms of $P_{PF}$\ (upper panel) 
and $P_{HR}$\ (lower panel). The red and
brown bars indicate type I and type II AGNs, respectively. 
The vertical black dashed lines in both panels indicate
the value of 5\%, our variability criterion. The leftmost 
column indicates all sources with $P_{PF}$\ ($P_{HR}$) 
$<0.1\%$. 
Photon-flux variability is generally more 
common than hardness-ratio variability.
}
\label{fig:P_hist}
\end{figure}

The histograms of $P_{PF}$\ and 
$P_{HR}$\ are presented in Figure~\ref{fig:P_hist}.
Using \hbox{$P=5\%$}\ as the 
threshold for variability (\hbox{$\sim 68\times 5\% \lsim 4$} 
false positives expected), 90\% (61/68) and 16\% 
(11/68) of our sources display $PF$\ variability and $HR$\ variability, 
respectively. All of our six sources with the most counts are 
variable (see Figure~\ref{fig:light_curv}).
Changing the threshold to \hbox{$P=1\%$}\ 
(\hbox{$\sim 68\times 1\% \lsim 1$} false positive expected) leads to 
a $PF$-variable fraction and a $HR$-variable fraction
of 84\% (57/68) and 9\% (6/68), respectively. This $PF$-variable 
fraction (84\%) is higher than that of optically selected 
Sloan Digital Sky Survey (SDSS, \citealt{york00}) quasars
under the same confidence level (i.e., 65\%; the sample is from 
G12 with the same count constraint of 600 applied).
This result supports an intrinsic anticorrelation between variability
amplitude and \xray\ luminosity (see Section~\ref{sec:sigma_vs_Lx_nH}),
considering those SDSS quasars are generally 
\hbox{$\approx 10$}\ times more \xray\ luminous 
than our sources (see Table~\ref{tab:var_study} and
Section~\ref{sec:identify}). However, this behavior 
might also be a result of different rest-frame time samplings 
(see Figure~\ref{fig:Lx_vs_det_t}).  
With the \hbox{$P=5\%$} threshold, almost all (23/24) type~I AGNs 
and 86\% (38/44) of type~II AGNs are $PF$ variable;
13\% (3/24) of type~I AGNs and 18\% (8/44) of type~II 
AGNs are $HR$\ variable. 
The higher $PF$-variable source fraction for type~I
sources might be due to their higher numbers of counts 
(Figure~\ref{fig:cnt_hist}). 
Type~II sources are more 
likely to be $HR$\ variable despite their 
smaller numbers of counts
However, Fisher's exact 
test shows that the dependences of variable source fractions
on optical spectral type are not statistically significant.

\subsubsection{Variability Within Each Epoch}\label{sec:variability_within}
We also use each \chandra\ observation as 
a bin to analyze the 
variability within each epoch ($\lsim 1$~yr, observed-frame). 
The observation lengths range from about 30 to 100~ks (see L16). 
We calculate $P_{PF}$\ and $P_{HR}$\ for each source within each 
of the four epochs. The method is the same as described in 
Section~\ref{sec:method}. $PF$\ variability
is still significant: the $PF$-variable 
(i.e., $P_{PF}<5\%$) source fractions are 49\%, 29\%,
56\%, and 38\% for each bin, respectively.\footnote{
Besides random fluctuations, there are many 
factors, including total exposure time and epoch bin width,
that can affect the detected variable source fraction 
for each epoch.
}
We find the variable source fraction is higher for 
sources with more counts, likely due to their higher 
S/N. 
We do not detect statistically significant $HR$\ 
variability: the $HR$-variable (i.e., $P_{HR}<5\%$) 
source fractions are 3\%, 3\%,
1\%, and 3\%, respectively, consistent with the 
expected false-positive rate. This result is likely due to
the low S/N in this dataset compared to that of the 
binned data. 

More detailed analyses based on such a binning strategy 
will be presented in another paper 
(X.~C.~Zheng et al., in preparation).
Those analyses suggest that short-term (days-to-months) 
variability of our sources by large amplitudes 
(e.g., more than a factor of two) is rare, similar to our 
results on long time scales (Sections~\ref{sec:flux_15} 
and \ref{sec:sigmaexc}). 

\subsection{Spectral Variability}\label{sec:spec_analyses}
\subsubsection{Method}\label{sec:spec_fit_method}
The photometric analyses in Section~\ref{sec:PFHR} roughly 
evaluate the spectral normalization and shape changes.
To characterize the spectral variability more accurately
as well as gain improved physical insights, we perform 
spectral fitting for each source. The fitting is based on
full-band (observed-frame \hbox{0.5--7}~keV) spectra to maximize 
the available counts. Given our high-quality \xray\ 
spectra (with a median full-band S/N \hbox{$\approx 17$}\ in
each epoch) 
owing to the good angular resolution and low source-cell 
background of \chandra, 
we can still do reliable spectral
fitting even though the counts per epoch (median \hbox{$\sim350$})
are moderate. Note  that we are typically accessing 
penetrating rest-frame \xray s up to \hbox{$\approx10-30$~keV}.

\begin{figure}
\includegraphics[width=\linewidth]{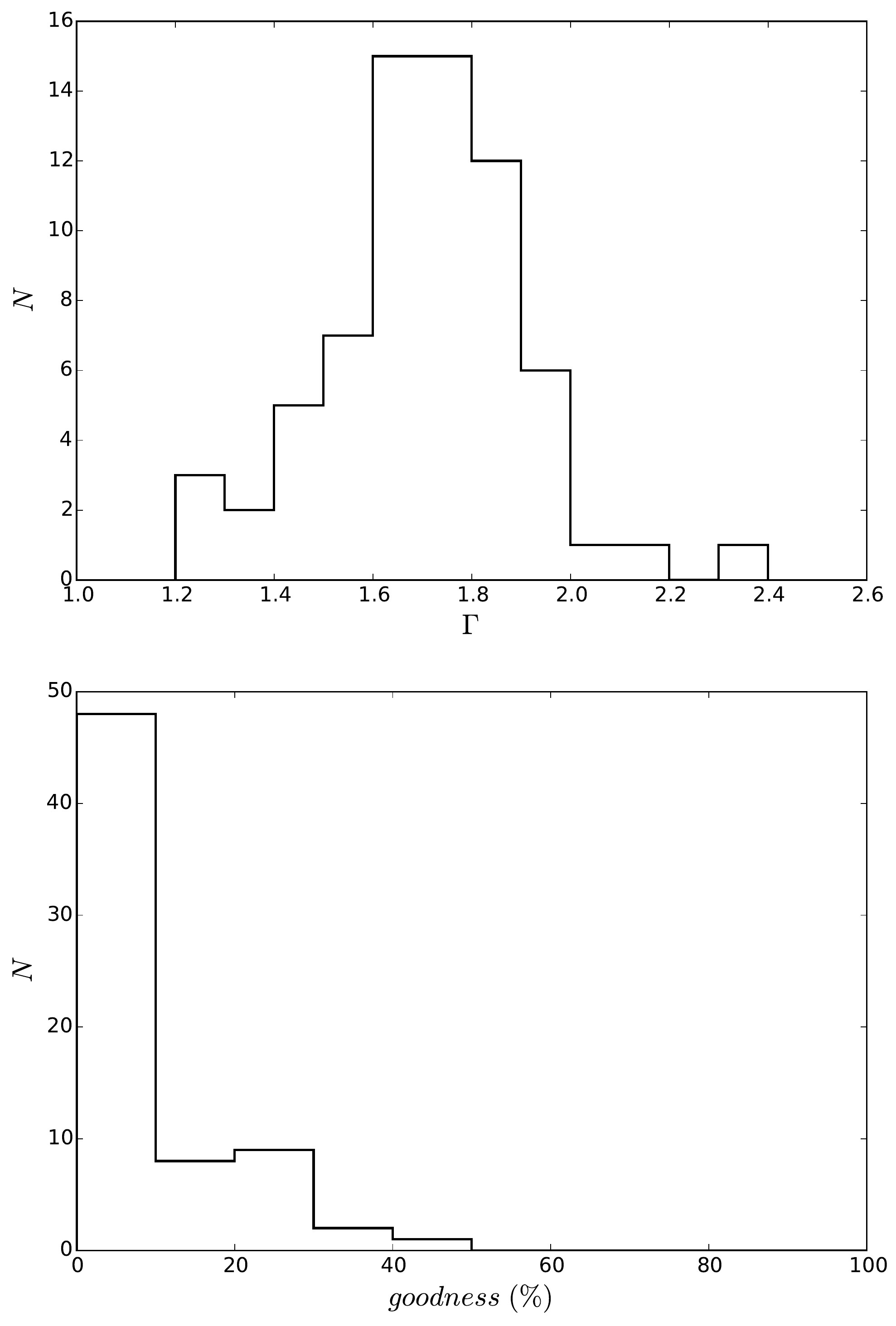}
\caption{Upper panel: best-fit $\Gamma$\ 
histogram. Lower panel: $goodness$\  
histogram. The $goodness$\ distribution 
indicates the $wabs \times zwabs \times 
pegpwrlw$\ model 
generally describes our data acceptably.}
\label{fig:gam_goodness}
\end{figure}

We use XSPEC v12.9.0i \citep{arnaud96} to carry out spectral  
fitting. Counts are not binned over different ACIS
pulse height amplitude (PHA) channels, 
and the Cash statistic \citep{cash79} is used for fitting. 
Considering the available counts, 
we adopt a simple fitting model of
a power law with Galactic and intrinsic absorption
$wabs \times zwabs \times pegpwrlw$\
(see \citealt{morrison83} for the 
$wabs$\ and $zwabs$\ models).
We adopt $pegpwrlw$\ 
(normalized over a finite energy band)
instead of the widely used $powerlaw$\
(normalized at observed-frame 1~keV)
for the reasons described in Appendix~\ref{app:err}.
We fix the redshift ($zwabs$) as the adopted 
value in Section~\ref{sec:sample}. 
The allowed ranges of \nh\ ($zwabs$) and $\Gamma$\ 
($pegpwrlw$) are set to $10^{19}-10^{24}\ \rm cm^{-2}$ 
and $1.2-2.4$, respectively.  
The counts of our sources in each epoch (median \hbox{$\sim350$}) generally 
cannot constrain $\Gamma$\ effectively; e.g., \cite{brightman13}
found that even a simple $powerlaw$\ model would require 
$\gsim 1000$~counts to obtain an uncertainty 
(90\% confidence level) of $\Gamma$\ 
within 0.2. 
Studies of quasars and local luminous AGNs show that 
spectral variability generally follows a ``softer when brighter'' behavior 
\citep[e.g.,][]{sobolewska09, gibson12, sarma15, connolly16}.\footnote{
This behavior likely changes when the Eddington ratio is low 
(i.e.,$\lsim 10^{-3}$; e.g., \citealt{yang15, connolly16}),
but the Eddington ratios for our sources are likely to be higher 
than this threshold, considering their luminosities 
(see Figure~\ref{fig:z_Lx_dis}) and 
that they are the \xray\ brightest AGNs in the \cdfs. }
However, based on the empirical \xray\ \hbox{flux--$\Gamma$}\ relations 
given by those studies, the $\Gamma$\ variability amplitudes 
for our sources are expected to be small and not detectable given
the counts we have in each epoch. For example, 
flux variability by a factor of two (about the upper limit of our 
flux-variability amplitudes; see Section~\ref{sec:flux_15}) only produces 
\hbox{$\Delta \Gamma \approx 0.1$}.
Motivated by this expected near constancy of 
$\Gamma$, we thus simultaneously fit the spectra 
of all four epochs by linking their 
photon indexes $\Gamma$ ($pegpwrlw$), 
assuming no substantial $\Gamma$\ variability.
Considering the low source-cell backgrounds and high count
numbers for our sources, we do not specifically model the 
background but employ the XSPEC default background modeling 
strategy.\footnote{See 
https://heasarc.gsfc.nasa.gov/xanadu/xspec for details.}
The $norm_i$\ (i.e., normalization of $pegpwrlw$) 
and \nh$_{,i}$ ($zwabs$) are set free and not linked across epochs, 
where subscripts $i$\ denote the epoch indexes. 
This spectral fitting yields best-fit model 
parameters, i.e., $\Gamma_{\mathrm{fit}}$, 
$norm_{i}$, and \nh$_{,i}$, and the errors of $norm_{i}$.
The errors of  \nh$_{,i}$\ are estimated with the method detailed in 
Appendix~\ref{app:err}. This method is critical for obtaining 
accurate errors of \nh$_{,i}$ for the reasons explained in 
Appendix~\ref{app:err}.
The $\Gamma_{\mathrm{fit}}$ distribution is shown 
in Figure~\ref{fig:gam_goodness}. 
The distribution is 
similar to that found by \cite{tozzi06}, except that 
our dispersion is smaller likely due to our larger 
numbers of counts.
We set both Galactic and intrinsic absorption to none 
and calculate the absorption-corrected \xray\ luminosity
in the rest-frame \hbox{2--10}~keV band for each epoch. 
The unweighted-mean luminosity (absorption column 
density) of the four epochs is denoted as \lx\ (\nh) 
and widely used throughout this paper.
 
Since we fit the unbinned spectra with the
Cash statistic, the spectral-fit quality 
cannot be inferred directly from the best-fit
statistic, e.g., $\chi^2/\rm dof$\ in the 
minimum $\chi^2$\ fitting case.
Instead, we perform goodness-of-fit Monte Carlo 
simulations for each source in XSPEC.\footnote{
For more details, see https://heasarc.gsfc.nasa.gov/xanadu
/xspec/manual/XSappendixStatistics.html and 
\hbox{http://xraygroup.astro.noa.gr/Webpage-prodec/}\\
documentation.html\#good} 
XSPEC simulates 1000 spectra from the best-fit parameters, 
and calculates the fraction ($goodness$) of simulated spectra with 
KS statistic less than that of the observed spectrum.  
Here, the KS statistic is used to describe the 
``similarity'' between a spectrum and the model, similar 
to $\chi^2$\ in the minimum $\chi^2$\ fitting case.
In other words, $goodness$\ can be interpreted
as the confidence level to reject the model.
The distribution of $goodness$\ is presented 
in Figure~\ref{fig:gam_goodness}. 
The fits are acceptable 
($goodness < 50\%$, e.g., \citealt{corral15}) 
for all our sources.
The worst case is \xidctk\ with \hbox{$goodness=44\%$}, 
a Compton-thick candidate reported in the literature 
\citep{tozzi06, comastri11}.
A detailed discussion of this source is presented 
in Appendix~\ref{sec:ct_AGN}.

\subsubsection{Flux Variability over 15 Years}\label{sec:flux_15}
Long-term \xray\ variability may affect physical 
inferences about \xray\ source populations; e.g., the 
\hbox{star-formation} vs.\ \hbox{AGN-activity} connection 
may be affected by \hbox{$\sim$Myr} timescale AGN variability  
(e.g., \citealt{hickox14}). It is thus of interest to assess whether 
\xray\ variability on the longest observable time scales (i.e., between 
epoch 1 and epoch 4) substantially affects the appearance 
of the overall \xray\ source population. 

\begin{figure}
\includegraphics[width=\linewidth]{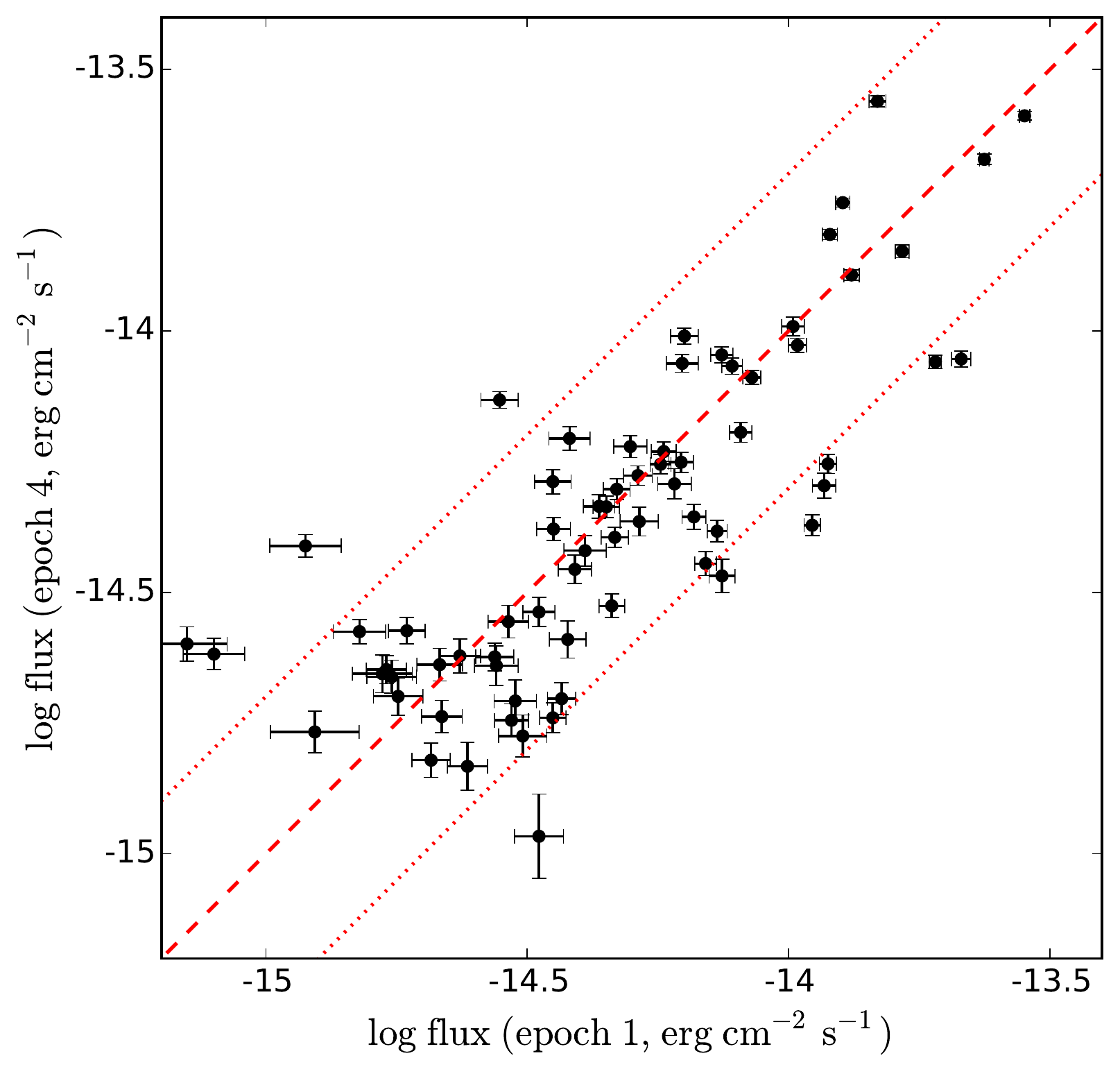}
\caption{Flux (epoch 4) vs.\ flux (epoch 1).
The fluxes are calculated from our spectral-fitting results 
(see Section~\ref{sec:spec_fit_method}), and 
are not corrected for Galactic or intrinsic absorption. 
The energy band is observed-frame \hbox{0.5--7}~keV (i.e., full-band).
The red dashed line indicates no flux change; 
the red dotted lines indicate flux changes by a factor of 2. 
The ranks of fluxes are similar in the two epochs. 
Thus, \xray\ variability over 15~years does not 
significantly affect the appearance of the overall \xray\ source 
population.\\
}
\label{fig:flux_vs_flux}
\end{figure}

Figure~\ref{fig:flux_vs_flux} displays the full-band fluxes of 
epoch 1 and epoch 4. Applying Spearman's test on 
the fluxes of epoch 1 and epoch 4 yields \hbox{$\rho=0.82$}\ and a \hbox{$p$-value} 
of $10^{-17}$; the large value of $\rho$\ indicates a good association 
of ranks. In general, the brightest (faintest)
sources remain the brightest (faintest) after 15~years. Only 11 
(\hbox{$11/68=16\%$}) sources have flux changes greater than a factor 
of two. Qualitatively similar results are obtained when we pair
any two of our available epochs.

The rarity of extremely variable sources in the distant universe
over long time scales 
is broadly consistent with a recent study focusing on low-redshift 
AGNs \citep{strotjohann16}.
Our results prove that, at least on a 15-year 
observed-frame time scale, basic 
inferences about distant AGN source populations are not strongly affected by
variability, e.g., the expected \hbox{star-formation} vs.\ 
\hbox{AGN-activity} connection is unlikely to be hidden by AGN 
variability on a rest-frame time scale of \hbox{$\sim 5-10$~years}.
These results also show that \xray\ changing-look AGNs are rare, 
and they constrain the frequency of other types of novel 
long-term \xray\ variability (see Section~\ref{sec:intro}).
Setting direct \xray\ constraints on much longer time scales
is unfortunately limited by, e.g., the age of \xray\ astronomy.
Cosmic \xray\ surveys often bin observations 
performed over \hbox{1--15}~years; our results indicate that long-term 
variability is not likely to affect the interpretation of such 
survey results greatly.

\subsubsection{Identification of \lx- and \nh- Variable Sources}
\label{sec:identify}
We identify \lx- and \nh-variable sources using the 
Akaike information criterion \citep[AIC,][]{akaike74}.
The AIC is based on information theory and does not 
assume a particular model-fitting technique or distribution
of uncertainties \citep[e.g.,][]{burnham02, kelly07b, buchner14}. 
AIC is defined as \hbox{$\mathrm{AIC}=C+2k$}, 
where $C$\ is the fitting statistic 
(i.e., the Cash statistic in our case) and $k$\ is the 
number of free parameters in the model.\footnote{
Note that we do not need to use the AICc, the 
corrected AIC designed for the cases where sample size $n$\ is 
\hbox{$\approx k^2$}. This is because our \hbox{$n=$ 
number} of spectral PHA bins for all four epochs together 
(\hbox{$\approx 2000$})\ is much greater than $k^2$\ (\hbox{$\approx 50$}).} 
Models with smaller AIC are considered to be more probable.

For each source, we calculate AIC (denoted as $\rm{AIC_0}$)
for the spectral-fitting results in Section~\ref{sec:spec_fit_method}.
To evaluate the significance of \lx\ variability, we 
link $norm_i$\ in all spectra of the four epochs,\footnote{
Hereafter, we use parameter $norm$\ to evaluate \lx\ variability, 
since \lx\ is proportional to $norm$\ for a given source assuming 
no $\Gamma$\ variability.}
and redo the fitting. Other settings of the model are 
the same as in Section~\ref{sec:spec_fit_method}. 
We calculate AIC for the new fitting results, denoted as 
AIC$_1$. We use the difference 
\hbox{$\Delta \mathrm{AIC}_L = \rm AIC_1 - AIC_0$}\
to evaluate the significance of \lx\ variability. 
If \hbox{$\Delta \mathrm{AIC}_L>4$}\ 
\citep[e.g.,][]{burnham02},\footnote{
We have also performed classic $\chi^2$\ tests for variable 
source identification, assuming our errors are 
Gaussian. For both \lx\ and \nh\ variability, 
we find good correlations between $\Delta$AIC and 
$\chi^2_{\rm dof=3}$, with \hbox{$\Delta$AIC$=4$}\ corresponding to 
\hbox{$\chi^2_{\rm dof=3}\approx 10$} (i.e., $p$-value \hbox{$\approx 2\%$}).
However, we prefer the AIC approach because it does not 
rely on the assumption of Gaussian errors.}\label{foot:aic_vs_chi2}
we assign this source as an \lx-variable source. Similarly, 
we identify \nh-variable sources by calculating the AIC 
difference $\Delta$AIC$_N$ between 
\nh-linked and unlinked models and comparing it with
the threshold of 4. 
The $\Delta \mathrm{AIC}_L$\ and $\Delta \mathrm{AIC}_N$\
for each source are listed in Table~\ref{tab:src_info}.
 
The resulting \lx- and \nh-variable source fractions are 
74\% (50/68) and 16\% (11/68), respectively. 
Four of the 11 \nh-variable sources are  
$HR$-variable (Section~\ref{sec:PFHR}).\footnote{s
Our AIC method selects variable 
sources at a significance level of \hbox{$\approx 98\%$} 
(see Footnote~\ref{foot:aic_vs_chi2}). Only 6 sources show $HR$\ 
variability at this significance level (i.e., $P_{HR}<2\%$),
and all 4 \nh-variable and $HR$-variable sources are included 
among these 6.
}
The other 7 sources have large $P_{HR}$\ values ($>10\%$). 
One reason for this result is likely to be that 
the definition of HR (Equation~\ref{equ:hr}) 
uses observed-frame 2~keV as the boundary between the soft and hard bands.  
This choice of 2~keV is just by convention. It corresponds to different 
rest-frame energies for different sources and might 
not be sensitive in selecting some \nh-variable sources.
We checked the 4-epoch spectra of the 7 sources, and found that 
2~keV as a boundary is either too high or too low to detect 
their \nh\ variability effectively.

\begin{figure}
\includegraphics[width=\linewidth]{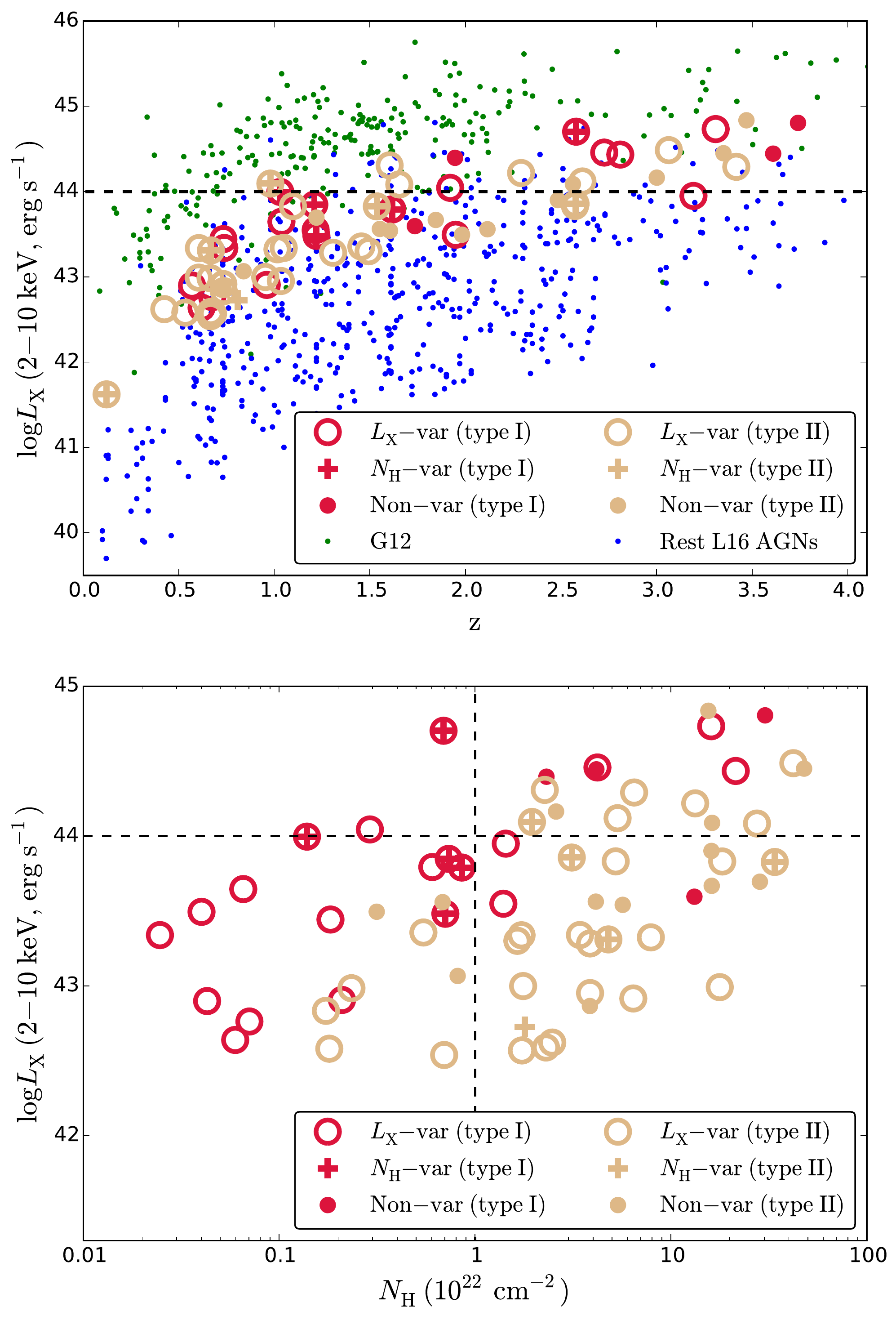}
\caption{Upper panel: \lx\ vs.\ $z$. The open circles, 
crosses and solid circles indicate \lx-variable,
\nh-variable, and non-variable sources, respectively. 
A source can be both \lx-variable and \nh-variable.
Red and brown colors indicate type I and type II sources, respectively.
The green and the blue points indicate G12 quasars and the rest
of the AGNs in L16. 
At a given redshift, our sources 
are generally less luminous than G12 quasars and more luminous
than the other L16 AGNs. 
The dashed horizontal line indicates our definition
of quasars (i.e., \hbox{\lx$>10^{44}\ \mathrm{erg\ s^{-1}}$}, 
see Section~\ref{sec:sample}). Lower panel: \lx\ vs.\ \nh. The 
symbols have the same meanings as for the upper panel. The 
horizontal and vertical dashed lines indicate our definition
of quasars and the common definition of 
\xray\ obscured AGNs (i.e., \hbox{\nh$>10^{22}\ \mathrm{cm^{-2}}$}), 
respectively. 
}
\label{fig:Lx_vs_z_and_Lx_vs_nH}
\end{figure}

Most (10/11) \nh\ variable sources are also \lx\ variable. 
The median $\Delta \mathrm{AIC}_L$\ 
($\Delta \mathrm{AIC}_N$) for \lx-variable (\nh-variable) 
sources is 43 (6.2). Therefore, our \lx\ variability is 
generally more significant than \nh\ variability.
The BAL quasar \xidbal\ shows \lx\ variability 
(\hbox{$\Delta$AIC$_L=4.9$}) but not \nh\ variability
(\hbox{$\Delta$AIC$_N=-2.7$}).
We investigate three significantly variable sources,
\xidone, \xidtwo, and \xidthr, as illustrative examples 
in Appendix~\ref{sec:sig_var}.
Notably, \xidthr\ transitions from an \xray\ unobscured
to obscured state. 
The positions of the identified variable sources 
in the \lx-$z$ and \lx-\nh\ planes are indicated in 
Figure~\ref{fig:Lx_vs_z_and_Lx_vs_nH}. 
G12 performed similar variability analyses but for
optically selected SDSS quasars. Their quasars are included
in the \lx-$z$ plot to demonstrate the differences between 
their sample and ours. Luminosities for the G12 objects are 
estimated assuming a power law with \hbox{$\Gamma=1.8$}.  
The typical luminosity of our sources is generally about one order of 
magnitude lower than that of the SDSS quasars at a given redshift.
The majority (\hbox{$\approx 60\%$}) of the \xray\ obscured quasars
(i.e., \hbox{\nh $>10^{22}\ \mathrm{cm^{-2}}$}\ 
and \hbox{\lx $>10^{44}$~erg~s$^{-1}$}) in our sample are 
\lx-variable, supporting the idea that we are observing their 
central \xray\ emitting regions directly. 
Also, 7 of our 11 optically classified type II quasars show 
\lx\ variability (Figure~\ref{fig:Lx_vs_z_and_Lx_vs_nH}).
The well-studied optically classified type II quasar at $z=3.70$\ 
in the \cdfs\ 
\citep[e.g.,][]{norman02, comastri11} is not 
included in our sample due to its limited number of 
available counts (see Section~\ref{sec:sample}).

\subsubsection{Relation Between \lx\ and \nh\ Variability}
\label{sec:lx_vs_nh_var}
If the observed \nh\ variability were caused by changes 
of the ionization parameter of the obscuring matter, an 
anticorrelation between \lx\ and \nh\ variability might be expected; 
when \lx\ rises, the obscuring matter would become 
more ionized and generally less opaque. We performed a 
Spearman's test on $L_{\mathrm{X},i}/L_{\mathrm{X}}$\
and $N_{\mathrm{H},i}/N_{\mathrm{H}}$\ for the 10 sources 
together that show both \lx\ and \nh\ variability, 
but do not find a significant anticorrelation. There is also no 
significant anticorrelation produced if we expand the Spearman's 
test to all sources. 
Therefore, the \nh\ variability is not likely to be primarily 
driven by changes of ionization parameter. 
Nevertheless, some ionization-driven \nh\ variability might still exist 
if there are year-scale time delays between \lx\ and \nh\ 
variability due to, e.g., a low density of the absorber 
\citep[e.g.,][]{krolik95, collinge01}.

\subsubsection{Variability Amplitude Estimation}\label{sec:var_scale}
\label{sec:sigmaexc}
In this section, we illustrate and evaluate our method to quantify 
variability amplitude, and the results are used in all subsequent material.
We use the normalized excess 
variance (\sigmaexc, e.g., \citealt{turner99, vaughan03}) to estimate 
intrinsic variability scale for \lx\ (\nh) of each source. 
\sigmaexc\ is calculated as
\begin{equation}\label{equ:sigma_sq}
\sigma_{\rm exc}^2 = \frac{1}{N \langle x \rangle^2}
	\sum_{i=1}^{N} [(x_i-\langle x \rangle)^2-(\delta x_i)^2],
\end{equation}
where $N$\ is the number of epochs (i.e., 4); 
$x_i$\ and $\delta x_i$\ are the best-fit
$norm_i$\ (\nh$_{,i}$) and its 1$\sigma$ error 
$\delta norm_i$\ ($\delta N_{\mathrm{H},i}$), respectively;
$\langle x \rangle$\ is the unweighted mean of $norm_i$\ (\nh$_{,i}$).
The error of \sigmaexc\ is estimated as $s_D/(\langle x \rangle^2 \sqrt{N})$,
where
\begin{equation}\label{equ:sigma_sq_err}
	s^2_D = \frac{1}{N-1}
	\sum_{i=1}^{N} [(x_i-\langle x \rangle)^2-(\delta x_i)^2 - 
		  	\sigma^2_{\rm exc} \langle x \rangle^2 ]^2.
\end{equation}
is the variance of the summed terms 
$(x_i-\langle x \rangle)^2-(\delta x_i)^2$\ in 
Equation~\ref{equ:sigma_sq}. 
We calculate the \lx\ \sigmaexc\ (\sigmaexcL) for each source. 
We only derive the \nh\ \sigmaexc\ 
(\sigmaexcN) for the 35 sources with all four epochs having  
\nh$_{,i}$ fractional errors $\delta N_{\mathrm{H},i}/N_{\mathrm{H},i}<0.4$\
(Appendix~\ref{app:err}). 

The results are listed in Table~\ref{tab:src_info}.
While \sigmaexcL\ and \sigmaexcN\ are designed to
measure intrinsic variability under ideal conditions, 
they may also be affected by the available counts 
\citep[e.g.,][]{allevato13}. 
However, the biases are minimized for high S/N, as is the case
for our data (Section~\ref{sec:spec_fit_method}).
Indeed, Spearman's test demonstrates no significant dependence 
of \sigmaexcL\ and \sigmaexcN\ on the number of counts.

\subsubsection{Variability Dependence on Luminosity, 
\xray\ Absorption, and Redshift}
\label{sec:sigma_vs_Lx_nH}
\begin{figure}
\includegraphics[width=\linewidth]{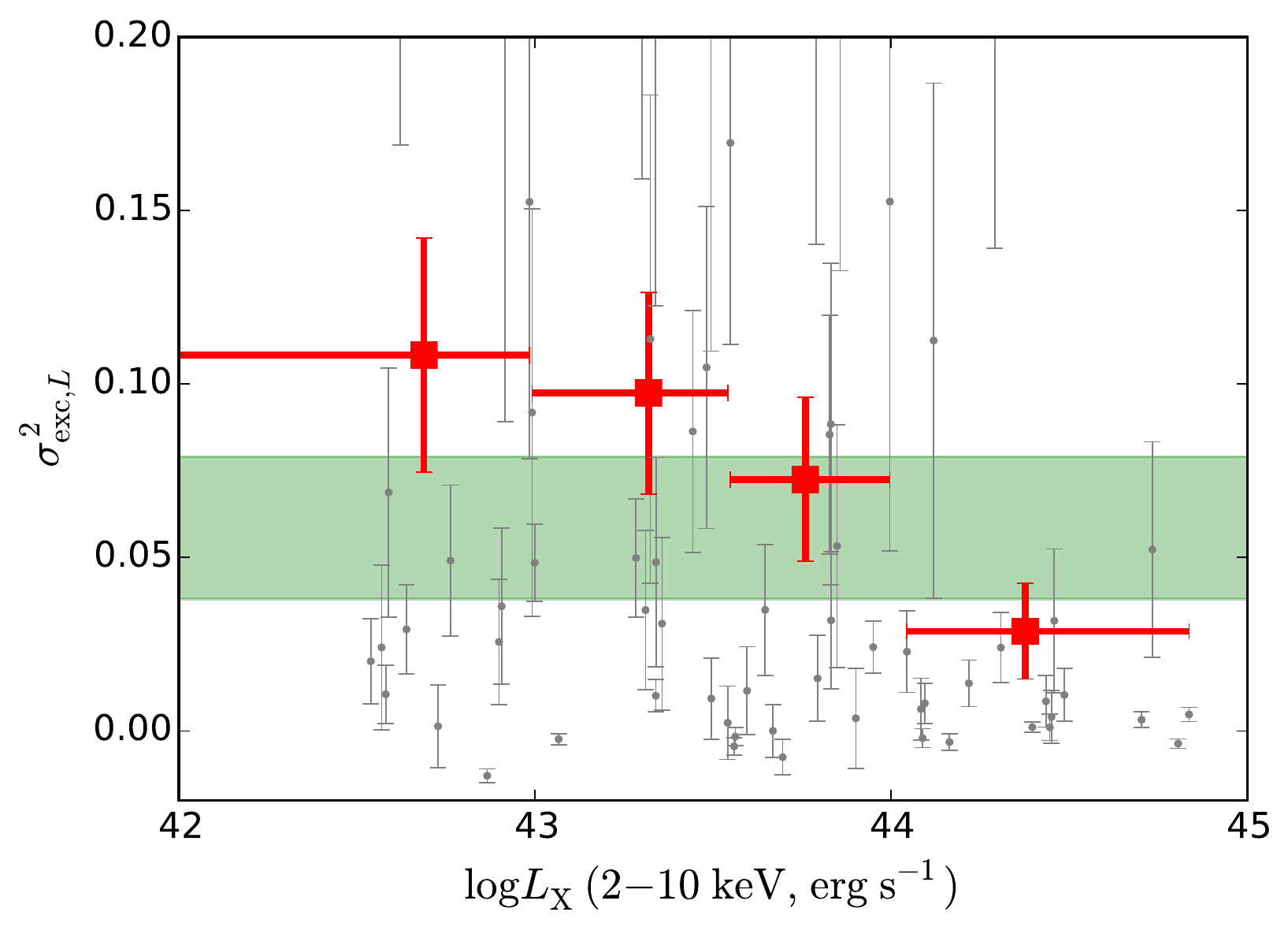}
\caption{
\sigmaexcL\ vs.\ \lx. \lx\ values are the
mean values in Section~\ref{sec:spec_fit_method}. 
Gray points indicate each individual source.
The red squares
indicate the mean value of each bin. 
The red horizontal error 
bars indicate the bin widths; 
the red vertical error bars indicate
the errors on the mean values calculated analytically 
(see Section~\ref{sec:sigma_vs_Lx_nH}).
The green shaded region indicates the \sigmaexcL\ expected from
a power-law PSD with a slope of $-1$\ (see Section~\ref{sec:psd}).
Our \sigmaexcL\ decreases at high luminosity. 
This is likely caused by the fact that our highest sampling 
frequency exceeds the PSD break frequency of the luminous sources
(Section~\ref{sec:psd}). 
}
\label{fig:sigma_vs_Lx_nH}
\end{figure}

It has been well established that the strength of AGN \xray\ 
flux variability decreases as luminosity increases on
time scales from minutes to about a year
\citep[e.g.,][]{paolillo04, ponti12}. Also, studies 
of local Seyfert galaxies show that more obscured sources 
tend to be less variable on short time scales 
\citep[minutes to hours; e.g.,][]{turner97, hernandez15}. 
Therefore, it is of interest to investigate the dependence of
long-term (years) variability on luminosity 
and absorption level.
Figure~\ref{fig:sigma_vs_Lx_nH} shows the dependence of 
\sigmaexcL\ on \lx. 
The unweighted mean of \sigmaexcL\ decreases toward
higher luminosity. 
For each bin, the error on the mean is calculated as the standard
deviation of individual \sigmaexcL\ divided by $\sqrt{N}$, where 
$N$\ is the number of sources in the bin [i.e., Equation~13 
of \cite{allevato13}; but note that a power index of 
2 is missing in their summed term].
The mean value in each bin can well 
represent the typical variability amplitude of the sources 
in the bin, despite the large error bars for individual
sources \citep[e.g.,][]{allevato13}; the median has 
not been established to represent the typical variability amplitude.
Spearman's test applied to the individual sources shows a 
significant anticorrelation between \sigmaexcL\ 
and \lx\ (Spearman's \hbox{$\rho=-0.31$}, \hbox{$p$-value$=0.009$}).
This anticorrelation is mainly caused
by the \sigmaexcL\ difference between quasars 
(\hbox{\lx $> 10^{44}$~erg~s$^{-1}$}) and other AGNs. After 
removing quasars (only 19 sources), we find no significant
relation between \lx\ and \sigmaexcL. 
There is also tentative evidence of an anticorrelation
between \sigmaexcL\ and \nh: 
Spearman's \hbox{$\rho=-0.23$}, \hbox{$p$-value$=0.06$}.
However, this result might be a byproduct of the \sigmaexcL-\lx\
relation, since in our counts-limited sample 
(i.e., \hbox{$>600$} net counts required) heavily obscured 
sources tend to be more luminous compared
to less-obscured sources 
(see Figure~\ref{fig:Lx_vs_z_and_Lx_vs_nH}).
This interpretation is supported by the fact that no 
significant \hbox{\sigmaexcL-\nh}\ relation is found for 
luminosity-controlled samples
(\hbox{\lx$\leq 10^{44}\ \mathrm{erg\ s^{-1}}$}\ and 
\hbox{\lx$> 10^{44}\ \mathrm{erg\ s^{-1}}$}). 

The anticorrelation between \sigmaexcL\ 
and \lx\ is unlikely to be a bias caused by the available 
counts, since we are in a high S/N regime
(see Section~\ref{sec:sigmaexc}).
Another bias might 
exist considering that we are probing generally shorter 
time scales for more luminous sources, since they have relatively
high redshifts (see Figure~\ref{fig:Lx_vs_z_and_Lx_vs_nH}),
increasing the effects of time-dilation. To test this point,
we drop the first-epoch data for the low-redshift sample 
(\hbox{$z<2$}, median \hbox{$z=1.0$}), while keeping all data for 
the high-redshift sample (\hbox{$z\geqslant2$}, median 
\hbox{$z=2.8$}). 
Hence, the observed-frame total time spans are about 8
and 15 years for the low- and high-redshift samples, 
respectively (see Table~\ref{tab:obs}). The median 
values of the rest-frame total time spans are both 
about 4 years [low-redshift: 8/(1+1.0), high-redshift: 
15/(1+2.8)]. We calculate \sigmaexcL\ for 
this data set and find Spearman's 
\hbox{$\rho=-0.29$}, 
\hbox{$p$-value$=0.015$}, similar to the 
results derived from the original data set. 
The results are also similar if we drop the 
fourth-epoch data instead of the first-epoch data 
for the low-redshift sample. We conclude 
the anticorrelation between 
\sigmaexcL\ and \lx\ is not caused by a
bias of different rest-frame time spans. 
A bias might also arise from the fact that our 
observed-frame full energy band corresponds to
different rest-frame energy bands. There is 
some evidence of energy-dependent 
variability reported in studies of Seyfert galaxies 
\citep[e.g.,][]{markowitz04, ponti12}. 
To estimate this effect, we fit the 
spectra of the rest-frame \hbox{2--10 keV} band for each 
source. 
We calculate the luminosity excess standard deviation 
(\sigmaexcLr)\ and its
uncertainty. The derived \sigmaexcLr\ matches well with 
\sigmaexcL, with only six cases showing 2$\sigma$\
or higher deviation from 
\hbox{\sigmaexcLr$=$\sigmaexcL}, i.e.,
\begin{equation}
|\sigma^2_{\mathrm{exc},L,\mathrm{rest}}-\sigma^2_{\mathrm{exc},L}| > 
2\mathrm{max}(\delta\sigma^2_{\mathrm{exc},L,\mathrm{rest}},\
\delta\sigma^2_{\mathrm{exc},L}),
\end{equation}
where $\delta$\sigmaexcLr\ and 
$\delta$\sigmaexcL\ are the uncertainties of 
\sigmaexcLr\ and \sigmaexcL,
respectively. In agreement with \sigmaexcL, 
\sigmaexcLr\ is also anticorrelated with \lx:
Spearman's \hbox{$\rho=-0.30$}, 
\hbox{$p$-value$=0.012$}.

Several previous studies suggest that at a given 
luminosity level, sources at higher redshifts tend to have stronger 
\xray\ variability \citep[e.g.,][]{almaini00, manners02, paolillo04}.
To assess this point, we perform Spearman's test on \sigmaexcL\ 
and redshifts for luminosity-controlled samples 
(\lx$\leq 10^{44}\ \mathrm{erg\ s^{-1}}$\ 
and \lx$> 10^{44}\ \mathrm{erg\ s^{-1}}$). 
The results show no significant correlation between \sigmaexcL\
and redshift for both samples.
However, the majority of the \
\lx$\leq 10^{44}\ \mathrm{erg\ s^{-1}}$\ and 
\lx$> 10^{44}\ \mathrm{erg\ s^{-1}}$\ sources 
are at $z\lsim2$\ and $z\gsim 2$\ (Figure~\ref{fig:Lx_vs_z_and_Lx_vs_nH}),
respectively. Thus, we cannot test the redshift dependence 
over a wide redshift range.

We do not find significant \sigmaexcN\ dependence 
on \lx, \nh, or $z$\ for the 35 sources with \sigmaexcN\
calculated (i.e., the sources with all four epochs having  
$\delta N_{\mathrm{H},i}/N_{\mathrm{H},i}<0.4$).

\subsubsection{Variability Dependence on Optical Spectral Type and Color}
\label{sec:opt_type}
\begin{figure}
\includegraphics[width=\linewidth]{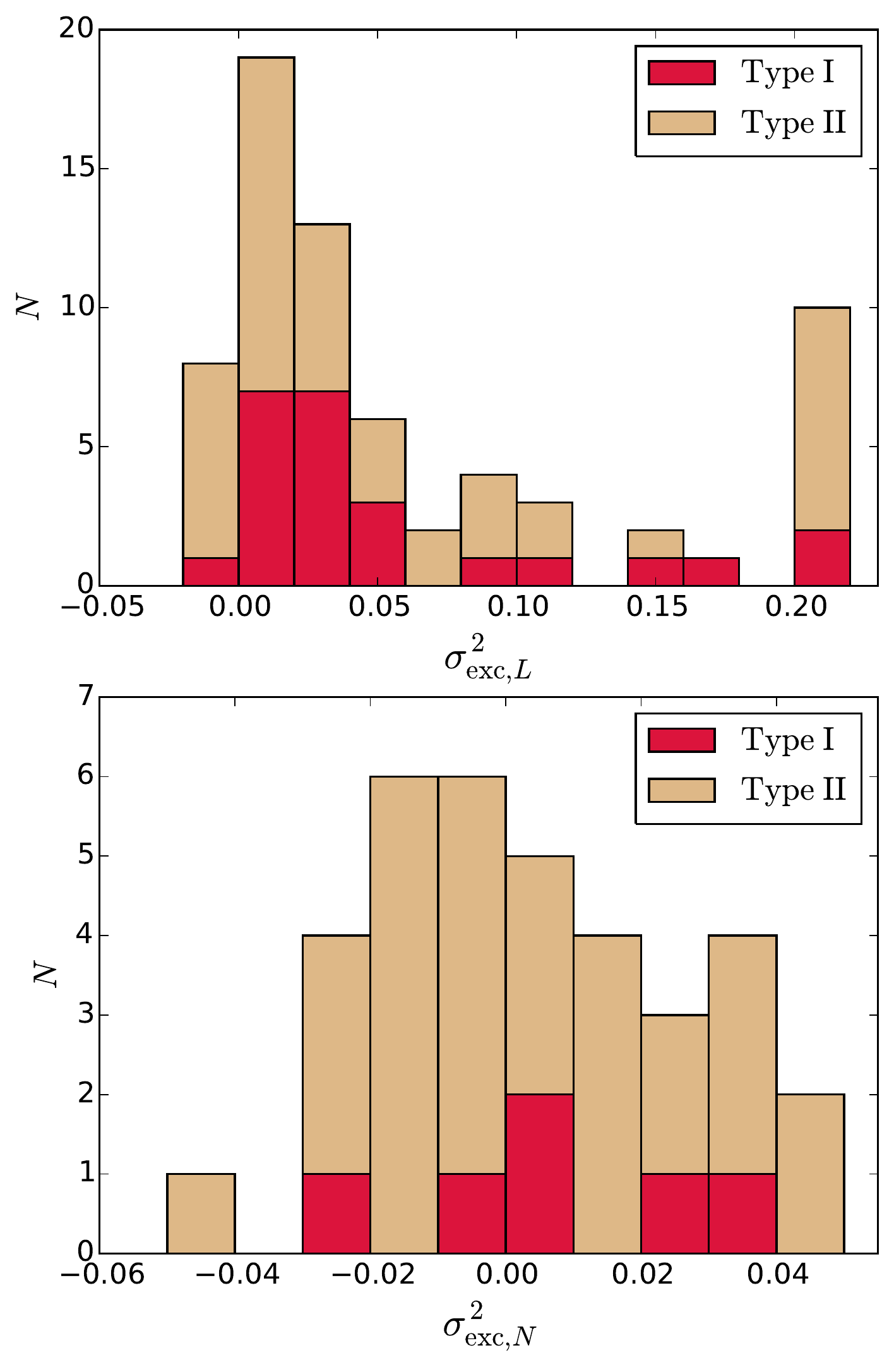}
\caption{Upper panel: the \sigmaexcL\ 
histogram. The red and brown colors indicate type I
and type II AGNs, respectively. 
A KS test shows the dependence of \sigmaexcL\ 
on optical type is not significant.
The rightmost column indicates all sources with 
\sigmaexcL\ $>0.2$.
Lower panel: \sigmaexcN\ histogram. 
Only the 35 sources with \sigmaexcN\ calculated 
are shown here (Section~\ref{sec:opt_type}). 
The rightmost column indicates all sources with 
\sigmaexcN\ $>0.04$.
}
\label{fig:sigma_hist}
\end{figure}

Type I AGNs have a higher \lx\ variable source fraction 
than their type II counterparts (83\% vs.\ 68\%), likely due
to their higher numbers of counts (Figure~\ref{fig:cnt_hist}).
Also, type I AGNs have a higher fraction 
of \nh\ variable sources (21\% vs.\ 14\%).
Figure~\ref{fig:sigma_hist} displays histograms of 
the \sigmaexcL\ and \sigmaexcN\ distributions for different 
optical spectral types. 
There are very few \hbox{type I} AGNs in the 
\sigmaexcN\ histogram, because 
most \hbox{type I} AGNs are \xray\ unobscured. 
For the \sigmaexcL\ distributions, a KS test shows no 
apparent differences between \hbox{type I} and \hbox{type II} AGNs.
Since optical spectral type is often related to 
the rest-frame optical color (e.g., $u-g$; see 
Section~\ref{sec:sample}), we also check the 
\sigmaexcL\ dependence on rest-frame $u-g$\ color  
but do not find a significant relation.
The similarity of long-term (years) \xray\ variability 
between \hbox{type I} and \hbox{type II} AGNs has also been found 
by previous studies of both distant AGNs 
\citep[e.g.,][]{lanzuisi14} and local Compton-thin Seyfert galaxies 
\citep[e.g.,][]{turner97, hernandez15}, though
\hbox{Seyfert IIs} tend to be less variable than \hbox{Seyfert Is} 
on short time scales (minutes to hours)  
\citep[e.g.,][]{turner97, awaki06}. 

\subsubsection{Variability Dependence on Time Scale}
\label{sec:nh_var_vs_t}
Considering that our light curves are sparsely sampled 
(with four epochs), we cannot investigate the variability 
dependence on time scale for each source individually.
Instead, we consider our entire sample as an ensemble 
and investigate its \lx\ (\nh) variability on 
different time scales. We calculate the 
``structure function'' ($SF$, i.e., \hbox{ensemble-averaged}
fractional variability amplitude between two observations) 
as a function of rest-frame time interval ($\trest$).  

\begin{figure}
\includegraphics[width=\linewidth]{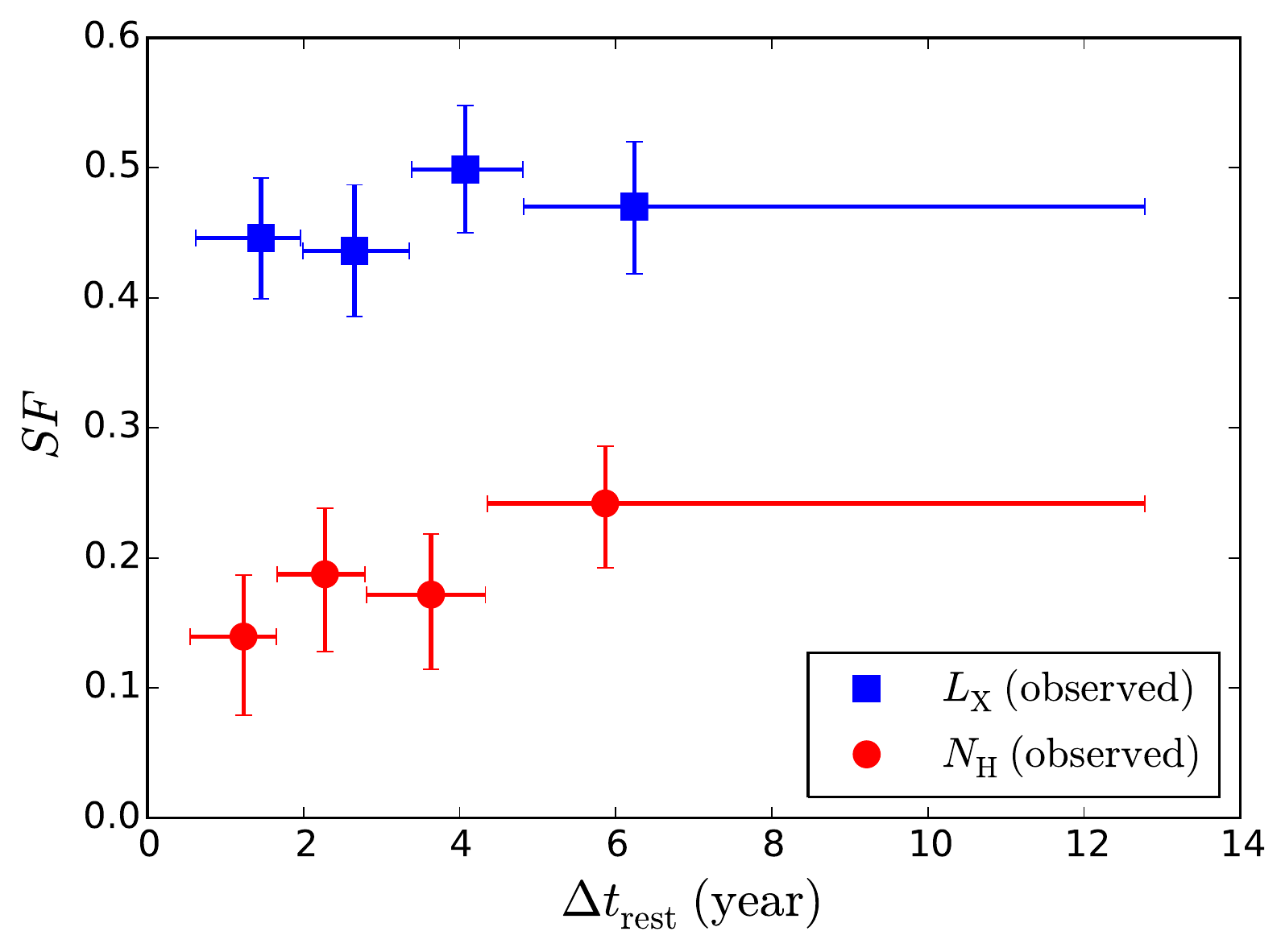}
\caption{
Structure functions ($SF$s) as a function of  
rest-frame time interval, $\trest$. The blue square 
points and the red circle points indicate the observed 
$SF$s for \lx\ and \nh, respectively. 
The $SF$\ for \lx\ is calculated based on 
non-quasar (\lx\ $<10^{44}$~erg~s$^{-1}$) sources only 
to avoid a bias (Section~\ref{sec:nh_var_vs_t}). 
The error bars of the $SF$s\ indicate 1$\sigma$\ 
uncertainties calculated from 
bootstrapping; the error bars of $\Delta t_{\rm rest}$\
indicate the bin width. 
}
\label{fig:f_vs_det_t}
\end{figure}

First, for each epoch pair of a source (6 pairs 
in total for each source) we obtain a variability factor  
\begin{equation}
f_v
= \frac{\Delta x}{\overline{x}}
= \frac{x_2 - x_1}
{( x_2 + x_1 )/2};
\end{equation}
and its uncertainty from error propagation
\begin{equation}
\delta f_v
= \frac{ x_1 x_2}{\overline{x}^2} \times
\sqrt{  \left(\frac{\delta x_1}{x_1}\right)^2 + 
	\left(\frac{\delta x_2}{x_2}\right)^2  },
\end{equation}
where $x_1$\ and $x_2$\ are the best-fit 
$norm_i$\ (\nh$_{,i}$)
of two different epochs; $\delta x_1$\ and $\delta x_2$\ 
are their 1$\sigma$\ errors.
Each $f_v$\ is associated with a 
$\trest$\ between two epochs. We then bin $f_v$\ 
with similar $\trest$\ and calculate the $SF$\ 
\citep[e.g.,][]{berk04} as
\begin{equation}
SF
= \sqrt{ \frac{\pi}{2} \langle|f_v|\rangle^2
		- \langle(\delta f_v)^2\rangle },
\end{equation}
where the angle brackets denote average values in
the $\trest$\ bin.  
We estimate the uncertainties of the $SF$\
from bootstrapping,\footnote{We calculate the 
confidence interval as the range between the 
16th and 84th percentiles.} 
and the results are displayed in Figure~\ref{fig:f_vs_det_t}. 

Our 19 quasars generally have weaker variability 
(Section~\ref{sec:sigma_vs_Lx_nH}) and shorter
$\trest$\ (due to cosmological time dilation; 
see Figure~~\ref{fig:Lx_vs_z_and_Lx_vs_nH}). 
Hence, they might cause the \lx\ $SF$\ to be lower at shorter 
$\trest$. To avoid this bias, we do not include them 
when calculating the \lx\ $SF$. 
The resulting \lx\ $SF$\ is relatively flat as a 
function of time scale, with perhaps a suggestion 
of rising toward longer time scales.

The fractional variability amplitude of \nh\ 
is generally smaller than that of \lx, and appears to 
increase as $\trest$\ increases. 
In the \nh\ $SF$\ calculations, 
we only include the 35 sources with all four epochs having 
$\delta N_{\mathrm{H},i}/N_{\mathrm{H},i}<0.4$\
(Appendix~\ref{app:err}).

\section{Summary, Discussion, and Future Prospects}\label{sec:conclusion}
In this paper, we have performed long-term 
(up to \hbox{$\approx 15$~yr}, observed-frame) \xray\ 
variability analyses for the 68 \xray\ brightest 
radio-quiet AGNs in the uniquely deep \cdfs; most of these objects are 
at redshifts of \hbox{0.6--3.1}, providing access to 
penetrating rest-frame \xray s up to \hbox{$\approx 10-30$~keV}.
AGNs like those studied here produce a significant fraction of cosmic
accretion power; in this sense, 
they are the typical AGNs of the Universe.
We have performed both photometric and spectral variability analyses, and
studied the dependence of variability on source properties and time scale.
We summarize our main results in Section~\ref{sec:summ}. In 
Section~\ref{sec:psd}, we interpret the \lx\ variability in the 
context of AGN PSD. 
We present practical future extensions to this work 
in Section~\ref{sec:future}.

\subsection{Summary of Main Results}\label{sec:summ}
The main results are the following:
\begin{enumerate}

\item Photometric analyses (Section~\ref{sec:PFHR}) show 
that at above a 95\% confidence level, 90\% (61/68) and 16\% 
(11/68) of our sources are variable in photon flux ($PF$) and 
hardness ratio ($HR$), respectively.
Our results confirm the prevalence of \xray\ $PF$\ variability
for typical AGNs in the distant universe
\citep[e.g.,][]{paolillo04, lanzuisi14}.
A large fraction of sources (\hbox{$\sim 50\%$})
is also found to be $PF$-variable within single epochs. 
However, $HR$\ variability is generally
insignificant within single epochs ($\lsim 1$~yr, observed-frame).

\item Spectral analyses (using a 
$wabs \times zwabs \times pegpwrlw$\
model; see Section~\ref{sec:spec_analyses}) 
demonstrate that the \lx- and \nh-variable 
source fractions are 74\% (50/68) 
and 16\% (11/68), respectively. 
Among the \xray\ obscured quasars, the 
\lx-variable source fraction is also high (\hbox{$\approx60\%$});
this includes the BAL quasar (\xidbal).
Large-amplitude flux variability is rare; most 
sources (84\%) have flux changes within a factor of 2
over 15~yr (observed-frame, see Section~\ref{sec:flux_15}). We 
do not find a significant anticorrelation between \lx\ and
\nh\ variability, as might be expected for a photoionized absorber
(see Section~\ref{sec:lx_vs_nh_var}).

\item We have quantified the fractional variability scale by 
calculating the normalized excess variance
(\sigmaexc) for each source (see Section~\ref{sec:var_scale}). 
Quasars with \lx\ $>10^{44}$~erg~s$^{-1}$\
generally have smaller variability amplitudes 
than less-luminous AGNs.
We have not found any significant dependence of \lx\ 
variability amplitudes on optical spectral type, 
consistent with the results of \cite{lanzuisi14}. 
Therefore, we appear to be 
observing the \xray\ emission of most type II AGNs directly 
from the central engine; this can occur if
\xray s are able to penetrate the obscuring material.

\item We have calculated $SF$s\ to illustrate
the variability dependence on rest-frame time scale
(Section~\ref{sec:nh_var_vs_t}).
The \lx\ $SF$\ is relatively flat;
the \nh\ $SF$\ appears to rise toward longer time scales.

\item A Compton-thick AGN candidate \citep{comastri11}
in our sample (\xidctk) shows notable \xray\ variability. 
Motivated by \cite{comastri11}, we used 
a reflection-dominated $wabs \times zwabs \times pexmon$\
model to perform spectral analyses. The results indicate that
both the reflection flux and the \nh\ are
variable (see Appendix~\ref{sec:ct_AGN}). 
The variability time scale ($\approx$\ a year) 
indicates the size of the reflecting 
material is \hbox{$\lsim 0.3$~pc}. Nevertheless, it 
is also possible that the observed \xray\ flux 
is a combination of both transmitted and reflected 
radiation, and the observed high-energy flux 
variability is mainly caused by the variable 
transmitted component.

\item We have identified a source in our sample (\xidthr) that 
transitions from \xray\ unobscured to obscured states over a 
\hbox{$\approx3$~yr} rest-frame time scale 
(see Appendix~\ref{sec:sig_var}). 
The source is a type I object at \hbox{$z=1.21$} with 
\hbox{\lx$\approx 7\times 10^{43}\ \rm erg\ s^{-1}$}. 
Its \lx\ is higher when it is less \xray\ obscured. 
The angular size of an \xray\ eclipsing cloud is 
estimated to be several degrees (viewed from the central SMBH).
However, there is no corresponding optical spectral-type transition,
suggesting that the \xray\ eclipsing material is too
small to block most of the broad-line emission.

\end{enumerate}

\subsection{Interpretation from Power Spectral Density}\label{sec:psd}
The observed low \lx\ variability amplitudes of quasars 
can be plausibly explained by the AGN PSD,
which describes variability power as a function of frequency.
An AGN PSD can often be well 
modeled as a broken power law 
\citep[e.g.,][]{uttley02}
\begin{equation}\label{equ:psd}
\mathrm{PSD}(\nu)= \begin{cases}
		   \psdamp \nu^{-1},\ \mathrm{if}\ 
		   \nu < \nubf ; \\
		   \psdamp \nubf \nu^{-2} 
        	   ,\ \mathrm{if}\ \nu \geq \nubf .
		   \end{cases}
\end{equation}
The normalization $\psdamp$\ is roughly 
constant ($0.017\pm0.006$), according to 
studies of Seyfert galaxies \citep{papadakis04}.
The low-frequency power law extends at least to
several-year time scales in local Seyfert galaxies \citep[e.g.,][]{zhang11}.
$\nubf$\ is related to both SMBH mass ($M_\mathrm{BH}$) and bolometric 
luminosity ($L_{\rm bol}$) as \citep{mchardy06}
\begin{equation}\label{equ:nubf}
\begin{split}
\nubf &\sim 50
	 \left( \frac{M_{\rm BH}}{10^8 M_\sun} \right)^{-2} 
	 \left( \frac{L_{\rm bol}}{10^{45}\ \mathrm{erg\ s^{-1}}} \right)\ 
	 \mathrm{yr^{-1}}\\
      &\sim 10 
	 \left( \frac{L_{\rm X}}{10^{44}\ \mathrm{erg\ s^{-1}}} \right)^{-1}
	 \left( \frac{k_{\rm bol}}{50} \right)^{-1}
	 \left( \frac{\lambda_{\rm Edd}}{0.1} \right)^{2}
	\mathrm{yr^{-1}}
\end{split}
\end{equation}
where $\lambda_\mathrm{Edd}$\ and $k_\mathrm{bol}$\ are the Eddington ratio 
and bolometric correction factor for \lx\ (\hbox{2--10}~keV), respectively. 
Given a PSD of a source, the \sigmaexcL\ can be estimated as
\citep[e.g.,][]{papadakis04, papadakis08}
\begin{equation}\label{equ:sigma_psd}
\begin{split}
\sigma_{\mathrm{exc},L}^2 &= \int^{\nu_{\rm hf}}_{\nu_{\rm lf}} 
					\mathrm{PSD}(\nu) \mathrm{d}\nu \\
			  &= \psdamp \ln\left(\frac{\nu_{\mathrm hf}}
			{\nu_{\mathrm lf}}\right),\ \mathrm{if}\ 
			\nu_{\rm hf} < \nubf.
\end{split}
\end{equation}
The integral bounds for our study are
\begin{equation}\label{equ:nu_range}
\begin{split}
\nu_{\rm lf} &\sim \frac{1+z}{t_{\rm span}}\ \mathrm{and} \\
\nu_{\rm hf} &\sim \frac{1+z}{t_{\rm bin}}, \\
\end{split}
\end{equation}
respectively, where \hbox{$t_{\rm span}= 14.3$~yr} is the 
observed-frame total observation span and \hbox{$t_{\rm bin}= 0.46$~yr} 
is the observed-frame median bin width of the four epochs. 
For an AGN with \hbox{\lx $=10^{44}$~erg~s$^{-1}$}\ and 
\hbox{$\lambda_\mathrm{Edd} \sim 0.1$},
the typical bolometric correction factor is \hbox{$k_\mathrm{bol}\sim 50$}\
\citep[e.g.,][]{hopkins07}, and the typical redshift is \hbox{$\approx2$}
(see Figure~\ref{fig:Lx_vs_z_and_Lx_vs_nH}). Thus, Equations~\ref{equ:nubf} and
\ref{equ:nu_range} yield \hbox{$\nubf \sim \nu_{\rm hf} \sim 10$~yr$^{-1}$}.
For AGNs with higher luminosity (i.e., quasars), both $k_\mathrm{bol}$
and $z$\ tend to be larger. $\nu_{\rm hf}$\ and $\nubf$\ will 
increase and decrease, respectively (assuming the same 
$\lambda_\mathrm{Edd}$); e.g., for a typical quasar at \hbox{$z\sim 3$}\ with 
\hbox{\lx$\sim 3\times 10^{44}$~erg~s$^{-1}$} (see Figure~\ref{fig:Lx_vs_z_and_Lx_vs_nH}), 
$\nubf$\ will increase by a factor of \hbox{$\sim 1.3$}\ due to $z$, and 
$\nu_{\rm hf}$\ will decrease at least 3 times due to \lx\ and 
$k_\mathrm{bol}$.\footnote{
The \lx\ (and related $k_\mathrm{bol}$)\ has a stronger effect than
$z$. This is why we attribute \lx\ rather than $z$\ to be a major factor 
affecting variability in Section~\ref{sec:sigma_vs_Lx_nH}.
}
Therefore, $\nu_{\rm hf}$\ is likely to be 
higher than $\nubf$\ for quasars (Equations~\ref{equ:nubf} and \ref{equ:nu_range}), 
i.e., the integral range in Equation~\ref{equ:sigma_psd}\ 
covers the power law with slope $-2$\ that drops strongly toward high frequency. 
The resulting $\sigma_{\mathrm{exc},L}$\ for quasars 
should thus be smaller than for other
AGNs, consistent with observations (Section~\ref{sec:sigma_vs_Lx_nH}).
For a non-quasar AGN, we might be sampling only the low-frequency part
of its PSD with slope $-1$\ (i.e., $\nu_{\rm hf} < \nubf$); 
Equations~\ref{equ:sigma_psd} and \ref{equ:nu_range} 
result in an approximately constant 
\sigmaexcL\ value of $0.059\pm0.021$, regardless of source properties.
This \sigmaexcL\ value is generally lower than the observed
values for our non-quasar AGNs (see Figure~\ref{fig:sigma_vs_Lx_nH}), 
casting doubt on the universality of the constant-amplitude\ PSD model
(e.g., \citealt{ponti12}; M.~Paolillo et al., in preparation).  

\subsection{Future Work}\label{sec:future}
Considering reasonably in-depth studies of the long-term \hbox{X-ray} 
variability of typical distant AGNs, it will be difficult to 
surpass greatly the present work for a considerable period of 
time; this is primarily due to the unmatched \hbox{CDF-S} exposure 
obtained over the extended period of \hbox{$\approx 15$~yr}. Additional 
\hbox{X-ray} variability studies should be done for the large population 
of \hbox{X-ray} fainter \hbox{CDF-S} AGNs 
(see Figure~\ref{fig:Lx_vs_z_and_Lx_vs_nH}; 
e.g., X.~C.~Zheng et al., in preparation), although it will be more 
difficult to characterize these systems individually in depth. 
If \chandra\ continues to operate for another \hbox{$\approx 10$~yr}, 
as appears plausible \citep[e.g.,][]{wilkes15}, obtaining additional 
\hbox{CDF-S} exposure in several years could lengthen our time 
baseline to up to \hbox{$\approx 25$~yr} in total. The {\it Advanced Telescope 
for High Energy Astrophysics\/} (\athena; e.g., \citealt{barcons15}), 
planned for launch in \hbox{$\approx 13$~yr}, has the best current prospects 
for substantially advancing long-term \hbox{X-ray} variability studies 
of typical AGNs in the distant universe. Owing to its greatly
improved photon collecting area, it will obtain much better 
photon statistics for its deep-field AGNs. With suitable
observation scheduling, it could efficiently perform a study 
similar to that in this work but for many more 
objects and with tens of epochs of 
observations spanning a wide range of timescales. The prime
deep-survey field for \athena\ is arguably the \hbox{CDF-S}, and 
\athena\ variability studies could build upon the long-term 
baseline of \chandra\ \cdfs\ observations utilized in this work.

%======================Acknowledgements=========================
\acknowledgments
\section*{acknowledgments}
We thank the referee for helpful feedback that improved this work.
We thank Johannes Buchner, Michael Eracleous, Eric Feigelson, Brandon Kelly, 
Wanjun Liu, Kirpal Nandra, Michael Nowak, Piero Rosati, Paolo Tozzi, Phil Uttley, 
and Ningxiao Zhang for helpful discussions, 
and Scott Croom, Giorgio Lanzuisi, and James Mullaney for providing relevant data.
G.Y, W.N.B, and F.V acknowledge support from \chandra\ \xray\ Center
grant GO4-15130A. Y.Q.X, M.Y.S, and X.C.Z acknowledge support from 
the National Thousand Young Talents program, the 973 Program 
(2015CB857004), NSFC-11473026, NSFC-11421303, the Strategic 
Priority Research Program ``The Emergence of Cosmological 
Structures'' of the Chinese Academy of Sciences (XDB09000000), 
and the Fundamental Research Funds for the Central Universities.
F.E.B. and S.~Schulze acknowledge support from CONICYT-Chile grants 
Basal-CATA PFB-06/2007 and the Ministry of Economy, Development, 
and Tourism's Millennium Science Initiative through grant IC120009, 
awarded to The Millennium Institute of Astrophysics, MAS. Furthermore, 
F.E.B. acknowledges support from FONDECYT Regular 1141218 and ``EMBIGGEN'' 
Anillo ACT1101, and S.~Schulze acknowledges support from FONDECYT grant 
3140534. S.~Kim acknowledges support from FONDECYT grant 3130488.
J.X.W acknowledges support from NSFC 11233002 and the 973 program 2015CB857005.
This research has made use of Astropy, a community developed core 
Python package for Astronomy (Astropy Collaboration, 2013),
and the VizieR catalogue access tool, CDS, 
Strasbourg, France. 

%======================Appendix=================================
\appendix
\section{Error Estimation of Model Parameters}
\label{app:err}
As described in Section~\ref{sec:spec_fit_method}, 
we fit spectra of the four epochs simultaneously 
with a $wabs\times zwabs \times pegpwrlw$\ model.
The photon index, $\Gamma$, is linked across the  
four epochs, assuming no $\Gamma$\ variability
(see Section~\ref{sec:spec_fit_method} for 
the justification of this assumption).
In this Appendix, we first explain our choice 
of $pegpwrlw$\ over $powerlaw$, and then describe our method to 
estimate the errors of \nh$_{,i}$.

\begin{figure}
\includegraphics[width=\linewidth]{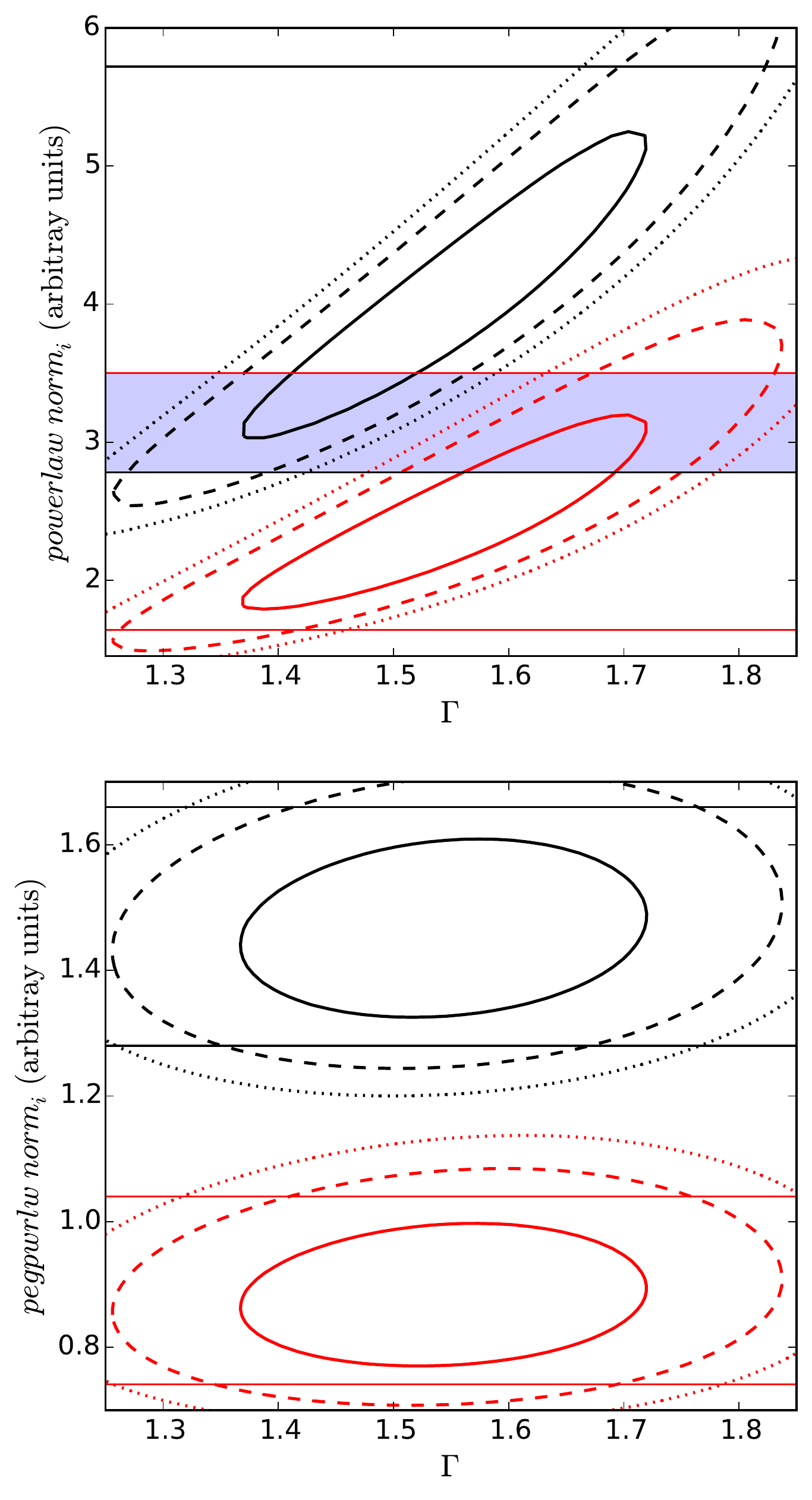}
\caption{$norm_i-\Gamma$\ confidence contours
of the first two epochs of \hbox{J033217.1--275220}
resulting from modelling with
$wabs\times zwabs \times powerlaw$\ (upper panel) and 
$wabs\times zwabs \times pegpwrlw$\ (lower panel). 
Black and red colors indicate epochs 1 and 2, respectively. 
The solid, dashed, and dotted curves indicate 1$\sigma$\ (68\%),
2$\sigma$\ (95\%), and 99\%\ confidence contours, respectively.
The horizontal solid lines indicate the 2$\sigma$\ 
uncertainty ranges of $norm_i$.
In the upper panel, the 99\% confidence contours 
do not overlap, indicating highly significant variability 
of $norm_i$\ (i.e., $>99.99\%$\ confidence level; see 
Appendix~\ref{app:err}). However, the two projected 
$norm_i$\ uncertainty ranges have an overlapping interval 
(blue shaded region), and a $\chi^2$\ test shows that 
the significance level of $norm_i$\ variability is only 93\%.
Thus, the errors of $norm_i$\ are 
overestimated under the assumption of constant $\Gamma$. 
In the lower panel, this issue does not occur.
}
\label{fig:norm_gam_cont}
\end{figure}

%To illustrate the merit of $pegpwrlw$\ over $powerlaw$, 
As an illustrative example, we show the $norm_i$-$\Gamma$\ 
confidence contours of the first two epochs of 
\hbox{J033217.1--275220} resulting from fitting 
with a model of $wabs\times zwabs \times powerlaw$\
(see the upper panel of Figure~\ref{fig:norm_gam_cont}).
The contours, as expected, show positive correlations 
between $norm_i$\ and $\Gamma$.
Since no overlapping region exists between the two 99\% confidence 
contours, the probability of the two epochs having both 
the same $norm_i$\ and $\Gamma$\ is very low 
[\hbox{$<(1-99\%)^2=0.01\%$}].
Therefore, the $norm_i$\ variability must be very significant
(\hbox{$>1-0.01\%=99.99\%$})
under our assumption of constant $\Gamma$.
However, if we evaluate the $norm_i$\ variability by checking 
the 1-dimensional (1D) errors ($2\sigma$) of $norm_i$\ 
(the projected range on the $y$-axis),
the variability seems to be less significant due to the 
existence of an overlapping interval (blue shaded region).
More quantitatively, if we perform a $\chi^2$\ test of 
the $norm_i$\ variability using the best-fit values and 
1D errors, the resulting significance of variability 
is only 93\%. 
This apparent discrepancy occurs because when calculating 
the 1D errors of $norm_i$\ by projecting the 2D contours, 
the positive $norm_i-\Gamma$\ correlations and 
our underlying assumption of constant $\Gamma$\ are ``forgotten''.
This is evident since the blue-shaded region only covers 
the two dashed contours at very different $\Gamma$, i.e.,
the same $norm_i$\ can be achieved only when the 
assumption of constant $\Gamma$\ is violated. 
Reading Figure~\ref{fig:norm_gam_cont}, $\Gamma$\
would need to change by $\Delta \Gamma \approx 0.2-0.3$,
and this is larger than any physically expected $\Gamma$\
change (see Section~\ref{sec:spec_fit_method}). 
Therefore, the errors of $norm_i$\ 
are overestimated under this assumption of constant $\Gamma$.
This problem is prevalent when using $powerlaw$, since 
positive correlations exist for almost all sources.
However, the positive correlations can
be mostly eliminated by replacing $powerlaw$\ with $pegpwrlw$\ 
(see the lower panel of Figure~\ref{fig:norm_gam_cont}). 
$pegpwrlw$\ differs from $powerlaw$\ by having its normalization 
based on intrinsic flux in a given finite band 
rather than the flux density at observed-frame 1~keV.
We find that setting the normalization band 
as \hbox{$E_{peg}-7$~keV} produces nearly horizontal 
\hbox{$norm_i$-$\Gamma$}\ contours for all sources, 
where $E_{peg}$\ is the minimum between 0.5 keV and 
the observed-frame $e$-folding energy ($E_{\rm fold}$) 
caused by intrinsic photoelectric absorption. 
Technically, we obtain the photoelectric
cross section $\sigma_{\rm photo}$\ as a function of energy 
\citep{morrison83}, and solve the equation 
$\langle N_{\rm H} \rangle \times 
\sigma_{\rm photo}[E_{\rm fold}\times(1+z)]=1$\ 
to obtain $E_{\rm fold}$,\footnote{
We first fit the spectrum 
using an arbitrary normalization band and calculate the 
$\langle N_{\rm H} \rangle$\ from the best-fit \nh$_{,i}$. 
The best-fit \nh$_{,i}$\ is independent of the exact choice 
of energy band. 
} where the factor $1+z$\ is to convert observed-frame 
to rest-frame energy. 

\begin{figure}
\includegraphics[width=\linewidth]{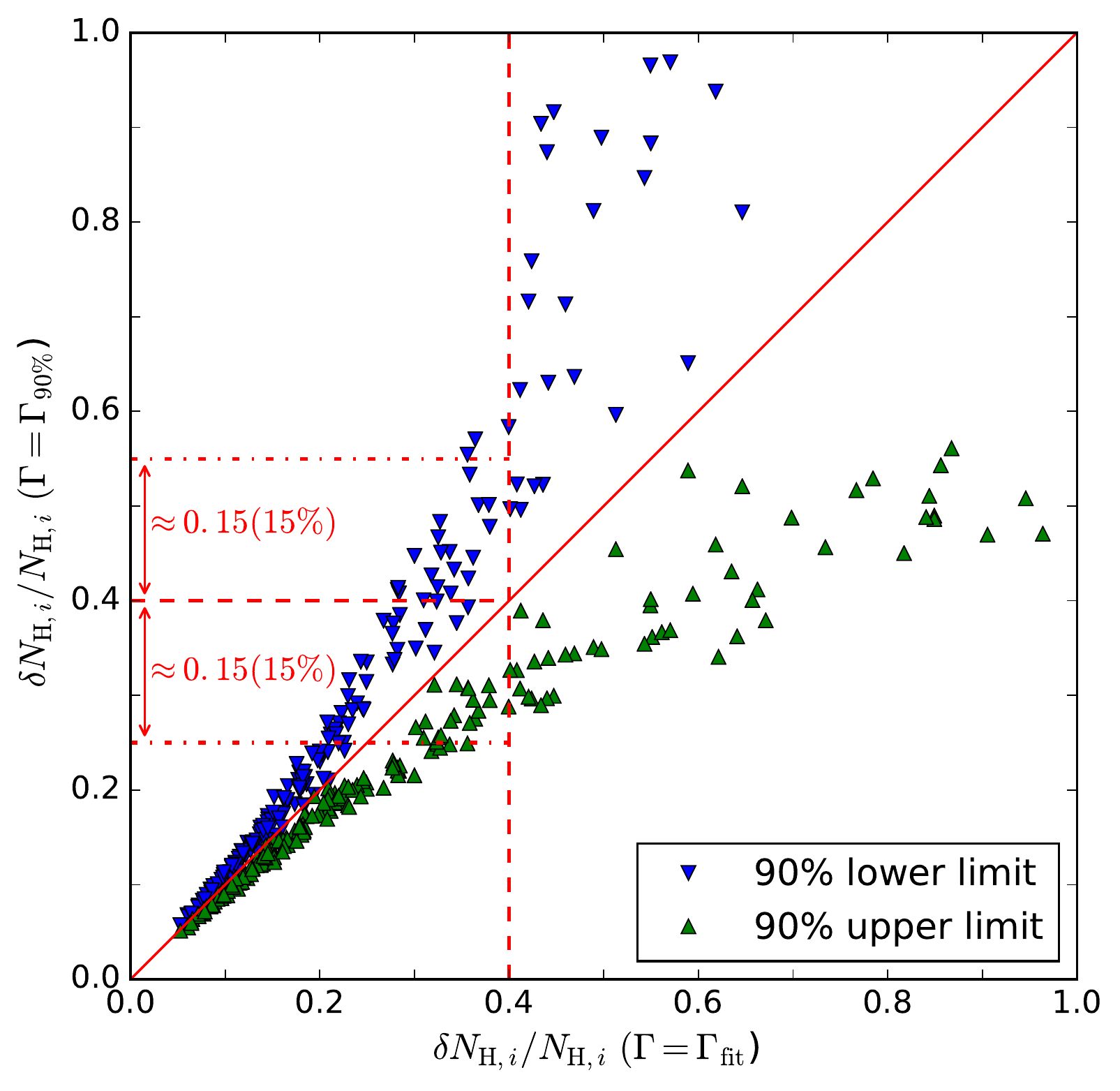}
\caption{Fractional errors 
$\delta N_{\mathrm{H},i}/N_{\mathrm{H},i}$\ obtained
by fixing $\Gamma$\ at the 90\% confidence limits of 
$\Gamma_{\rm fit}$\ vs.\ those obtained by 
fixing $\Gamma$\ at the best-fit value. 
The green upward-pointing and blue downward-pointing
triangles indicate the fractional errors obtained by fixing $\Gamma$\ at 
the upper and lower limits of $\Gamma_{\rm fit}$, 
respectively. The red solid line indicates 
the ideal situation, i.e., the estimated fractional errors are not 
affected by fixing $\Gamma$\ at different values. 
The vertical red dashed line indicates our
threshold, i.e., only sources with all four epochs having 
$\delta N_{\mathrm{H},i}/N_{\mathrm{H},i}<0.4$\ are included 
in the analyses where errors on \nh$_{,i}$\ are being used. 
Its intersection points with the 
approximate outer envelopes of the blue and green
triangles are indicated 
by the horizontal red dash-dotted 
lines, and their deviations from 
$\delta N_{\mathrm{H},i}/N_{\mathrm{H},i}=0.4$\ 
(the horizontal red dashed line)\ are $\approx 0.15$\ 
(15\%, as marked).
Our sample consists 68 sources with each of them 
having 4 epochs. Thus, there are 272 ($68 \times 4$) pairs 
of triangles plotted (though some are located out 
of the plotted ranges and are not shown).
Note that for each source, the 
$\delta N_{\mathrm{H},i}/N_{\mathrm{H},i}$\ (\hbox{$x$-axis}) are 
different for different epochs. 
}
\label{fig:err_dif_gam}
\end{figure}

Similar positive correlations are also, as expected,  
prevalent in the \hbox{\nh$_{,i}$-$\Gamma$}\
contours. However, we are not aware of any model choices that 
can eliminate these correlations 
(as for the choice of $pegpwrlw$\ in the $norm_i$\ case). 
Alternatively, if intrinsic $\Gamma$\ ($\Gamma_{\rm intr}$) 
were given, one could fix \hbox{$\Gamma=\Gamma_{\rm intr}$} 
and estimate the the errors of \nh$_{,i}$. 
But in reality, $\Gamma_{\rm intr}$\ is not perfectly known.
We thus adopt an approximation, i.e., fixing  
\hbox{$\Gamma=\Gamma_{\rm fit}$.}
The idea is to approximate $\Gamma_{\rm intr}$ 
as $\Gamma_{\rm fit}$.  
Admittedly, 
compared to fixing \hbox{$\Gamma=\Gamma_{\rm intr}$},
this approximation might 
overestimate or underestimate the errors of \nh$_{,i}$.
To evaluate this possible issue, 
we fix $\Gamma$\ at the 90\% confidence
lower and upper limits of $\Gamma_{\rm fit}$, 
respectively, and then estimate the errors.
The lower and upper limits approximate the boundaries of 
possible $\Gamma_{\rm intr}$\ values. If fixing $\Gamma$\ at 
these boundaries results in similar errors as fixing $\Gamma$\
at $\Gamma_{\rm fit}$, we conclude that our approximation 
(i.e., fixing \hbox{$\Gamma=\Gamma_{\rm fit}$}) gives accurate 
errors of \nh$_{,i}$\ for a given source.
Figure~\ref{fig:err_dif_gam}\ shows the results.
The dependence of estimated fractional error 
$\delta N_{\mathrm{H},i}/N_{\mathrm{H},i}$ on $\Gamma$\ is 
stronger when the error is larger.
In Figure~\ref{fig:err_dif_gam}, at 
$\delta N_{\mathrm{H},i}/N_{\mathrm{H},i}=0.4$\ (the 
vertical dashed line), fixing $\Gamma$\ at its 
90\%-confidence lower 
and upper limits of $\Gamma_{\rm fit}$ results in 
$\delta N_{\mathrm{H},i}/N_{\mathrm{H},i} \lsim 0.55$\ 
and $\gsim 0.25$, respectively (the dash-dotted lines). 
Since $\Gamma_{\rm intr}$\ is very 
likely within the range of the 90\% limits of $\Gamma_{\rm fit}$, 
fixing $\Gamma=\Gamma_{\rm intr}$\ 
(if it were perfectly known) should give 
$\delta N_{\mathrm{H},i}/N_{\mathrm{H},i}$ within the range of 
$\approx 0.25-0.55$.
Therefore, fixing $\Gamma=\Gamma_{\rm intr}$\ 
would lead to $\delta N_{\mathrm{H},i}/N_{\mathrm{H},i}$\ 
deviating from that obtained by our approximation 
(fixing \hbox{$\Gamma=\Gamma_{\rm fit}$}) by 
$\lsim 15\%$\ (i.e., $15\%=0.15=0.55-0.4=0.4-0.25$).
For epochs with $\delta N_{\mathrm{H},i}/N_{\mathrm{H},i}<0.4$,
this value should be even lower. 	
In the analyses where errors on \nh$_{,i}$\ are being used
(Sections~\ref{sec:sigma_vs_Lx_nH} and \ref{sec:nh_var_vs_t}), 
we only include the 35 sources with all four epochs having 
$\delta N_{\mathrm{H},i}/N_{\mathrm{H},i}<0.4$. 
This criterion guarantees that the \nh$_{,i}$\ fractional 
errors estimated by fixing \hbox{$\Gamma=\Gamma_{\rm fit}$} differ from
those estimated by fixing \hbox{$\Gamma=\Gamma_{\rm intr}$}\ by  
\hbox{$\lsim 15\%$}.
Though not ideal, this criterion is a practical solution 
for our \nh\ variability analyses, balancing between the 
accuracy of error estimation and the sample size. 

\section{A Highly Variable Compton-Thick Candidate}
\label{sec:ct_AGN}
\begin{figure}[htb]
\centering
\begin{tabular}{cc}
\includegraphics[width=\linewidth]{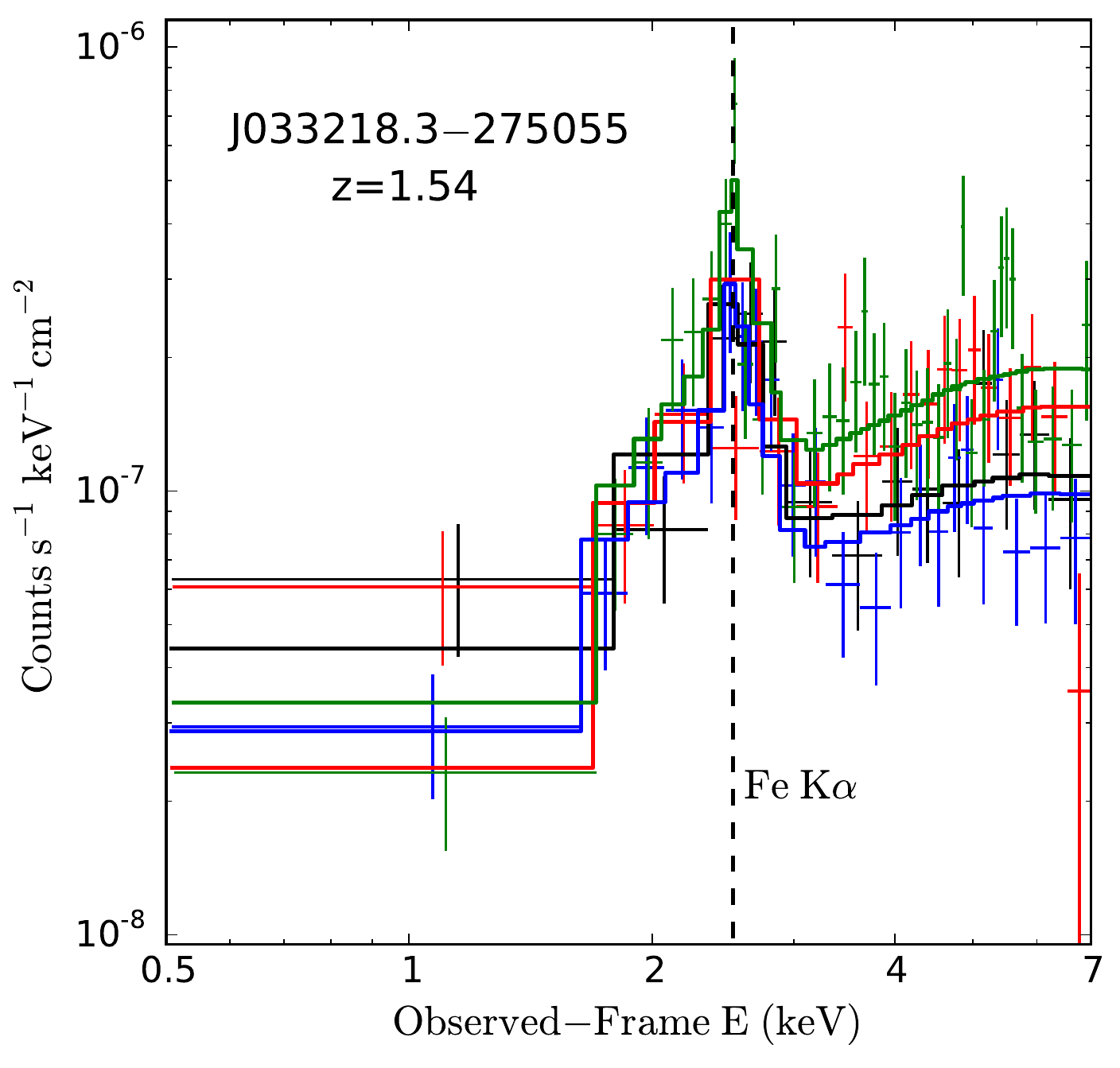} \\
\includegraphics[width=\linewidth]{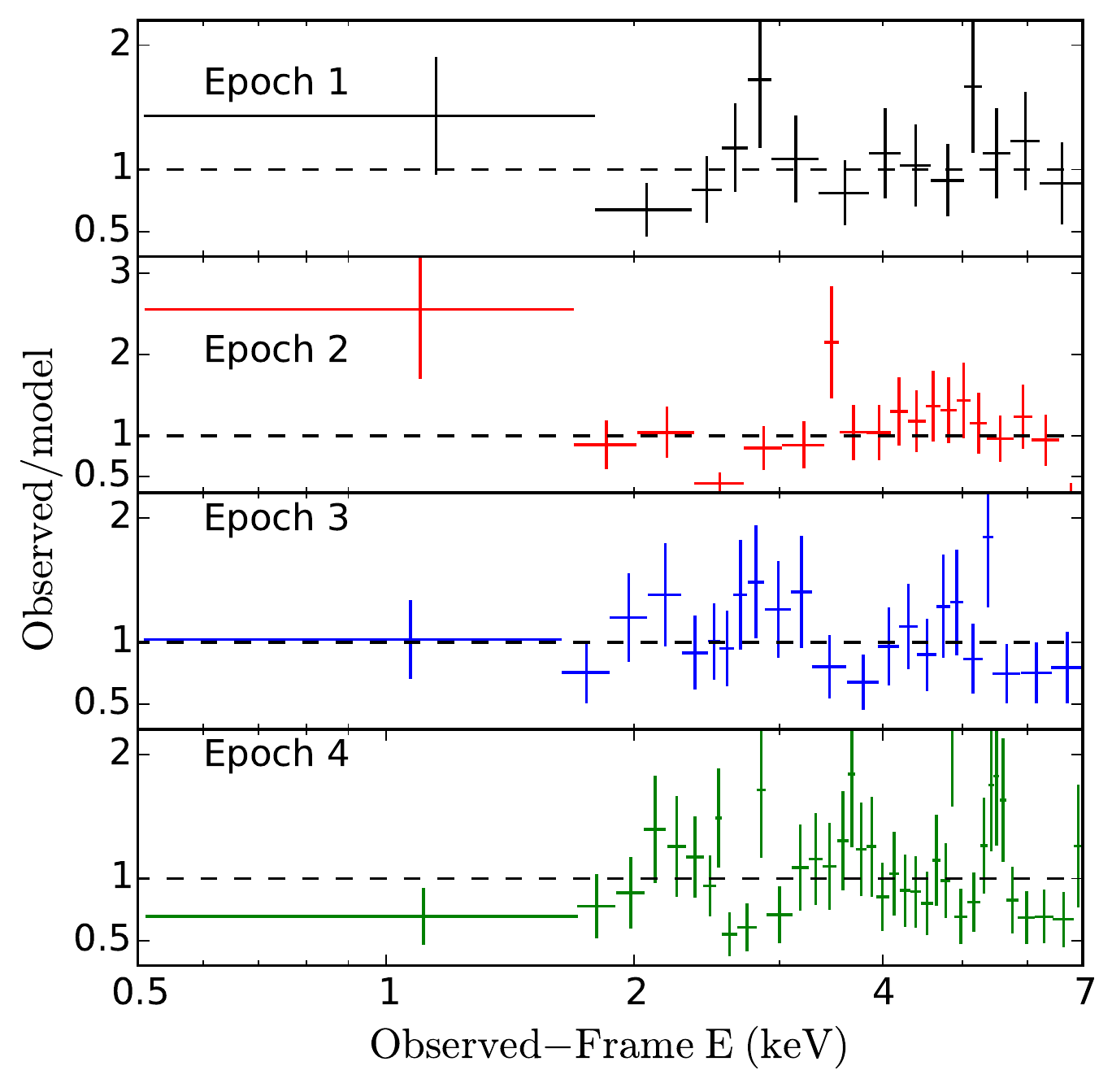} \\
\caption{
Upper panel: unfolded (i.e., intrinsic) spectra for the 
Compton-thick candidate \xidctk\ using fitting model 
$wabs\times zwabs \times pexmon$. 
The solid lines indicate best-fit model 
photon-flux density. 
Different colors indicate different epochs. 
The vertical dashed line indicates the energy 
of Fe K$\alpha$\ line (i.e., rest-frame 6.4 keV).
Data are binned for display purposes only.
Note that the flux in the high-energy band 
(less affected by absorption) varies significantly.
Lower panel: ratio of observed and model photon flux.
}
\label{fig:CT_spec}
\end{tabular}
\end{figure}

\Xidctk\ is a bright Compton-thick candidate AGN 
at \hbox{$z=1.54$}\ reported in the literature 
\citep{tozzi06, comastri11}. 
It is classified as an optical type II object.
It shows a strong 
Fe K$\alpha$\ emission line with a rest-frame equivalent width 
(REW) of \hbox{$\approx 1.2$~keV} \citep{comastri11}.
Our spectral fitting confirms its highly obscured 
nature (\hbox{$N_{\mathrm{H}}=3\times 10^{23}\ \mathrm{cm^{-2}}$}).
Both its \lx\ and \nh\ values are variable 
(\hbox{$\Delta \mathrm{AIC}_L=38$}\ and 
\hbox{$\Delta \mathrm{AIC}_N=5.1$}). 
However, its high $goodness$\ (44\%)
and low $\Gamma$\ (1.2, i.e., our allowed lower limit, 
see Section~\ref{sec:spec_fit_method}) indicate the model 
$wabs \times zwabs \times pegpwrlw$\ is likely inappropriate
(see Figure~\ref{fig:gam_goodness}).
Thus, motivated by \cite{comastri11}, we use the 
model $wabs\times zwabs\times pexmon$\ 
\citep[for $pexmon$, see][]{magdziarz95, nandra07} 
to fit the data. We set the $pexmon$\ 
model to be fully reflection dominated 
and fix the inclination angle at 60\degree. 
We link the $\Gamma$\ ($pexmon$) of 
all four epochs, and set $norm_i$\ ($pexmon$)
and \nh$_{,i}$ ($zwabs$) free (not linked). 
The fitted spectra of the reflection-dominated model are shown 
in Figure~\ref{fig:CT_spec}, and the detailed fitting results 
are presented in Table~\ref{tab:CT_spec}.  

\begin{figure}
\includegraphics[width=\linewidth]{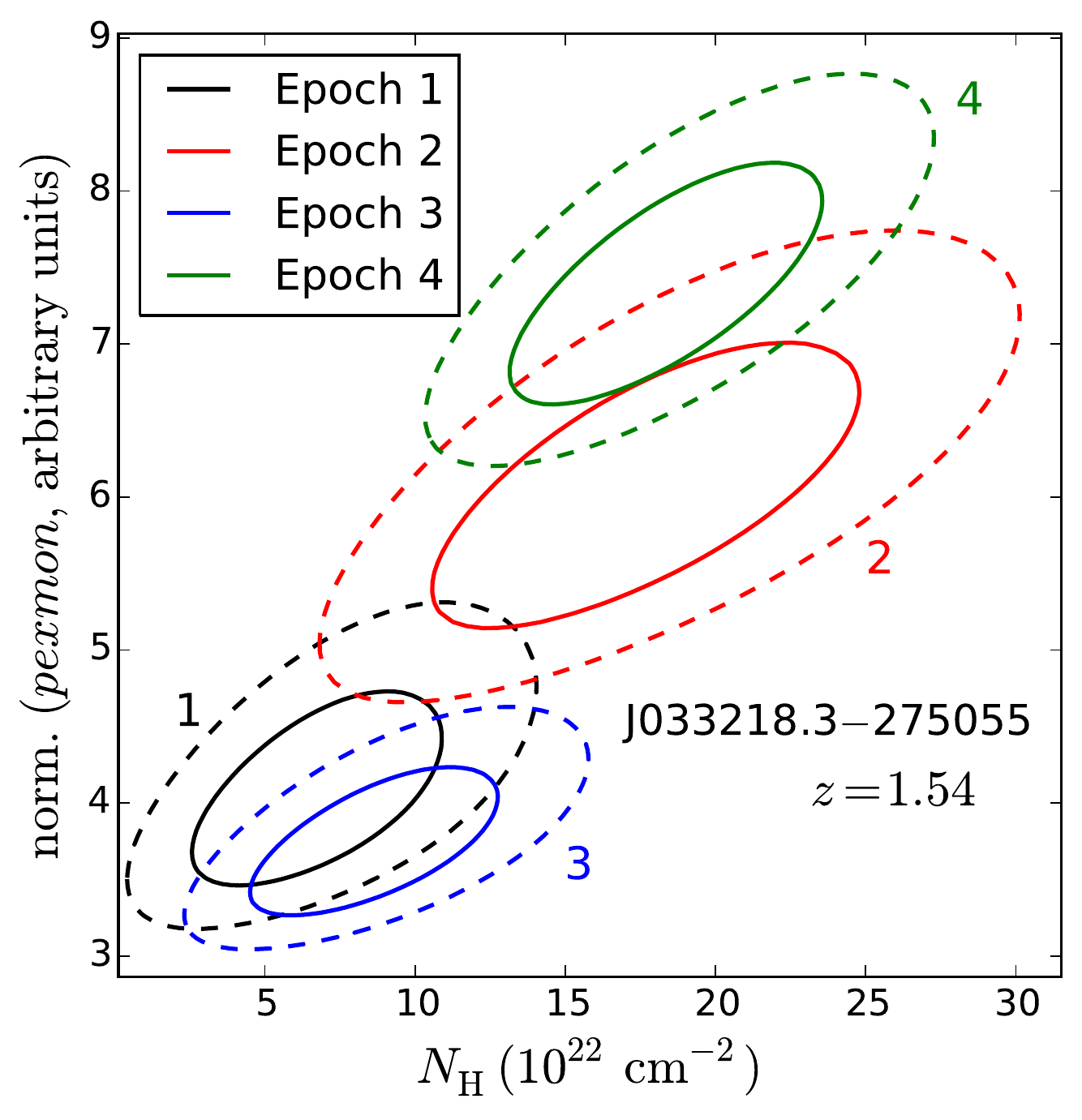}
\caption{Confidence contours of the normalizations of 
$pexmon$\ and \nh\ for source \xidctk, the Compton-thick
candidate. The solid and dashed curves indicate
$1\sigma$\ and $2\sigma$\ confidence contours, respectively. 
Different colors indicate different epochs, and the 
epoch indexes are also labeled beside the corresponding 
contours.
}
\label{fig:CT_contour}
\end{figure}

Despite having the same number of degrees of freedom 
as the previous model ($wabs \times zwabs \times pegpwrlw$), 
the new fitting results in a 
\hbox{$goodness=8\%$}\ and best-fit \hbox{$\Gamma=1.80$}, values that are
more common among the $goodness$\ and $\Gamma$\ distributions 
for our overall sample 
(see Figure~\ref{fig:gam_goodness}).
Also, the new model has AIC much smaller than 
the previous model (\hbox{$\Delta\rm AIC=54$}), indicating a significant 
improvement in the fit quality.
Thus, we consider this reflection-dominated model to be 
more physically plausible than the simple transmission-dominated model. 
Following Section~2.4.1 of \cite{nandra07}, 
this $\Gamma$\ value (i.e., 1.80) results in an REW 
of Fe K$\alpha$\ ($pexmon$) of \hbox{$\approx 1.4$~keV}, consistent 
with \cite{comastri11}.
Similar to the approach in Section~\ref{sec:identify}, we test
the significance of reflection-flux and \nh\ variability by 
linking $norm_i$\ ($pexmon$) and \nh$_{,i}$, respectively. 
The increased AIC values (i.e., 39 and 4.2) are both greater than 4, 
indicating both the reflection flux and \nh\ are variable 
(Section~\ref{sec:identify}).
The confidence contours are shown in Figure~\ref{fig:CT_contour}. 
Both the absorption and continuum are weak in epoch 1; they 
rise in epoch 2 and then drop in epoch 3; in epoch 4, they rise 
again. The amplitude of flux variability is large; e.g., the 
flux in epoch 4 is almost twice that in epochs 1 and 3. 
The variability time scale ($\approx$\ a year)
constrains the size of the reflecting 
material to be $\lsim 0.3$~pc.

However, there is a possible alternative explanation for the observed
high-energy flux variability of this source, though our 
reflection-dominated model explains the data satisfactorily.  
The observed \xray\ emission might be a combination of both 
transmitted and reflected radiation. In this scenario, the transmission 
component is variable and results in the observed high-energy flux variability,  
while the reflection component is stable.  
To test this scenario, we use a 
$wabs \times zwabs \times (pegpwrlw+pexmon)$\ model. 
We assume that the $pegpwrlw$\ and $pexmon$\ components share the same 
non-variable $\Gamma$, and the normalization of $pexmon$\ is 
the same across epochs. The \nh\ ($zwabs$) and the normalization 
of $pegpwrlw$\ are allowed to vary across epochs. 
Other parameters of the $pexmon$\ component 
are the same as for the reflection-dominated model. This composite 
model yields \hbox{$\Gamma=1.5$}\ and \hbox{$goodness=5\%$}, similar to that
of the reflection-dominated model. The resulting 
transmitted flux almost completely shuts down at epoch 1
and epoch 3 (more than an order of magnitude smaller 
than at epoch 2 and epoch 4). Such strong variability is not 
likely to be realistic, considering the general variability 
amplitudes of our sources (see Section~\ref{sec:flux_15}). 
Nevertheless, more complex transmission-reflection hybrid 
models might produce more physically plausible
results, though they cannot be constrained well owing 
to the available number of counts.

\begin{table}
\begin{center}
\caption{Spectral Fitting$^{\mathrm{a}}$ Results of the Compton-Thick Candidate (\xidctk)}
\label{tab:CT_spec}
\begin{tabular}{cccc}\hline\hline
Epoch & $\Gamma$ & \nh					   &   flux$^{\mathrm{b}}$ \\ 
      &          & ($10^{22}\rm\ cm^{-2}$) & ($10^{-15}\rm\ erg\ cm^{-2}\ s^{-1}$)  \\ \hline
 1 &1.80 &$5.61^{+2.88}_{-2.38}$ &$4.16^{+0.34}_{-0.40}$ \\
 2 &-- &$17.00^{+4.80}_{-4.40}$ &$5.47^{+0.44}_{-0.46}$ \\
 3 &-- &$7.81^{+2.89}_{-2.52}$ &$3.71^{+0.26}_{-0.26}$ \\
 4 &-- &$17.80^{+3.60}_{-3.30}$ &$6.64^{+0.33}_{-0.35}$ \\ \hline
\end{tabular}
\end{center}
{\sc Note.} --- \\
a.\ The spectral fitting model is $wabs\times zwabs \times pexmon$\ (see Section~\ref{sec:ct_AGN}). \\
b.\ Full-band (\hbox{0.5--7}~keV) model flux, not corrected for Galactic or intrinsic absorption.\\
\end{table}

\section{Three Significantly Variable Sources}
\label{sec:sig_var}
\begin{figure*}[htb]
\centering
\begin{tabular}{ccc}
\includegraphics[width=0.33\linewidth]{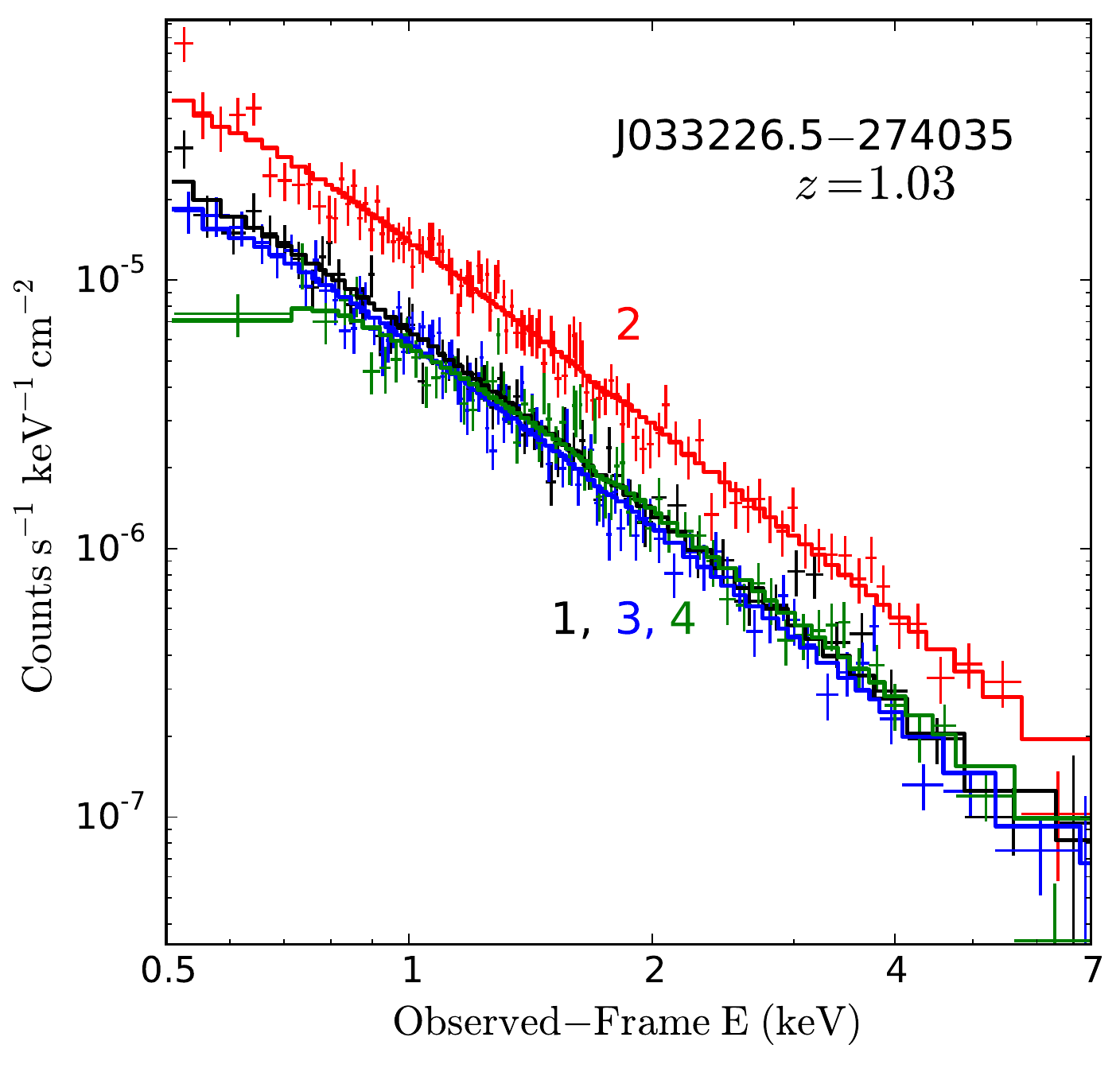} 
\includegraphics[width=0.33\linewidth]{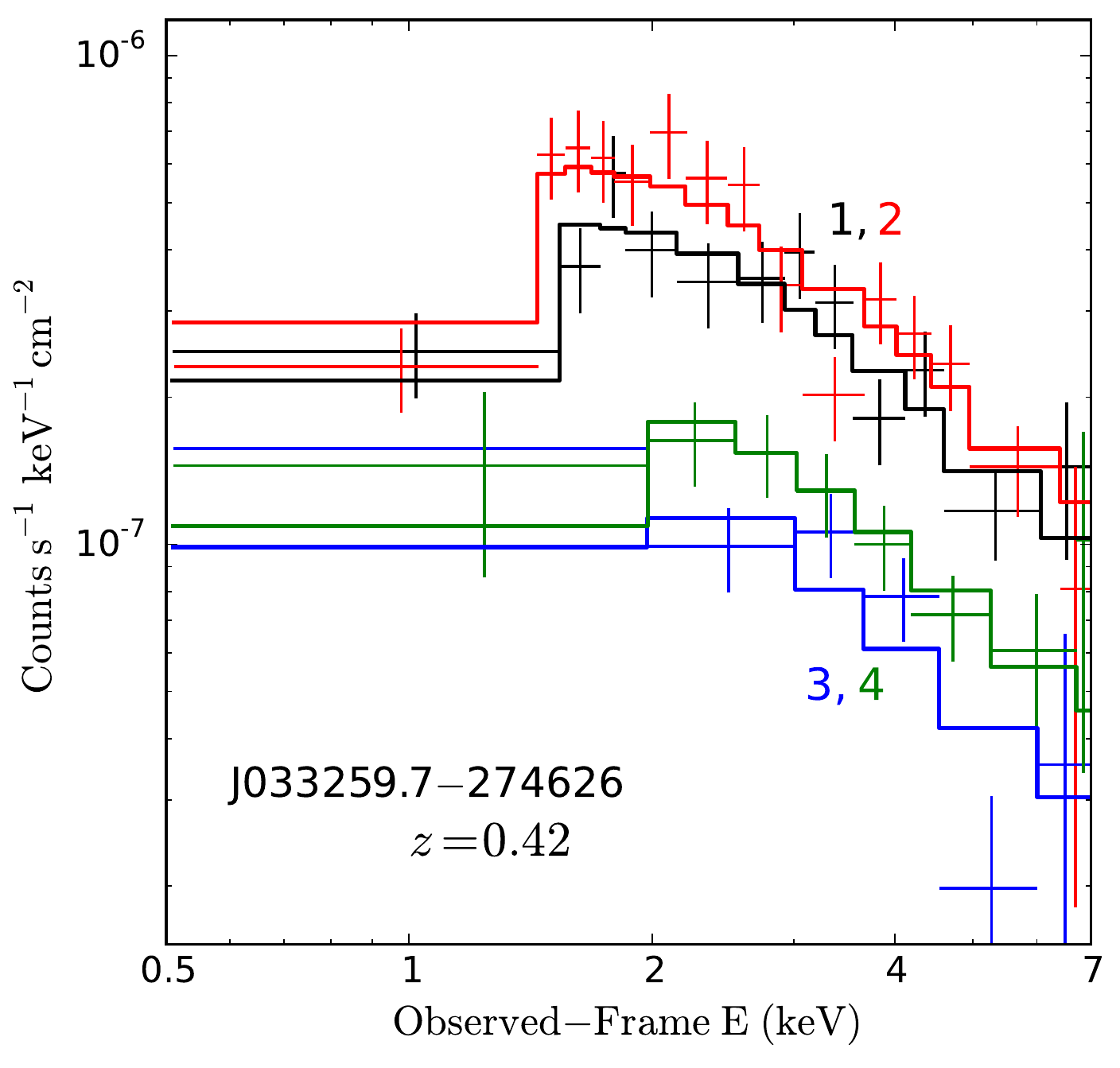}
\includegraphics[width=0.33\linewidth]{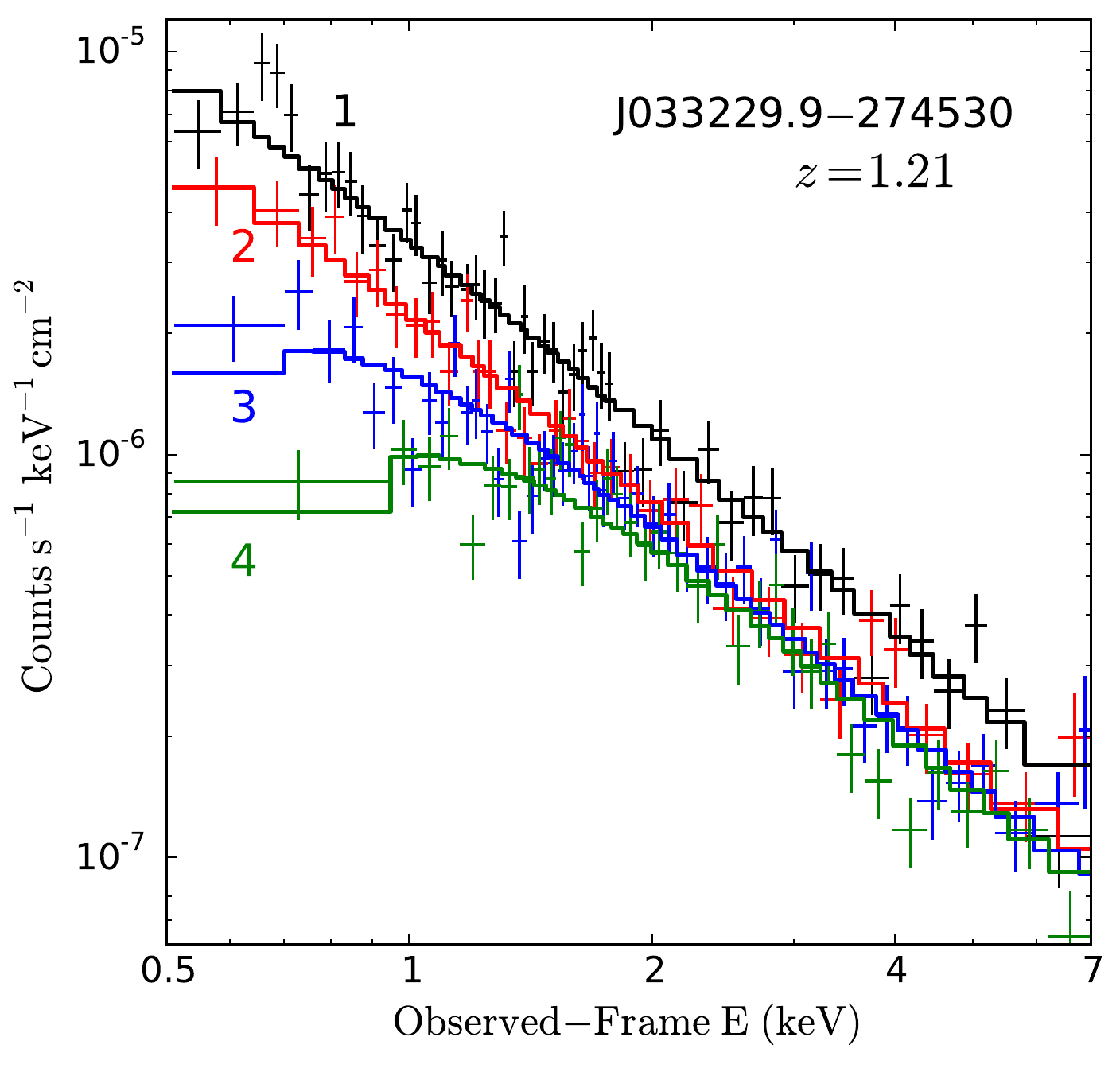} \\
\includegraphics[width=0.33\linewidth]{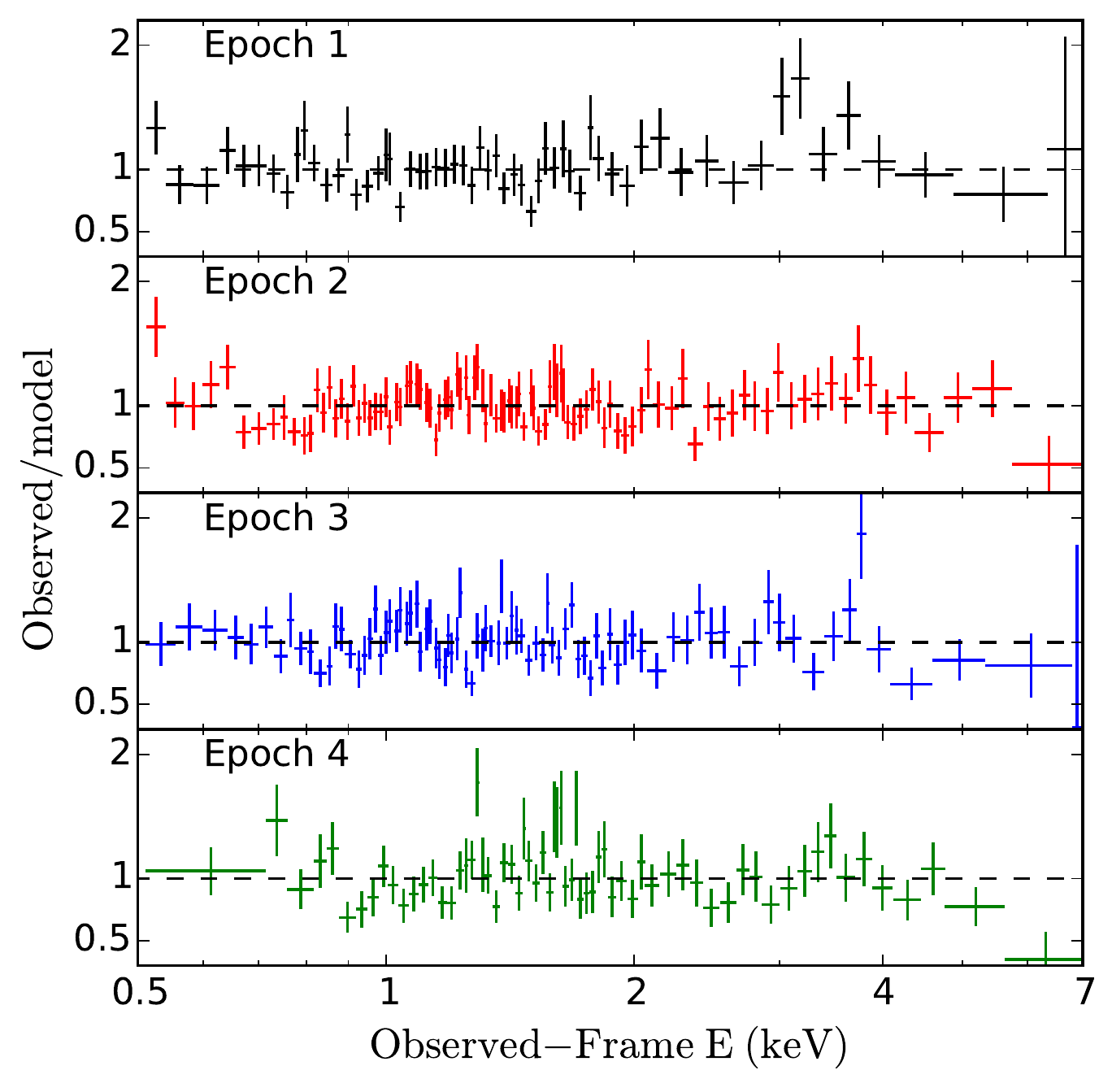} 
\includegraphics[width=0.33\linewidth]{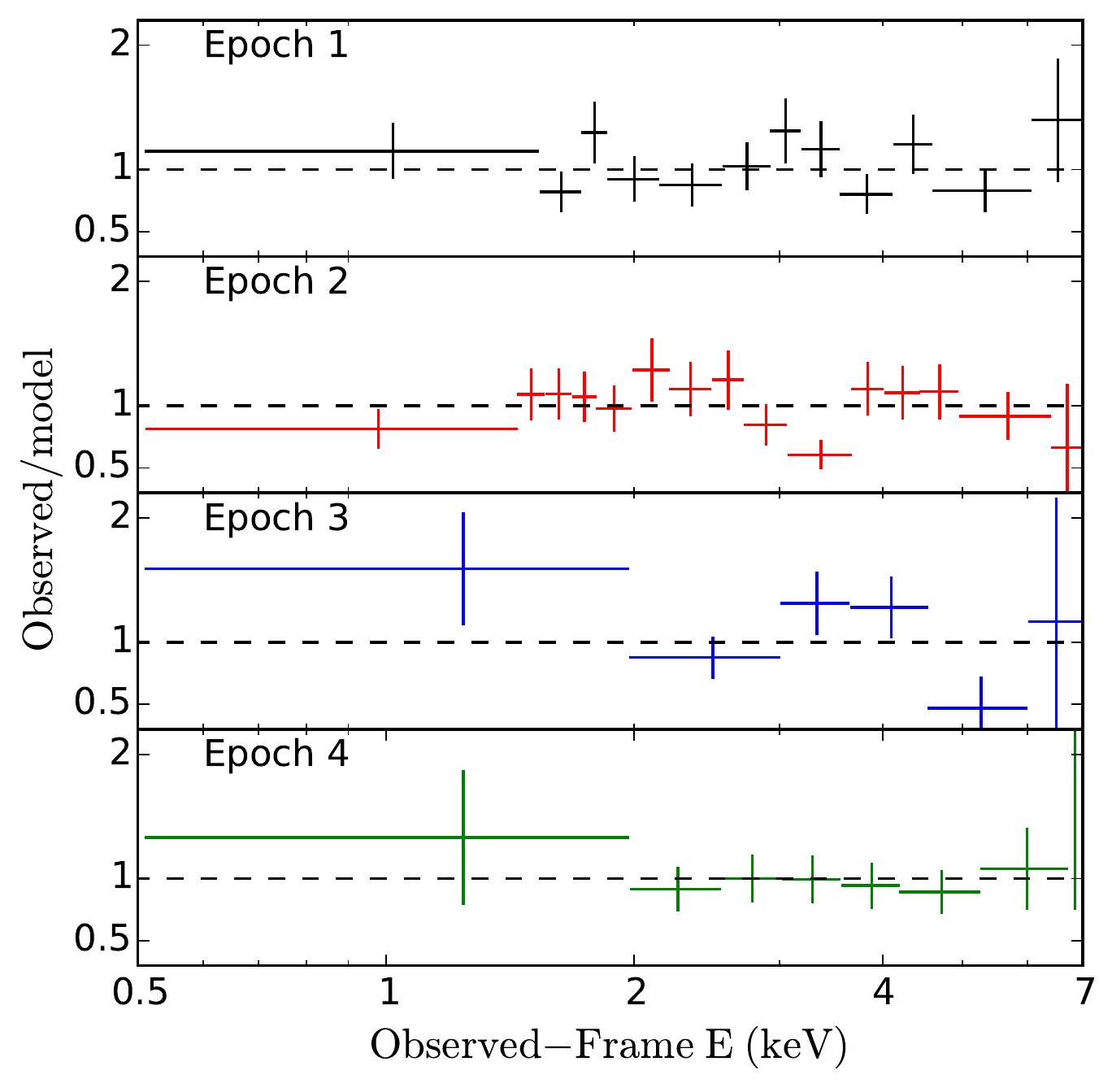}
\includegraphics[width=0.33\linewidth]{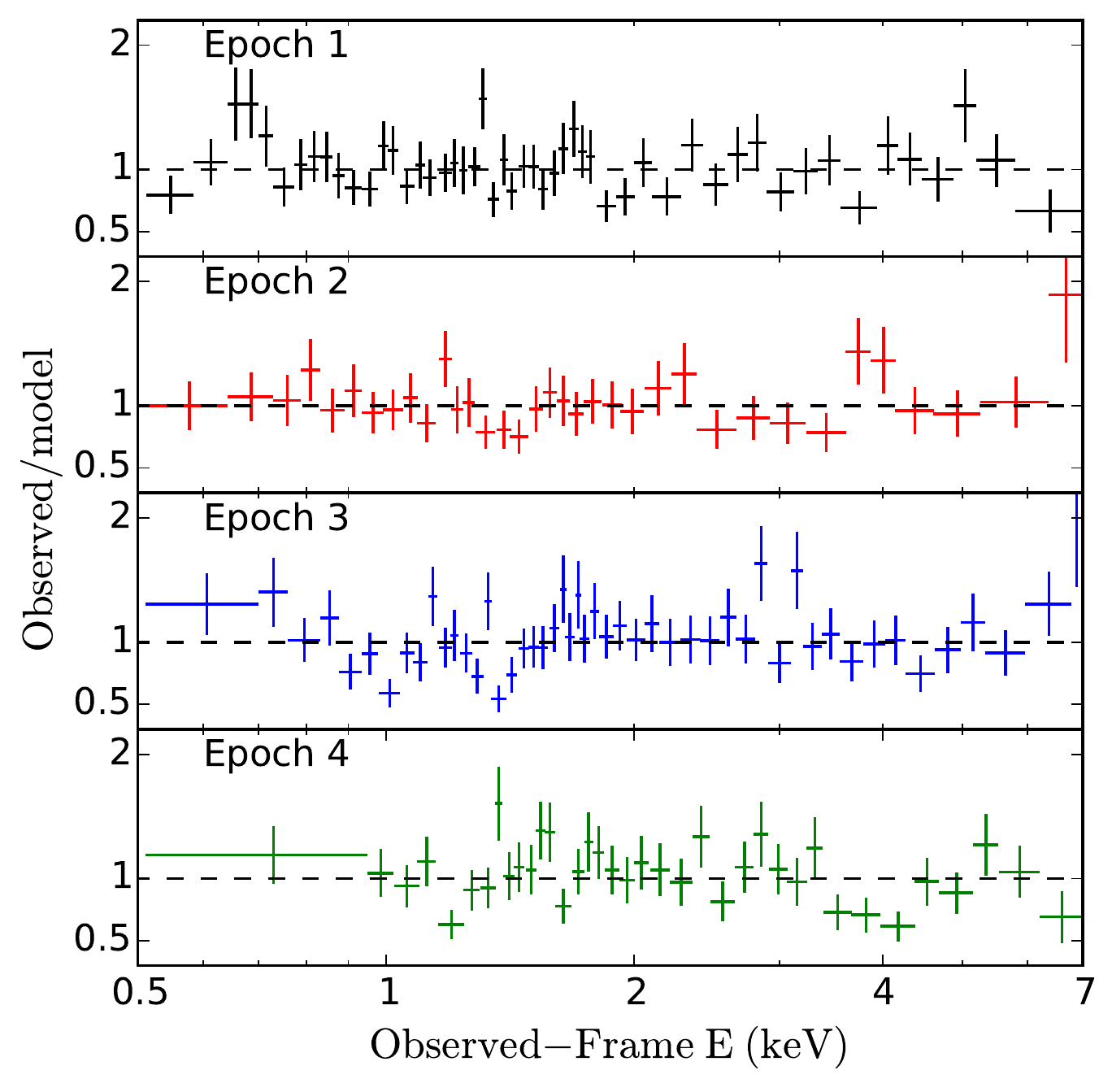} \\
\includegraphics[width=0.33\linewidth]{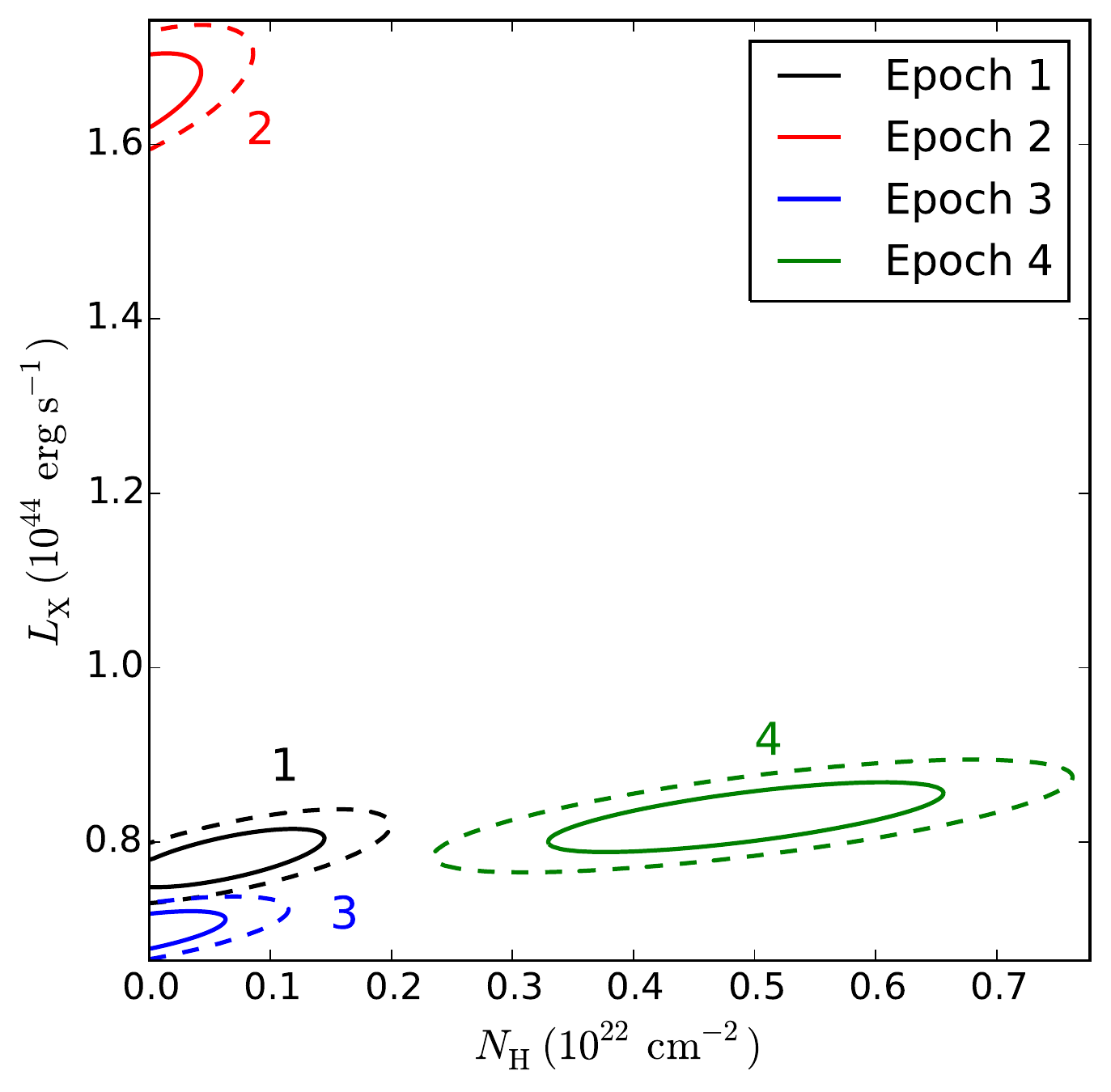} 
\includegraphics[width=0.33\linewidth]{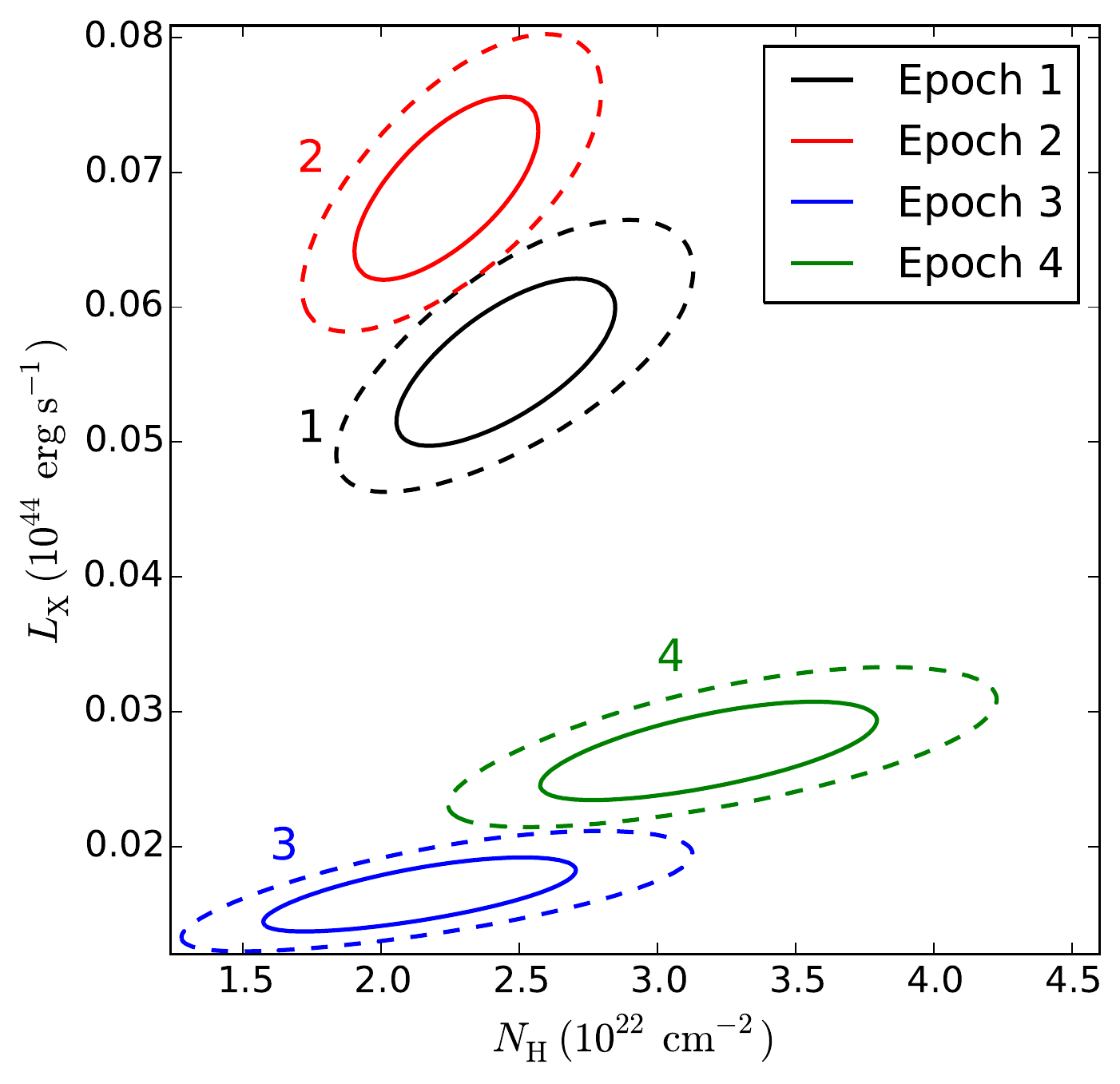}
\includegraphics[width=0.33\linewidth]{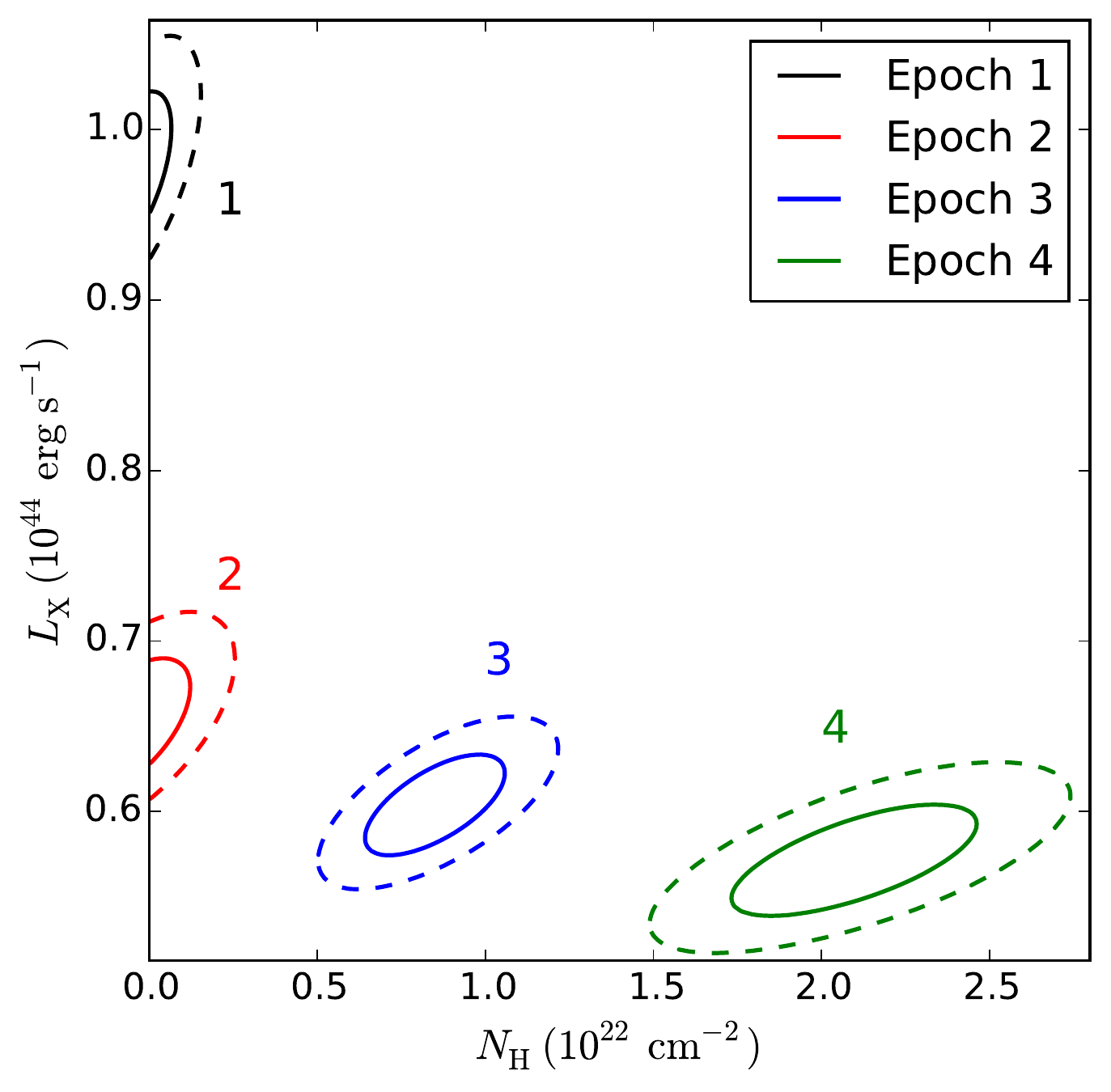} \\
\caption{
Upper panels: unfolded (i.e., intrinsic) spectra for \xidone, \xidtwo, 
and \xidthr, respectively. The solid lines indicate best-fit model 
photon flux. Different colors indicate different epochs, and the epoch 
indexes are also labeled beside the corresponding spectra.
Data are binned for display purposes only. 
Middle panels: the corresponding ratios of observed and model 
photon-flux density.
Lower panels: the corresponding \lx-\nh\ confidence contours.
\xidone\ has significantly higher luminosity in epoch 2.
The luminosity of \xidtwo\ drops by a large factor (\hbox{$\approx 3$}) since 
epoch 3. For \xidthr, the spectral shape changes obviously among epochs.
}
\label{fig:spec_fit}\label{fig:contour}
\end{tabular}
\end{figure*}

As illustrative examples, we investigate 
three significantly variable sources. 
The first source, \xidone\ (\hbox{$z=1.03$}), is 
our brightest source; it also has the most significant 
\lx\ variability. 
The second source, \xidtwo\ (\hbox{$z=0.42$}), is
also \lx-variable, but has counts typical in our sample;
this source is shown as a representative sample member.
The third source, \xidthr\ (\hbox{$z=1.21$}), 
has the most significant \nh\ 
variability; it transitions from an \xray\ unobscured 
to obscured state. 
Their fitted spectra and \hbox{\lx--\nh} confidence contours 
are shown in Figure~\ref{fig:spec_fit}. 

\Xidone\ has the largest number of total net counts 
(\hbox{$\approx$11000}). It has 
\hbox{$\Delta \mathrm{AIC}_L=1059$}\ and 
\hbox{$\Delta \mathrm{AIC}_N=15$}, 
indicating both \lx\ and \nh\ variability. 
The $\Delta \mathrm{AIC}_L$\ is the largest among those 
of our sources.
Its \lx\ in the second epoch is about two times the 
\lx\ in the other epochs. Its \nh\ is generally low: 
\nh\ only has upper limits
in the first three epochs and rises
to \hbox{$\approx 4\times 10^{21}\ \mathrm{cm^{-2}}$} in the last epoch. 
\xidone\ and another bright (net counts\ \hbox{$\approx 6500$}) 
and variable source (\hbox{$\Delta \mathrm{AIC}_L=38$}, 
\hbox{$\Delta \mathrm{AIC}_N=5.1$}),
\hbox{J033218.3--275055}, are also identified as optically 
variable sources by \cite{falocco15} based on their $r$-band
variability.

\Xidtwo\ has total net counts of \hbox{$\approx$1300}, 
similar to the median counts of our sample (1399). 
It has \hbox{$\Delta \mathrm{AIC}_L=165$}, indicating
significant \lx\ variability. 
Its \lx\ values in the first two epochs are about three times 
higher than those in the last two epochs. 
It is \xray\ obscured (i.e., \hbox{\nh$>10^{22}$~cm$^{-2}$}) with no 
significant \nh\ variability detected (\hbox{$\Delta \mathrm{AIC}_N=-0.9$}).

\begin{figure}
\includegraphics[width=\linewidth]{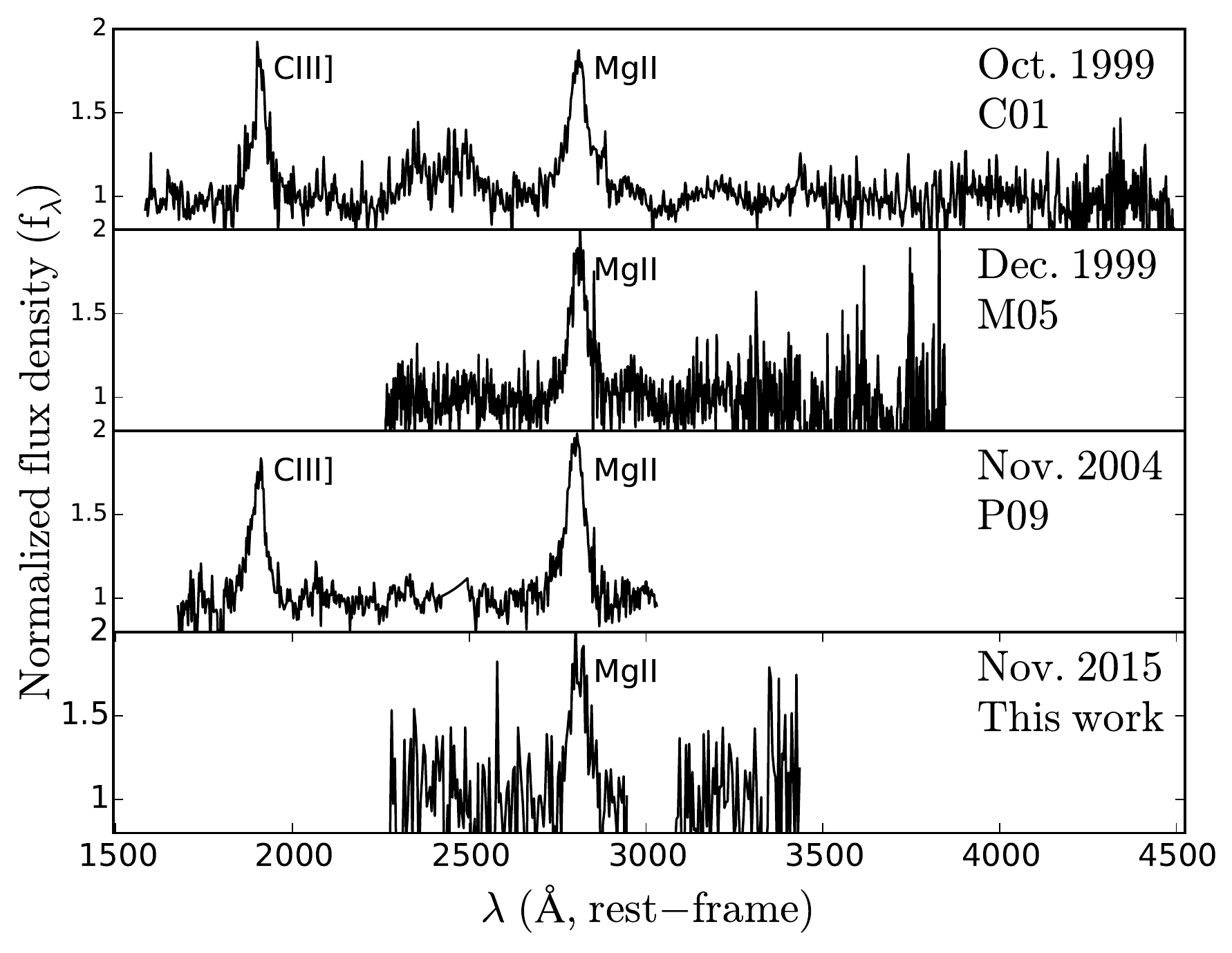}
\caption{Normalized optical spectra of \xidthr\ at \hbox{$z=1.21$}. 
The observation dates and the references are listed in the 
upper-right corner of each panel 
(C01: \citealt{croom01}; M05: \citealt{mignoli05}; 
P09: \citealt{popesso09}). 
The Mg\,{\sc ii}\ $\lambda$2798\ broad emission line
is present in all four spectra.
} 
\label{fig:optical_spec_xid508}
\end{figure}

\Xidthr\ has \hbox{$\Delta \mathrm{AIC}_L=144$}\ and 
\hbox{$\Delta \mathrm{AIC}_N=83$}\ with a total 
net counts of \hbox{$\approx 5000$}.
It is the most significantly \nh\ variable source 
(i.e., has maximum $\Delta \mathrm{AIC}_N$).
The \nh\ values in the first two epochs are low 
(consistent with zero). The \nh\ rises in \hbox{epoch 3} and reaches
\hbox{$\approx 2\times 10^{22}\rm\ cm^{-2}$} in \hbox{epoch 4}. Its \lx\ in
\hbox{epoch 1} is \hbox{$\approx 2$}\ times higher than \lx\ in the other 
epochs. 
The \xray\ type transition happens between epochs 2 and 4, 
and thus corresponds to a rest-frame time scale 
\hbox{$t_{\rm tran} \sim 3$~year}.
If we interpret the transition as an ``eclipse'' event, this long time 
scale indicates the eclipsing material is located at a distance larger
than that of the BLR from the central engine. This is because 
BLR-cloud eclipses are likely to happen on much shorter 
time scales (hours to days; e.g., \citealt{maiolino10, wang10}).  
Assuming that the eclipsing material is in a single ``cloud'' within 
the inner torus region, its distance ($r$) 
from the central engine is \hbox{$\sim 0.1$~pc}.\footnote{
The distance is estimated from the empirical relation 
between the inner torus radius and \xray\ luminosity obtained from dust 
reverberation-mapping studies \citep{koshida14}.}
Applying Kepler's 3rd law, we can calculate its orbital period
\begin{equation}\label{equ:t_orbit}
\begin{split}
t_{\mathrm{orbit}} &= 2\pi \frac{r^{\frac{3}{2}}}{ (M_\mathrm{BH} G)^{\frac{1}{2}} } \\
  &\sim 300 
	\left( \frac{M_\mathrm{BH}}{10^8\ M_{\odot}} \right)^{-\frac{1}{2}}\ \mathrm{yr}.
\end{split}
\end{equation}
Then we can estimate the angular size of the cloud, 
$\theta$\ (viewed from the central SMBH), as 
\begin{equation}\label{equ:theta}
\begin{split}
\theta &\sim \frac{t_{\rm tran}}{t_\mathrm{orbit}} \times 360^{\circ} \\
       &\sim  4^\circ 
		\left( \frac{M_\mathrm{BH}}{10^8\ M_{\odot}} \right)^{\frac{1}{2}}.
\end{split}
\end{equation}
To investigate if the optical spectral type also changes, 
we have compiled three optical spectra from the literature
\citep{croom01, mignoli05, popesso09} 
and obtained a new spectrum on November 8, 2015. The 
new observation was performed using the IMACS Short-Camera of the
6.5m Magellan Telescope.
The four spectra are presented in Figure~\ref{fig:optical_spec_xid508}.
In all four spectra, the broad Mg\,{\sc ii}\ $\lambda$2798 line is detected. 
In the first and third spectra, the C\,{\sc iii]}\ $\lambda$1909 line is also 
detected.\footnote{We do not perform 
quantitative analyses, since all of the spectra cannot be reduced uniformly.}
Thus, the optical spectral type of this source remains type I in all 
four spectra. 
The lack of optical spectral-type transitions indicates 
the \xray\ eclipsing material is not large enough to block 
most emission from the BLR.
This result is consistent with recent studies of optical spectral-type 
transition AGNs that suggest significant changes (\hbox{$\sim 10$~times}) 
in luminosity as the main cause of transitions 
\citep[e.g.,][]{lamassa15, runnoe16}.

\bibliography{all.bib}
\bibliographystyle{apj}
\end{document}